\documentclass{article}

\usepackage{amsthm}
\usepackage{amsmath}
\usepackage{amssymb}
\usepackage{bbm}
\usepackage{braket}
\usepackage{enumerate}
\usepackage{stmaryrd}
\usepackage{listliketab}
\usepackage{graphicx}
\usepackage{pstricks}
\usepackage{pst-solides3d}
\usepackage{pst-poly}
\usepackage{makeidx}
\usepackage{pst-3dplot}
\usepackage[linktocpage]{hyperref}

\DeclareMathOperator{\tr}{tr}
\DeclareMathOperator{\spa}{span}
\DeclareMathOperator{\Herm}{Herm}

\DeclareMathOperator{\cone}{cone}
\DeclareMathOperator{\conv}{conv}
\DeclareMathOperator{\ext}{ext}
\DeclareMathOperator{\aff}{aff}
\DeclareMathOperator{\supp}{supp}

\newcommand{\rst}[1]{\ensuremath{{\mathbin\upharpoonright}%
\raise-.5ex\hbox{$#1$}}}

\newtheoremstyle{thm}
     {10pt}
     {10pt}
     {}
     {}
     {\bfseries}
     {:}
     {.5em}
     {}

\theoremstyle{thm}
\newtheorem{thm}{Theorem}
\newtheorem{prop}[thm]{Proposition}

\newtheorem{lemma}[thm]{Lemma}
\newtheorem{cor}[thm]{Corollary}
\newtheorem*{notation}{Notation}

\newtheoremstyle{defi}
     {10pt}
     {10pt}
     {\itshape}
     {}
     {\bfseries}
     {:}
     {.5em}
     {}

\theoremstyle{defi}

\newtheorem{defi}[thm]{Definition}
\newtheorem{ass}{Assumption}

\newtheoremstyle{princ}
     {10pt}
     {10pt}
     {\itshape}
     {}
     {\bfseries}
     {:}
     {.5em}
     {}

\theoremstyle{princ}

\newtheorem{post}{Postulate}

\newtheoremstyle{ex}
     {10pt}
     {10pt}
     {}
     {}
     {\bfseries}
     {:}
     { }
     {}

\theoremstyle{ex}

\newtheorem{ex}[thm]{Example}

\numberwithin{equation}{section}
\numberwithin{thm}{section}
\numberwithin{figure}{section}

\makeatletter
\newcommand\ackname{Acknowledgements}
\if@titlepage
  \newenvironment{acknowledgements}{%
      \titlepage
      \null\vfil
      \@beginparpenalty\@lowpenalty
      \begin{center}%
        \bfseries \ackname
        \@endparpenalty\@M
      \end{center}}%
     {\par\vfil\null\endtitlepage}
\else
  \newenvironment{acknowledgements}{%
      \if@twocolumn
        \section*{\abstractname}%
      \else
        \small
        \begin{center}%
          {\bfseries \ackname\vspace{-.5em}\vspace{\z@}}%
        \end{center}%
        \quotation
      \fi}
      {\if@twocolumn\else\endquotation\fi}
\fi
\makeatother

\makeindex

\begin{document}

\title{{\huge\bf One simple postulate implies that every polytopic state space is classical}\\[2cm]}
\author{{\LARGE Corsin Pfister*}\\[1cm]}
\date{{\Large Master Thesis\\Institute for Theoretical Physics\\ETH Z\"urich\\[0.7cm]
Conducted at\\[0.1cm]
Centre for Quantum Technologies\\
National University of Singapore\\[0.8cm]
Supervisors:\\[0.3cm]
Prof. Renato Renner\\[0.1cm]}
{\normalsize ETH Z\"urich}\\[0.1cm]
{\Large Prof. Stephanie Wehner\\[0.1cm]}
{\normalsize National University of Singapore\\[1cm]}
{\large December 22, 2011}\\[0.9cm]
\normalsize{*mail@corsinpfister.com}}

\maketitle
\thispagestyle{empty}


\newpage
\thispagestyle{empty}
\quad
\newpage
\thispagestyle{empty}

\begin{abstract}
Quantum theory shows many surprising features like the uncertainty principle, entanglement or nonlocality. In order to understand these features, several attempts have been made to formulate quantum theory within a more general framework of probabilistic theories. Such a framework allows to formulate postulates and study their consequences in a general setting. In the past, generalized probabilistic theories have mostly been studied to understand the nonlocality of quantum theory.

This thesis approaches quantum theory from a different perspective. It is dedicated to the study of the consequences of postulates concerning post-measurement states. This aspect of generalized probabilistic theories has gained very little attention in the literature so far. As the main result of this thesis, we show that one very simple postulate rules out all probabilistic theories with a polytopic state space except for classical theory. This postulate states that if the outcome of a measurement can be predicted with certainty, then this measurement does not alter the state, i.e. the post-measurement state coincides with the initial state. Since this postulate is satisfied by quantum theory, this result gives a partial answer to the question which physical principles distinguish quantum theory from other probabilistic theories.

To develop an understanding of this main result and of generalized probabilistic theories in general, we give an introduction to a framework which has been called the abstract state space formalism. Such an introduction has not been provided by the literature so far. This comprises the proof of many properties of convex sets and abstract state spaces. In particular, the characteristics of physical theories with a polytopic state space are investigated. As a side result, we show that within polytopic theories, classical theory can be characterized by three postulates which we will call repeatability, the subspace principle and the state discrimination principle.
\end{abstract}

\newpage
\thispagestyle{empty}
\quad
\newpage
\thispagestyle{empty}

\begin{acknowledgements}
I thank Prof. Renato Renner for supporting me and for giving me the opportunity to write this thesis by getting me in touch with Prof. Stephanie Wehner. Big thanks goes to Prof. Stephanie Wehner for inviting me to the National University of Singapore to write this thesis and for her instructive and very friendly supervision during my four months in Singapore. I thank Christian Gogolin and Paolo Perinotti for insightful discussions. I also thank Esther H\"anggi for proofreading. Finally, I thank my family for supporting me and for encouraging me in my efforts.
\end{acknowledgements}

\newpage
\thispagestyle{empty}
\quad
\newpage

\thispagestyle{empty}
\tableofcontents
\thispagestyle{empty}

\newpage

\setcounter{page}{1}

\section{Introduction}

\subsection{The role of generalized probabilistic theories}
\label{the-role-section}

Quantum theory has many physical features which, from an everyday life point of view, are very surprising: Heisenberg's uncertainty principle, the superposition principle, entanglement, nonlocality and contextuality, to name but a few. When a physicist is asked for an explanation of these features, the only honest answer he can give is that they arise from the mathematical structure of quantum theory. He can mention a few comparatively weak motivations for the mathematical framework of quantum theory, but he cannot fully derive it from physical principles: he cannot say \emph{why} quantum theory is the way it is.

To get a better understanding of this situation, it is helpful to compare quantum theory to a physical theory which does not have this problem. Special relativity can be treated in two different ways. The first way is to start with the mathematical definition of the Minkowski spacetime. Then all physical features of special relativity, like the frame-independence of the speed of light and the principle of relativity, arise from this mathematical structure of spacetime. This way of dealing with special relativity is comparable to the way in which quantum theory is treated. The starting point of the theory is a mathematical framework, and physics is deduced from this mathematical structure. The advantage of this approach is that it is a very clear way of formulating a theory. This is, however, not the natural way of deriving a physical theory, and historically, this is not the way special relativity was discovered. The second, more natural way to deal with special relativity is to start with \emph{physical} postulates. In this approach, there are no initial assumptions about the specific mathematical structure of spacetime. Instead, the mathematics of special relativity are derived from the physical postulates of the invariance of the speed of light and the principle of relativity. The advantage of this approach is that all explanations of physical features of the theory can be based on \emph{physical} assumptions that underlie the theory.

The unsatisfactory characteristic of quantum theory is that it lacks such a second approach. There is no commonly accepted complete derivation of the mathematical structure of quantum theory, based on undeniable, purely physical postulates. In the recent past, several attempts to fill this gap have been made, e.g. \cite{Hardy:2001p302} \cite{Masanes:2011p12058} \cite{DAriano:2010p12767}. They provide interesting insights concerning the question which aspects of quantum theory could be regarded as being fundamental, and they might be a big step towards a physical derivation of the mathematics of quantum theory. However, in seeking a full derivation of quantum theory, the assumptions that are made to achieve this goal are not beyond any doubt. They exhibit several deficits: some of them are more of a mathematical than of a physical nature, others seem to be rather arbitrary and unmotivated, and others again are very strong, assuming far-reaching principles instead of deriving them from weaker assumptions.

Instead of seeking a full derivation of quantum theory, interesting insights in partial aspects of quantum theory can be gained by considering less powerful assumptions about physical theories which are not intended to imply the framework of quantum theory. A way of thinking which has become more and more important in this issue is to consider quantum theory, or physical theories in general, from an information theoretical point of view. This approach suggests considering quantum theory in a broader context of probabilistic theories. The idea is that one should start with as few assumptions about the concrete nature of the theory as possible, assuming only that the theory is probabilistic. This means that in such a theory, the combination of a state and a measurement is not enough to predict a measurement outcome with certainty. Instead, it gives a probability distribution over the outcomes. This gives rise to a comparatively weak mathematical structure. This mathematical framework has occasionally been called the framework of \emph{generalized probabilistic theories}. Using this probabilistic framework as a basis, one can then make further assumptions about the concrete properties of the theory.

Generalized probabilistic theories have been considered in different contexts, provided with different additional structures depending on which particular aspect of quantum or classical theory is investigated. For example, one aspect of quantum mechanics which attracts much attention is nonlocality. A simple example of a theory in the framework of generalized probabilistic theories which is dedicated to the study of nonlocality is the theory of the PR-box (named after Popescu and Rohrlich), also called nonlocal box \cite{Popescu:1994p5991}. In this context, the additional structure under investigation is the mathematical rule of assigning multipartite state spaces and measurements to systems of multiple constituents. This allows for the study of features like steering and teleportation in a more general setting. Other aspects under consideration are uncertainty relations and entropy measures or the possibility of cryptographic and information processing tasks.

\subsection{The ideas of this thesis}
\label{thesis-idea-section}

The ideas presented in this thesis arose from the attempt to describe consecutive measurements (and therefore post-measurement states) in generalized probabilistic theories. This aspect has gained little attention so far. The motivation for a further investigation of this aspect is that it seems that there is no straightforward definition of post-measurement states in generalized probabilistic theories. In fact, it turns out that requiring rather simple conditions about post-measurement states rule out a broad class of generalized probabilistic theories already.

The focus of this thesis is on a particular class of generalized probabilistic theories which we call polytopic theories. There are two good reasons for restricting ones attention to polytopic theories. The first reason is that they are technically easier to deal with. Another reason which makes this class of theories attractive to deal with is the fact that most toy theories that have been ``invented'' for the study of generalized probabilistic theories belong to this class. A shortcoming of this restriction, however, is that quantum theory is not a polytopic theory. Nonetheless, all properties of generalized probabilistic theories considered in this thesis are satisfied by quantum theory. Therefore, quantum theory belongs to a subclass of generalized probabilistic theories satisfying these properties, whereas most polytopic theories do not.

We present the principles considered in this thesis in two groups. The first group consists of three principles. The first is that measurements are repeatable.\footnote{More precisely, we only assume repeatability for a particular type of measurements which we call \emph{pure} measurements. For more details, see Sections \ref{gpt-section} and \ref{result-1-section}.} This means that if we perform a measurement twice (and we do not assume any intermediate dynamics which is not due to the measurement), then we will get the same outcome. The second principle of this group assumes that the set of all possible states after a measurement shows a certain subspace structure. The third principle is what we call the state discrimination principle. Roughly speaking, it states the following. Suppose that $\Lambda_1$ and $\Lambda_2$ are sets of states. Assume that we can perfectly distinguish $\Lambda_1$ from $\Lambda_2$ by a measurement. In addition, assume that two subsets $\Lambda_3, \Lambda_4 \subset \Lambda_2$ are such that we can also perfectly distinguish $\Lambda_3$ from $\Lambda_4$ by a measurement. The state discrimination principle states that in this case, we can perfectly distinguish between the sets $\Lambda_1, \Lambda_3$ and $\Lambda_4$ by a measurement. We will show that the only polytopic theories which obey these three principles have a simplex structure, and therefore coincide with classical theory.

The second group of principles consists of only \emph{one} very simple principle. It states that if we know the outcome of a particular measurement in advance with certainty, then we can perform this measurement without altering the statistics of any subsequent measurement. In other words, if the state of a system has probability one for an outcome of a particular measurement, then performing this measurement does not disturb the state of the system: the post-measurement state of the system coincides with the initial state. As the main result of this thesis, we will show that surprisingly, this seemingly weak assumption rules out all polytopic theories except for the classical theories (i.e. the theories where the states form a simplex).

\subsection{Overview}

This thesis is organized in two parts. In Part \ref{introduction-part}, we give an introduction to convex sets and to generalized probabilistic theories and we develop most of the techniques that we use in this thesis. In Part \ref{result-part}, we apply these techniques to infer the results of this thesis.

Part \ref{introduction-part} is structured as follows. Section \ref{convex-sets-section} is an introduction to the mathematics of convex sets and their interpretation in physical theories. Although we also talk about convex sets in general, we will particularly focus on the study of polytopes, since this is the kind of convex set which will be important in Part \ref{result-part} of this thesis. We will infer many properties of polytopes which will be important in the proofs of the results. Section \ref{gpt-section} is an introduction to generalized probabilistic theories. We will introduce the mathematics necessary to treat physical theories in a generalized probabilistic framework. Then we will infer a particular framework for generalized probabilistic theories which is called the \emph{abstract state space} formalism. This framework will be illustrated by examples of theories.

In Part \ref{result-part}, we apply the techniques developed in Part \ref{introduction-part} to infer the results of this thesis. As we mentioned in Section \ref{thesis-idea-section}, our results split into two parts. Both parts particularly address to polytopic theories, i.e. theories where the set of states is a polytope. In Section \ref{result-1-section}, we show that every polytopic theory that satisfies repeatability, a subspace principle and a state discrimination principle is a classical theory. Section \ref{result-2-section} is dedicated to the main result of this thesis. It infers classical theory from polytopic theory from only one simple postulate. This postulate states that every measurement for which the outcome can be predicted with certainty does not alter the state.

We will conclude this thesis by some remarks and an outlook on possible generalizations in Section \ref{outlook-section}.

\newpage

\part{Introduction to the framework and the derivation of the techniques}
\label{introduction-part}

\section{Convex sets}
\label{convex-sets-section}

Instead of starting with an introduction to the framework of generalized probabilistic theories, we introduce convex sets first. This gives an advantage. When we introduce generalized probabilistic theories in Section \ref{gpt-section}, where the sets of states are given by convex sets, we can refer to a variety of examples that we introduce in this section. To avoid dealing with convex sets without a physical motivation in mind, we explain in Section \ref{randomization-section} how convex sums naturally arise in the context of random processes and how they give rise to convex subsets of vector spaces. In Section \ref{abstract-convexity-section}, we discuss how the notion of convexity could be generalized and what makes convex subsets of vector spaces special in this more general context. Section \ref{general-convex-section} is dedicated to the study of some aspects of convex sets in general as far as they are important for the present thesis. In Section \ref{polytope-section}, we will focus on a particular class of convex sets called polytopes. In part \ref{result-part} of this thesis, we will consider generalized probabilistic theories whose sets of states are polytopes.

\subsection{Probabilistic mixtures and convex sums}
\label{randomization-section}

In generalized probabilistic theories, the set of states is commonly assumed to be a convex subset of a real vector space.\footnote{Given that a quantum mechanical Hilbert space $\mathcal{H}$ is a complex vector space, it might be confusing that the vector space is assumed to be real. However, the unit vectors of a quantum mechanical Hilbert space $\mathcal{H}$ only encompass the \emph{pure} states. In full generality, the quantum state has to be treated as a \emph{density operator}. The set of density operators $\mathcal{S}(\mathcal{H})$ is contained in the \emph{real} vector space $\Herm(\mathcal{H})$ of Hermitian operators on $\mathcal{H}$ (c.f. Example \ref{set-of-density-operators}). The reader unfamiliar with the concept of a density operator is referred to \cite{Nielsen-Chuang}.} As we will see in Section \ref{general-convex-section}, this is a subset $C$ of a vector space $V$ such that for every two elements $x$ and $y$ of $C$, the line segment\index{line segment} $L_{x, y} := \{ \lambda x + (1-\lambda) y \mid 0 \leq \lambda \leq 1 \}$ which connects $x$ and $y$ is contained in $C$ as well. The goal of this subsection is to motivate this assumption.

Suppose that we are given a physical system with initial state $\omega$. Assume that it undergoes a random process $\mathcal{R}$ and that the state of the system after the random process depends on the outcome $k$ of that random process. Subsequently, we perform a measurement $\mathcal{M}$ on the system. To describe the statistics of the measurement outcomes, we need a description of the state of the system prior to the measurement, i.e. after the random process. However, it might be the case that we do not know the outcome $k$ of the random process (on which the state of the system depends). But if we know the statistics of the random process, i.e. the probabilities $p_k$ of the outcomes $k$ of $\mathcal{R}$, we can still make predictions about the outcome of the subsequent measurement by describing the state of the system as the \emph{probabilistic mixture} of the states $\{ \omega_k \}_k$.

As an example, consider a quantum system with an initial state which is described by a density operator $\rho \in \mathcal{S}(\mathcal{H})$. The random process $\mathcal{R}$ prior to the measurement $\mathcal{M}$ might be a measurement as well. Say that this measurement is described by a projective POVM $\mathcal{R} = \{P_k\}_k$. The probability of getting the outcome $k$ is given by $p_k = \tr(P_k \rho)$. In this case (if the outcome is $k$), after the $\mathcal{R}$-measurement, the system is in the state
\begin{align}
\rho_k = \frac{P_k \rho P_k}{\tr(P_k \rho)} \,. \nonumber
\end{align}
If the subsequent measurement $\mathcal{M}$ is described by the POVM $\mathcal{M} = \{Q_l\}_l$, the probability of getting the outcome $l$, conditioned on the outcome $k$ of the $\mathcal{R}$-measurement, is given by
\begin{align}
p_{l|k} = \tr(Q_l \rho_k) = \tr\left( Q_l \frac{P_k \rho P_k}{\tr(P_k \rho)} \right) \,. \nonumber
\end{align}
If we would \emph{not} know the outcome of the random process, i.e. the outcome $k$ of the $\mathcal{R}$-measurement, the probability that we would assign to the outcome $l$ of the $\mathcal{M}$-measurement would be the probabilistic mixture of the probabilities $p_{l|k}$,
\begin{equation}
\label{linear-equation}
p_l = \sum\limits_k p_k p_{l|k} = \sum\limits_k p_k \tr(Q_l \rho_k) \,.
\end{equation}
The linearity of equation (\ref{linear-equation}) allows us to represent $p_l$ as
\begin{align}
p_l = \sum\limits_k p_k \tr(Q_l \rho_k) = \tr \left( Q_l \left( \sum\limits_k p_k \rho_k \right) \right) = \tr(Q_l \widetilde \rho)\,, \nonumber
\end{align}
where
\begin{equation}
\label{mixture-state}
\widetilde \rho = \sum\limits_k p_k \rho_k \,.
\end{equation}
The state $\widetilde \rho$ is the probabilistic mixture of the states $\{ \rho_k \}_k$ with probabilities $\{ p_k \}_k$. It is easily verified that every operator of the form (\ref{mixture-state}) is a density operator. Any state which is a non-trivial mixture of other states is a \emph{mixed} state.

In a more general framework, we might assume that a physical theory describes the set of states $\Omega$ of a system as a subset of some real vector space $A$ (like the set $\mathcal{S}(\mathcal{H})$ of density operators on a Hilbert space $\mathcal{H}$ is a subset of the real vector space $\Herm(\mathcal{H})$ of Hermitian operators on $\mathcal{H}$, c.f. Example \ref{set-of-density-operators}). The probabilistic mixture of states $\{ \omega_k \}_k \subset \Omega$ with respect to probabilities $\{ p_k \}_k$ would then be given by
\begin{equation}
\label{mixture-state-2}
\widetilde \omega = \sum\limits_k p_k \omega_k, \quad \text{where} \quad \omega_k \in \Omega, \quad \sum\limits_k p_k = 1 \,.
\end{equation}
A linear combination of the form (\ref{mixture-state-2}) is called a \emph{convex sum} of the elements $\{ \omega_k \}_k$ of $\Omega$. In order to treat the mixture $\widetilde \omega$ as a state, we need the consistency requirement that $\widetilde \omega \in \Omega$. Since we might think of random processes with any probabilities $\{ p_k \}_k$ which prepare any states $\{ \omega_k \}_k$, we require that any mixture of the form (\ref{mixture-state-2}) is an element of $\Omega$. A subset $\Omega$ of a vector space $A$ which has this property is called a $\emph{convex subset}$ of the vector space.

The requirement that convex sums of arbitrarily (but finitely) many elements of the set $\Omega$ have to be contained in $\Omega$ can be reduced to the requirement that the convex sum of only two elements has to be contained in $\Omega$. Clearly, if $\Omega$ is a convex subset of $A$, then any convex sum of two elements of $\Omega$ is again an element of $\Omega$. On the other hand, suppose that $\Omega$ has the property that for any two elements of $\Omega$, any convex sum of the two elements is again an element of $\Omega$. Then, any convex sum of three elements is an element of $\Omega$ as well:
\begin{align}
&\alpha, \beta, \gamma \in [0,1], \quad \alpha + \beta + \gamma = 1, \quad \omega, \sigma, \nu \in \Omega \nonumber \\
\Rightarrow \quad &\alpha \omega + \beta \sigma + \gamma \nu = (1-\gamma) \underbrace{\left( \frac{\alpha}{1-\gamma} \omega + \frac{\beta}{1-\gamma} \sigma \right)}_{\in \Omega \text{ since } \frac{\alpha}{1-\gamma} + \frac{\beta}{1 - \gamma} = 1} + \gamma \nu \in \Omega \,. \nonumber
\end{align}
This argument extends to convex sums of arbitrarily many elements of $\Omega$. Therefore, we can characterize a convex subset $C$ of a real vector space $V$ by the property that for any two elements $x$ and $y$ of $C$, the line segment\index{line segment} $L_{x, y} = \{ \lambda x + (1-\lambda) y \mid 0 \leq \lambda \leq 1 \}$ which connects $x$ and $y$ is contained in $C$ as well. We will give this property a geometric picture in Section \ref{general-convex-section}.

\subsection{More abstract notions of convexity}
\label{abstract-convexity-section}

It is very common to assume that probabilistic mixtures of states are given by convex sums. In a more general setting, however, it might be that the set of states is not a subset of a vector space, so that probabilistic mixtures cannot be expressed by linear combinations (a convex sum of the form (\ref{mixture-state-2}) is a linear combination). It is interesting to examine how probabilistic mixtures could be generalized to this more general case. In other words, one might ask how restrictive it is to assume that probabilistic mixtures are given by convex sums. To this end, we recapitulate the ingredients that we put together to get a probabilistic mixture. We have a tuple $(\omega_k)_k$ of states (where all states belong to a common set of states $\Omega$) and a tuple $(p_k)_k$ of probabilities. These ingredients are combined to form a mixture $\widetilde \omega$ of states (for the moment, we do not assume anything about \emph{how} they are combined). If we assume that the random process has $n$ possible outcomes, we can regard this as an operation
\begin{align}
\begin{array}{cll}
\Omega^n \times \Delta_n & \rightarrow & \Omega \\
((\omega_1, \ldots, \omega_n), (p_1, \ldots, p_n)) & \mapsto & \widetilde \omega
\end{array} \nonumber
\end{align}
where
\begin{align}
\Delta_n := \left\{ (p_1, \ldots, p_n) \in [0,1]^n \ \middle\vert \ \sum\limits_{i = 1}^n p_i = 1 \right\} \,. \nonumber
\end{align}
As before in Section \ref{randomization-section} with convex sums, if we do not consider any position in this operation as being distinguished from the others, we can regard this $n$-ary operation as emerging from a set of binary operations
\begin{align}
\begin{array}{llll}
\{cc_\lambda\}_{\lambda \in [0,1]}: & \Omega \times \Omega & \rightarrow & \Omega \\
& (\omega_1, \omega_2) & \mapsto & \widetilde \omega
\end{array} \nonumber
\end{align}
which satisfy some compatibility requirements which allow for concatenating the binary operation in an associative way to get an $n$-ary operation. This idea is captured by the following definition.

\begin{defi}[\cite{Fritz:2009p12357}]
\label{convex-space-def}
A \textbf{convex space}\index{convex space} is a set $X$ equipped with a family $\{cc_\lambda\}_{\lambda \in [0,1]}$ of maps
\begin{align}
cc_\lambda: X \times X \rightarrow X \,, \nonumber
\end{align}
which is called the \textbf{convex combination}, satisfying the following conditions:
\begin{align}
&\bullet \quad cc_0(x,y) = y &&\forall x, y \in X \,, \label{cs1} \\
&\bullet \quad cc_\lambda(x,x) = x &&\forall x \in X, \forall \lambda \in [0,1] \,, \label{cs2} \\
&\bullet \quad cc_\lambda(x,y) = cc_{1-\lambda}(y,x) &&\forall x, y \in X, \forall \lambda \in [0,1] \,, \label{cs3} \\
&\bullet \quad cc_\lambda(cc_\mu(x,y),z) = cc_{\tilde \lambda}(x,cc_{\tilde \mu}(y,z)) &&\forall x, y, z \in X, \forall \lambda, \mu \in [0,1] \label{cs4} 
\end{align}
with
\begin{align}
\tilde \lambda = \lambda \mu \,, \quad \tilde \mu = \begin{cases}
\frac{\lambda (1-\mu)}{1 - \lambda \mu}  & \text{if }\lambda \mu \neq 1, \\
\text{arbitrary}, & \text{if }\lambda = \mu = 1.
\end{cases} \nonumber
\end{align}
\end{defi}

It is not hard to convince oneself of the fact that if $X$ is a convex subset of a real vector space, then the binary convex sum
\begin{align}
cc_\lambda(x, y) \mathrel{\widehat{=}} \lambda x + (1-\lambda) y \label{cc-equiv}
\end{align}
satisfies the properties (\ref{cs1}) -- (\ref{cs4}) and is therefore a convex space in the sense of Definition \ref{convex-space-def}. However, it turns out that there are convex spaces that cannot be realized as a convex subset of a vector space. Within a physical interpretation, they have, in a certain sense, a possibilistic rather than a probabilistic structure. They do not provide a quantitative measure for how likely it is that an event occurs but only give a qualitative ``yes or no''-structure which says whether or not an event is possible. For example, the two-element set $\{i, f\}$ together with the operation
\begin{align}
cc_\lambda(i, f) = \begin{cases} f & \text{if } \lambda = 0 \\ i & \text{if } \lambda \neq 0 \end{cases} \nonumber
\end{align}
satisfies all the axioms (\ref{cs1}) -- (\ref{cs4}) for a convex space. This example looks pathological, but it arises as a special case of a more natural class of spaces of a combinatorial or possibilistic type. We will not discuss such spaces here, since we are interested in theories that give us quantitative predictions about the probability of events and are therefore of a probabilistic nature. For examples and a detailed discussion of possibilistic spaces, we refer to \cite{Fritz:2009p12357}.

Instead, we want to attend to the question of how convex subsets of real vector spaces can be distinguished from other types of convex spaces. This question has a mathematically clear answer provided by the following theorem.

\begin{thm}[Stone\footnote{This theorem is originally by Stone \cite{Stone:1949p12477}. The version presented here is a modified, more modern version by Capraro and Fritz \cite{Capraro:2011p13170}.} \cite{Stone:1949p12477}, see \cite{Capraro:2011p13170}]
\label{stone-thm}
A convex space embeds into a real vector space with (\ref{cc-equiv}) if and only if the following cancellation property holds:
\begin{equation}
\label{cancellation}
cc_\lambda(x,y) = cc_\lambda(x,z) \quad \text{with} \quad \lambda \in (0,1) \quad \Longrightarrow \quad y = z \,.
\end{equation}
\end{thm}

This gives us an explicit criterion which separates convex subsets of vector spaces from other convex spaces. Applied to state spaces of generalized probabilistic theories, it seems that there is no immediate physical interpretation of the cancellation property (\ref{cancellation}). Nonetheless, it is good to be aware of the fact that from a very abstract point of view, the assumption that convexity is represented by convex subsets of vector spaces causes an (arguably small) loss of generality.

In the following, we will always assume that the set of states in a physical theory is a convex subset of a real vector space. We will not refer to the more general notion of convexity of Definition \ref{convex-space-def} anymore. Whenever we will talk about convexity, we refer to convex subsets of real vector spaces, which we will often simply call convex sets. Therefore, when we say convex combination, we mean a convex sum.

\subsection{Convex subsets of vector spaces}
\label{general-convex-section}

In this section, we introduce some general aspects of convex subsets of vector spaces. The presentation of convex sets that we give here is not to be understood as a standard introduction to the field. Instead, we discuss some aspects and prove some properties of convex sets which are important for our particular purpose.

\begin{defi}
A subset $C$ of a real vector space $V$ is a \textbf{convex subset}\index{convex subset|see{convex set}} or \textbf{convex set}\index{convex set} if $x, y \in C$ implies $\lambda x + (1-\lambda) y \in C$ for all $0 \leq \lambda \leq 1$.
\end{defi}

\begin{figure}[htb]
\centering

\begin{pspicture}[showgrid=false](-1,-0.8)(1,1)
\PstTriangle[PstPicture=false]
\end{pspicture}
\begin{pspicture}[showgrid=false](-1,-1)(1,1)
\PstSquare[PstPicture=false]
\end{pspicture}
\begin{pspicture}[showgrid=false](-1,-1)(3,1)
\pscircle[PstPicture=false](0,0){0.8}
\end{pspicture}
\begin{pspicture}[showgrid=false](-1,-1)(1,1)
\pswedge{1}{-145}{145}
\psdot(-0.6,-0.6)
\psdot(-0.6,0.6)
\psline(-0.6,-0.6)(-0.6,0.6)
\uput[0](-0.6,-0.6){$x$}
\uput[0](-0.6,0.6){$y$}
\uput[180](-0.6,0){$L_{x,y}$}
\end{pspicture}

\caption{The triangle, the square and the circular disk are convex, but the circular section depicted on the right is not.}
\label{convex-nonconvex}
\end{figure}
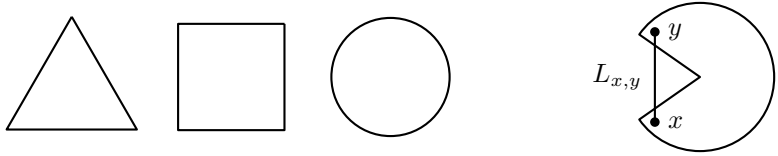

\label{line-segment}The property of being a convex set has a very simple geometric interpretation (see Figure \ref{convex-nonconvex}): For any two points $x, y \in C \subset V$, the \textbf{line segment}\index{line segment}
\begin{align}
L_{x, y} := \{ \lambda x + (1-\lambda) y \mid 0 \leq \lambda \leq 1 \} \nonumber
\end{align}
connecting the two points is contained in $C$ as well. By what we have discussed at the end of Section \ref{randomization-section}, this also implies that \emph{any} convex sum of elements in $C$ is contained in $C$ (and not just binary convex sums). It is easy to see that the intersection of two convex sets is convex.

One of the very central notions in the study of convex sets is the notion of \emph{extreme points}. The extreme points of a convex set $C$ are the elements which cannot be represented as a non-trivial convex combination of other elements of $C$.

\begin{defi}
\label{ext-point}
A point $e$ of a convex set $C$ is an \textbf{extreme point}\index{extreme point} of $C$ if $e = \alpha x + (1-\alpha) y$ with $x, y \in C$ and $0 < \alpha < 1$ implies $x = y = e$. The set of extreme points of a convex set $C$ is denoted by $\ext(C)$.
\end{defi}

For instance, the extreme points of a closed triangle (square) are the three (four) corners, whereas every point on the boundary of the closed circular disk is an extreme point. In contrast, an open ball or an open half-space in a vector space are both convex sets, but neither of them has any extreme point at all. We will see below (Theorem \ref{minkowski}) that this is related to the fact that those are non-compact sets.

Let us consider a more abstract and less trivial example of a convex set.

\begin{ex}[The set of density operators]
\label{set-of-density-operators}
For a finite-dimensional Hilbert space $\mathcal{H}$, the set $\mathcal{S}(\mathcal{H})$ of density operators on $\mathcal{H}$ (the set of \emph{states}) is defined by
\begin{align}
&\mathcal{S}(\mathcal{H}) := \{ \rho \in \Herm(\mathcal{H}) \mid \rho \geq 0, \tr\rho = 1\}, \text{ where} \nonumber \\
&\rho \geq 0 \ :\Leftrightarrow \ \langle \psi | \rho | \psi \rangle \geq 0 \quad \forall | \psi \rangle \in \mathcal{H} \,. \nonumber
\end{align}
A Hermitian operator $\rho$ with $\rho \geq 0$ is called a \emph{positive} operator. Thus, a density operator is a positive operator with unit trace. The set $\mathcal{S}(\mathcal{H})$ is convex:
\begin{align}
&0 \leq \alpha \leq 1, \quad \rho, \tau \in \mathcal{S}(\mathcal{H}) \nonumber \\
\Rightarrow \ &\langle \psi | \alpha \rho + (1-\alpha) \tau | \psi \rangle = \alpha \langle \psi | \rho | \psi \rangle + (1-\alpha) \langle \psi | \tau | \psi \rangle \geq 0 \,, \nonumber \\
&\tr( \alpha \rho + (1-\alpha) \tau) = \alpha \tr(\rho) + (1-\alpha) \tr(\tau) = 1 \nonumber \\
&\Rightarrow \alpha \rho + (1-\alpha) \tau \in \mathcal{S}(\mathcal{H}) \,. \nonumber
\end{align}
The extreme points of $\mathcal{S}(\mathcal{H})$ are the \emph{pure} states, i.e. the density operators of the form $\rho = | \psi \rangle \langle \psi |$ for some $| \psi \rangle \in \mathcal{H}$. In the case where $\mathcal{H}$ is a two-dimensional Hilbert space, the set $\mathcal{S}(\mathcal{H})$ can be visualized by the so called \emph{Bloch sphere}. The name Bloch ``sphere'' is slightly misleading since the Bloch sphere is actually not a sphere but a unit ball. This is a neat visualization since it shows the convexity of $\mathcal{S}(\mathcal{H})$ in a geometric way. The boundary of the ball is given by the extreme points and therefore the pure states of the system.
\hfill $\blacksquare$
\end{ex}

Another very central notion in the study of convex sets is the notion of a \emph{face} of a convex set. Roughly speaking, one might think of a face as some kind of convex ``extreme subset'' of a convex set.

\begin{defi}
\label{face-def}
A nonempty convex subset $F$ of a convex set $C$ is called a \textbf{face}\index{face} of $C$ if $\alpha x + (1-\alpha) y \in F$ with $x, y \in C$ and $0 < \alpha < 1$ imply $x, y \in F$. The set $F$ is a \textbf{proper face}\index{face!proper} of $C$ if $F$ is a face of $C$ and $F \neq C$.
\end{defi}

In other words, a face $F$ of a convex set $C$ is a face of $C$ if every line segment in $C$ with an interior point\footnote{We say that an element $z$ of a line segment $L_{x, y}$ is an \emph{interior point} of the line segment if $z = \lambda x + (1-\lambda) y$ for some $0 < \lambda < 1$.} in $F$ is completely contained in $F$. By definition of an extreme point, if $e \in \ext(C)$ is an extreme point of a convex set, then $\{e\}$ is a face of $C$. The faces of a triangle are given by the triangle itself, its edges and corners, the proper faces of a cube are its six square sides, its edges and its corners. Note that the requirement that a face has to be convex makes a difference. If this requirement would be dropped, then any subset of $\ext(M)$ would be a face, but obviously, not every subset of $\ext(M)$ is convex (e.g. the union of two corners of a square is not convex).

As we said above, when we compare the definition of a face with the definition of an extreme point, we could say that a face is some kind of convex ``extreme subset''. It is not only an extreme set in terms of binary convex combinations but in terms of arbitrary convex combinations. We state this more formally.

\begin{prop}
\label{extreme-set-prop}
Let $F$ be a face of a convex set $C$, let $x \in F$. Let $v_1, \ldots, v_n$ be points in $C$ such that there exists a convex combination of $v_1, \ldots, v_n$ with nonzero coefficients $\alpha_1, \ldots, \alpha_n$ which gives $x$, i.e.
\begin{align}
x = \sum\limits_{i = 1}^n \alpha_i v_i, \quad \text{for some } \alpha_i > 0 \text{ with } \sum\limits_{i=1}^n \alpha_i = 1 \,. \nonumber
\end{align}
Then $v_1, \ldots, v_n \in F$.
\end{prop}

\begin{proof}
We prove that for any $j \in \{1, \ldots, n\}$, we have that $v_j \in F$.
\begin{align}
x &= \sum\limits_{i = 1}^n \alpha_i v_i = \alpha_j v_j + \underset{i \neq j}{\sum\limits_{i = 1}^n} \alpha_i v_i = \alpha_j v_j + \underbrace{\left( \underset{i \neq j}{\sum\limits_{i = 1}^n} \alpha_i \right)}_{\widetilde \alpha} \underbrace{\frac{ \underset{i \neq j}{\sum\limits_{i = 1}^n} \alpha_i v_i }{ \underset{i \neq j}{\sum\limits_{i = 1}^n} \alpha_i }}_{\widetilde v} \nonumber \\
&= \alpha_j v_j + \widetilde \alpha \widetilde v = x \in F \quad \text{with} \quad \alpha_j, \widetilde \alpha > 0, \quad \alpha_j + \widetilde \alpha = 1. \label{tilde-equation}
\end{align}
The vector $\widetilde v$ is a convex combination of elements in $C$, so it is itself an element of $C$. Thus, by the definition of a face, (\ref{tilde-equation}) implies that $v_j \in F$.
\end{proof}

Note that Proposition \ref{extreme-set-prop} in particular applies to the case where the face $F$ consists of an extreme point, i.e. $F = \{e\}$ for some $e \in \ext(C)$. Another simple but very useful property of faces is the following.

\begin{prop}
\label{face-face-face}
For a convex set $C$, a face of a face of $C$ is itself a face of $C$.
\end{prop}

\begin{proof}
Let $F$ be a face of $C$, let $G$ be a face of $F$. Let $x, y \in C$ and $0 < \alpha < 1$ such that $\alpha x + (1-\alpha) y \in G$. Then $\alpha x + (1-\alpha) y \in F$, so $x, y \in F$ since $F$ is a face of $C$. This means that we have $x, y \in F$, $0 < \alpha < 1$ with $\alpha x + (1-\alpha) y \in G$. The set $G$ is a face of $F$, so this implies that $x, y \in G$. We have proved that $x, y \in C$ and $0 < \alpha < 1$ such that $\alpha x + (1-\alpha) y \in G$ implies $x, y \in G$, so we have proved that $G$ is a face of $C$.
\end{proof}

In the following, we will show a useful and intuitive property of faces of convex sets. Before we can state it, we have to introduce the \emph{affine hull} of a set.

\begin{defi}
\label{aff-hull-def}
Let $M$ be a subset of a real vector space $V$. The \textbf{affine hull}\index{affine hull} of $M$, denoted by $\aff(M)$, is defined by
\begin{align}
&\aff(M) := \left\{ \sum\limits_{i=1}^n \alpha_i v_i \ \middle\vert \ n \in \{0, 1, 2, \ldots \}, \ v_i \in M, \ \alpha_i \in \mathbb{R}, \ \sum\limits_{i=1}^n \alpha_i = 1 \right\} \,. \nonumber
\end{align}
\end{defi}
A few examples: The affine hull of a point is the point itself, the affine hull of two points is given by the straight line through the two points, and the affine hull of a triangle, square or circle disk is the plane which contains it. With the definition of the affine hull at hand, the property we want to prove reads as follows.

\begin{prop}
\label{face-cap-aff}
If $F$ is a face of a convex set $C$, then $F = \aff(F) \cap C$.
\end{prop}

\begin{proof}
The inclusion $F \subset \aff(F) \cap C$ is obvious. For the other inclusion, let $v$ be an element of $\aff(F) \cap C$. Our goal is to show that $v \in F$. There is an affine combination of finitely many elements of $F$ which gives $v$:
\begin{align}
&v = \sum\limits_{i \in I^+} \alpha_i v_i + \sum\limits_{j \in I^-} \alpha_j v_j \,, \text{ where} \nonumber \\
&\alpha_i > 0 \ \forall i \in I^+ \,, \quad \alpha_j < 0 \ \forall j \in I^- \,, \nonumber \\
&\sum\limits_{i \in I^+} \alpha_i + \sum\limits_{j \in I^-} \alpha_j = 1 \,. \label{alpha-sum}
\end{align}
We define
\begin{align}
v^+ := \frac{\sum\limits_{i \in I^+} \alpha_i v_i}{\sum\limits_{i \in I^+} \alpha_i} \,, \quad v^- := \frac{\sum\limits_{j \in I^-} \alpha_j v_j}{\sum\limits_{j \in I^-} \alpha_j} \,. \nonumber
\end{align}
This gives
\begin{align}
v = \left( \sum\limits_{i \in I^+} \alpha_i \right) v^+ + \left( \sum\limits_{j \in I^-} \alpha_j \right) v^- \,. \label{v-vplus-vminus}
\end{align}
The set $F$ is convex, so we have that both $v^+$ and $v^-$ are elements of $F$ since they are given by convex combinations of elements of $F$. If $I^-$ is empty, then by Equation (\ref{v-vplus-vminus}) one has that $v = v^+$ and therefore $v \in F$ (which is what we want to show). If $I^-$ is non-empty, then
\begin{align}
\sum\limits_{i \in I^+} \alpha_i > 1 \quad \Rightarrow &\quad 0 < \frac{1}{\sum\limits_{i \in I^+} \alpha_i} < 1 \quad \text{and} \label{interior-1} \\
&\quad 0 < 1 - \frac{1}{\sum\limits_{i \in I^+} \alpha_i} < 1 \,. \label{interior-2}
\end{align}
Note that
\begin{align}
1 - \frac{1}{\sum\limits_{i \in I^+} \alpha_i} = \frac{\left( \sum\limits_{i \in I^+} \alpha_i \right) -1}{\sum\limits_{i \in I^+} \alpha_i} \,. \label{useful-eq}
\end{align}
Inequalities (\ref{interior-1}) and (\ref{interior-2}) imply that the following is an interior point of the line segment $L_{v, v^-}$ from $v \in C$ to $v^- \in F \subset C$:
\begin{align}
&\frac{1}{\sum\limits_{i \in I^+} \alpha_i} v + \left( 1 - \frac{1}{\sum\limits_{i \in I^+} \alpha_i} \right) v^- \nonumber \\
\underset{(\ref{useful-eq})}{\overset{(\ref{v-vplus-vminus})}{=}} \ &v^+ + \frac{\sum\limits_{j \in I^-} \alpha_j v_j}{\sum\limits_{i \in I^+} \alpha_i} + \left( \frac{\left( \sum\limits_{i \in I^+} \alpha_i \right) -1}{\sum\limits_{i \in I^+} \alpha_i} \right)\frac{\sum\limits_{j \in I^-} \alpha_j v_j}{\sum\limits_{j \in I^-} \alpha_j} \,. \label{to-reform}
\end{align}
From (\ref{alpha-sum}), we get that
\begin{align}
\left( \sum\limits_{j \in I^+} \alpha_i \right) - 1 = - \sum\limits_{j \in I^-} \alpha_j \,. \nonumber
\end{align}
This allows us to reformulate the right-hand side of (\ref{to-reform}):
\begin{align}
\frac{1}{\sum\limits_{i \in I^+} \alpha_i} v + \left( 1 - \frac{1}{\sum\limits_{i \in I^+} \alpha_i} \right) v^- &= v^+ + \frac{\sum\limits_{j \in I^-} \alpha_j v_j}{\sum\limits_{i \in I^+} \alpha_i} - \frac{\sum\limits_{j \in I^-} \alpha_j}{\sum\limits_{i \in I^+} \alpha_i} \frac{\sum\limits_{j \in I^-} \alpha_j v_j}{\sum\limits_{j \in I^-} \alpha_j} \nonumber \\
&= v^+ \in F \,. \nonumber
\end{align}
The set $F$ is a face of $C$, so by the definition of a face, $v \in F$. The vector $v$ is an arbitrary element of $\aff(F) \cap C$, so we have shown that $\aff(F) \cap C \subset F$, which completes the proof.
\end{proof}

Next, we want to turn to a very central result in the study of convex sets. It states that a compact convex set is the convex hull of its extreme points. This needs some preparation. At first, we need to know what the convex hull is. For a subset $M$ of a real vector space $V$, the convex hull of $M$ can be characterized as the smallest convex subset of $V$ which contains $M$. This set can be obtained by taking all convex combinations of points in $M$. The following definition states this more formally.

\begin{defi}
For a subset $M$ of a real vector space $V$, the \textbf{convex hull}\index{convex hull} of $M$, denoted by $\conv(M)$, is defined by
\begin{align}
&\conv(M) := \left\{ \sum\limits_{i=1}^n \alpha_i v_i \ \middle\vert \ n \in \{0, 1, 2, \ldots \}, \ v_i \in M, \ \alpha_i \in [0,1], \ \sum\limits_{i=1}^n \alpha_i = 1 \right\} \,. \nonumber
\end{align}
\end{defi}

The definition of the convex hull reads similar to the definition of the affine hull (c.f. Definition \ref{aff-hull-def}). The only difference is that the coefficients in the sum are positive (instead of just real). Note that $\conv(M) \subset \aff(M)$ is true for any set $M$.

Another thing we have to understand is what it means for a subset of a vector space to be compact. Compactness is a topological property, and so far, we have not defined a topology. In finite-dimensional vector spaces, however, there is a canonical topology, as we will see below. This is very practical since we will restrict ourselves to the finite-dimensional case. Readers who are interested in the more general, infinite-dimensional case are referred to Appendix \ref{appendix-a}. We only state the following definition and theorem to show that we can refer to basic topological notions without explicitly defining a topology (we will refer to compactness and closedness of sets). We will not refer the notions of topological vector spaces or Hausdorff spaces again, so the reader unfamiliar with these concepts will not have any problems while reading this thesis.

\begin{defi}
\label{top-vs-def}
A \textbf{real topological vector space}\index{topological vector space} is a real vector space $V$ equipped with a topology such that the vector addition $V \times V \rightarrow V$ and the scalar multiplication $\mathbb{R} \times V \rightarrow V$ are continuous.
\end{defi}

\begin{thm}[see {\cite[Chapter 3]{Schaefer}}]
\label{unique-topo-thm}
For a finite-dimensional real vector space $V$, there is a unique Hausdorff topology on $V$ with respect to which $V$ is a real topological vector space.
\end{thm}

This means that in the case of a finite-dimensional vector space $V$, we can refer to topological properties of subsets of $V$ without explicitly specifying a topology on $V$. Now we are ready for the theorem.

\begin{thm}[Minkowski, see {\cite[Theorem 2.6.16]{Webster}}\footnote{In \cite[Theorem 2.6.16]{Webster}, this theorem is referred to as the Krein-Milman Theorem, which is not correct since it has been proved by Minkowski. As described in Appendix \ref{appendix-a}, the Krein-Milman Theorem is a statement about a more general case in infinite-dimensional vector spaces.}]
\label{minkowski}\index{Minkowski, theorem by}
Let $V$ be a finite-dimensional vector space and let $C$ be a compact convex subset of $V$. Then $C$ is the convex hull of its extreme points:
\begin{align}
C = \conv(\ext(C)) \,. \nonumber
\end{align}
In particular, $C$ has extreme points.
\end{thm}

The reader who wants to see how this theorem can be generalized to the infinite-dimensional case is referred to Appendix \ref{appendix-a}. Here, we restrict ourselves to the finite-dimensional case. Theorem~\ref{minkowski} states that a closed convex set is fully specified by its extreme points. When we apply this to a set of states $\Omega$ (which we will assume to be a compact convex subset of a finite-dimensional vector space), this gives us a physical interpretation (which we will discuss in Section \ref{maximal-knowledge-section}). The extreme points of $\Omega$ will be called \emph{pure states}, and they correspond to maximal knowledge about the system. According to Theorem \ref{minkowski}, all states of incomplete knowledge (i.e. the states which are not extreme) can be represented as a probabilistic mixture of states of maximal knowledge.

The next thing we want to learn is that a closed convex subset $C$ of Hilbert spaces $\mathcal{H}$ allows for a distance function $d( \ \cdot \ , C) : \mathcal{H} \rightarrow \mathbb{R}$. To define this function, we use the famous Hilbert Projection Theorem.

\begin{thm}[Hilbert Projection Theorem, see {\cite[Satz V.3.2]{Werner}}]
\label{hilbert-projection}\index{Hilbert Projection Theorem}
Let $\mathcal{H}$ be a Hilbert space, $C \subset \mathcal{H}$ closed and convex and $x \in H$. Then there is a unique $x_0 \in C$ such that $|| x_0-x || = \inf\limits_{y \in C} || y-x ||$. In this case, we define $d(x, C) := \inf\limits_{y \in C} || y - x ||$.
\end{thm}

\subsubsection{Convexity-preserving maps}

To conclude the introduction to general convex sets, we want to turn to the question when two convex sets are equivalent. The structure in question is the convexity structure of the two sets. To investigate whether two sets show the same convexity structure, it is convenient to introduce a map which conserves this structure. We call such a map a \emph{convex-linear} map. As we will see below, this is the same as an affine map. This kind of map will be important in the proof of our main result in Section \ref{result-2-result}. For the reader interested in the uniqueness of abstract state spaces (we will come back to this issue in Section \ref{abs-st-sp-section}), affine maps play a central in the proof of the equivalence of compact convex sets and abstract state spaces presented in Appendix \ref{appendix-b}.

We start with the definition of a \emph{convex-linear map}. Simply speaking, this is a map which commutes with the action of taking convex combinations, so it preserves the convexity-structure. In formal terms, this reads as follows.

\begin{defi}
\label{convex-linear-def}
A map $f: V \rightarrow W$ between finite-dimensional real vector spaces $V$ and $W$ is \textbf{convex-linear}\index{convex-linear map} if
\begin{align}
f(\lambda x + (1-\lambda) y) = \lambda f(x) + (1-\lambda) f(y) \quad \forall x, y \in V, \ \forall \lambda \in [0, 1] \,. \nonumber
\end{align}
Two convex subsets $S \subset V$ and $T \subset W$ are \textbf{convex-isomorphic}\index{convex-isomorphic} if there is a bijective map $\widetilde f: S \rightarrow T$ which extends to a convex-linear map $f: V \rightarrow W$.
\end{defi}

Convex-linearity exactly represents our intuition for the ``conservation of the convexity-structure''. However, it turns out that we could have defined the property that a map ``conserves the convexity-structure'' in a (seemingly) stronger way without loss of generality, as we see in the following.

\begin{defi}
A map $f: V \rightarrow W$ between finite-dimensional real vector spaces $V$ and $W$ is \textbf{affine}\index{affine map} if
\begin{align}
f(\lambda x + (1-\lambda) y) = \lambda f(x) + (1-\lambda) f(y) \quad \forall x, y \in V, \ \forall \lambda \in \mathbb{R} \,. \nonumber
\end{align}
\end{defi}

The difference to Definition (\ref{convex-linear-def}) is that the scalar $\lambda$ can be any real number instead of only an element of $[0,1]$.

\begin{prop}
\label{conv-lin-aff}
Every convex-linear map is affine.
\end{prop}

\begin{proof}
Let $f: V \rightarrow W$ be a convex-linear map, let $x, y \in V$ and let $\lambda \in \mathbb{R}$. If $\lambda \in [0,1]$, then
\begin{align}
f(\lambda x + (1-\lambda) y) = \lambda f(x) + (1-\lambda) f(y)\,. \nonumber
\end{align}
If $\lambda \notin [0,1]$ we can assume without loss of generality that $\lambda < 1$ (in the other case where $\lambda > 1$, we can simply interchange the role of $x$ and $y$). We can write $y$ as the following convex combination:
\begin{align}
y &= \underbrace{\frac{1}{1-\lambda}}_{\in [0, 1]} (\lambda x + (1 - \lambda) y) + \left( 1-\frac{1}{1-\lambda} \right) x\,. \nonumber
\end{align}
This allows us to write
\begin{align}
f(y) &= f\left( \frac{1}{1-\lambda} (\lambda x + (1 - \lambda) y) + \left( 1-\frac{1}{1-\lambda} \right) x \right)\,. \nonumber
\end{align}
The map $f$ is convex-linear, so
\begin{align}
&f(y) = \frac{1}{1-\lambda} f (\lambda x + (1 - \lambda) y) + \underbrace{\left( 1-\frac{1}{1-\lambda} \right)}_{-\frac{\lambda}{1-\lambda}} f(x)\,. \nonumber \\
\Leftrightarrow \quad &\frac{1}{1-\lambda} f (\lambda x + (1 - \lambda) y) = f(y) + \frac{\lambda}{1-\lambda} f(x) \nonumber \\
\Leftrightarrow \quad &f(\lambda x + (1-\lambda) y) = \lambda f(x) + (1-\lambda) f(y)\,. \label{conv-aff-proof}
\end{align}
We have proved that Equation (\ref{conv-aff-proof}) holds for all $x, y \in V$ and for all $\lambda \in \mathbb{R}$, so $f$ is affine.
\end{proof}

For practical purposes, as well as for the intuition for convex-linear maps, it is useful to see that a convex-linear map can always be represented by the action of a linear map followed by a translation. This is the statement of the following theorem.

\begin{thm}[{\cite[Theorem 1.5.2]{Webster}}]
\label{convex-linear-affine}
A map $f: V \rightarrow W$ between finite-dimensional real vector spaces is affine (by Proposition \ref{conv-lin-aff}, we can equivalently say convex-linear) if and only if it is of the form
\begin{align}
f(x) = L(x) + y \quad \text{for some } y \in W \text{ and for some linear map } L: V \rightarrow W \,. \nonumber
\end{align}
\end{thm}

Now we show two propositions which will be helpful for the proof of the main result in Section \ref{result-2-result}.

\begin{prop}
\label{aff-is-arbit-aff}
Let $f: A \rightarrow B$ be an affine or convex-linear map. Then $f$ commutes with arbitrary affine combinations. More precisely, for any $x_1, \ldots, x_n \in A$, on has that
\begin{align}
f \left( \sum\limits_{i=1}^n \alpha_i x_i \right) = \sum\limits_{i=1}^n \alpha_i f(x_i) \quad \text{for any real numbers $\alpha_i$ with} \quad \sum\limits_{i=1}^n \alpha_i = 1 \,. \nonumber
\end{align}
In particular, $f(\aff(M)) = \aff(f(M))$ (and since every convex combination is an affine combination, we also have $f(\conv(M)) = \conv(f(M))$) for any subset $M$ of $A$.
\end{prop}

\begin{proof}
Let $f: A \rightarrow B$ be affine, i.e.
\begin{align}
f(\lambda x + (1-\lambda) y) = \lambda f(x) + (1-\lambda)f(y) \quad \forall x, y \in A, \ \forall \lambda \in \mathbb{R} \,. \label{f-aff-help}
\end{align}
Let $\sum_{i=1}^n \alpha_i x_i$ be any affine combination of elements $x_1, \ldots, x_n \in A$. Then
\begin{align}
f \left( \sum\limits_{i=1}^n \alpha_i x_i \right) &= f \left( \alpha_1 x_1 + \left( \sum\limits_{i=2}^n \alpha_i \right) \left( \sum\limits_{j=2}^n \frac{\alpha_j x_j}{\left(\sum\limits_{i=2}^n \alpha_i \right)} \right) \right) \nonumber \\
&\overset{(\ref{f-aff-help})}{=} \alpha_1 f(x_1) + \left(\sum\limits_{i=2}^n \alpha_i \right) f \left( \sum\limits_{j=2}^n \frac{\alpha_j x_j}{\left(\sum\limits_{i=2}^n \alpha_i \right)} \right) \nonumber \\
&= \alpha_1 f(x_1) + \left( \sum\limits_{i=2}^n \alpha_i \right) f \left( \frac{\alpha_2}{\left( \sum\limits_{i=2}^n \alpha_i \right)} x_2 + \sum\limits_{j=3}^n \frac{\alpha_j x_j}{\left(\sum\limits_{i=2}^n \alpha_i \right)} \right) \nonumber \\
&= \alpha_1 f(x_1) + \left( \sum\limits_{i=2}^n \alpha_i \right) f \left( \frac{\alpha_2}{\left( \sum\limits_{i=2}^n \alpha_i \right)} x_2 \right. \nonumber \\
& \left. \qquad \qquad \qquad \qquad + \frac{\left(\sum\limits_{k=3}^n \alpha_k \right)}{\left( \sum\limits_{i=2}^n \alpha_i \right)} \left( \sum\limits_{j=3}^n \frac{\alpha_j x_j}{\left( \sum\limits_{k=3}^n \alpha_k \right)} \right) \right) \nonumber \\
&\overset{(\ref{f-aff-help})}{=} \alpha_1 f(x_1) + \underbrace{\left( \sum\limits_{i=2}^n \alpha_i \right) \frac{\alpha_2}{\left(\sum\limits_{i=2}^n \alpha_i \right)}}_{\alpha_2} f(x_2) \nonumber \\
& \qquad \qquad \qquad \qquad + \frac{\left(\sum\limits_{k=3}^n \alpha_k \right)}{\left( \sum\limits_{i=2}^n \alpha_i \right)} f \left( \sum\limits_{j=3}^n \frac{\alpha_j x_j}{\left( \sum\limits_{k=3}^n \alpha_k \right)} \right) \nonumber \\
&= \ldots \nonumber \\
&= \alpha_1 f(x_1) + \alpha_2 f(x_2) + \ldots + \alpha_n f(x_n) \,. \tag*{\qedhere}
\end{align}
\end{proof}

\begin{prop}
\label{aff-inj-prop}
Let $f: A \rightarrow B$ be an affine map. Then the following statements are equivalent:
\begin{enumerate}[(a)]
\item The map $f$ is injective.
\item The map $f$ maps affinely independent points to affinely independent points.\footnote{As we will see in Definition \ref{aff-indep-def}, a subset of a vector space is \emph{affinely independent}\index{affinely independent} if no element of the subset lies in the affine hull of the other elements of the subset.}
\end{enumerate}
\end{prop}

\begin{proof}
We prove the two implications separately.
\begin{itemize}
\item (a) $\Rightarrow$ (b): Let $f: A \rightarrow B$ be an injective affine map. We prove the contraposition: Let $x_1, \ldots, x_n \in A$ such that $f(x_1), \ldots, f(x_n) \in B$ are affinely dependent. This means that there exists a $k \in \{1, \ldots, n\}$ such that
\begin{align}
\underset{i \neq k}{\sum\limits_{i=1}^n} \alpha_i f(x_i) = f(x_k) \quad \text{for some real numbers $\alpha_i$ with} \quad \underset{i \neq k}{\sum\limits_{i=1}^n} \alpha_i = 1 \,. \label{f-aff-dep}
\end{align}
The map $f$ is affine, so we can rewrite Equation (\ref{f-aff-dep}):
\begin{align}
f \left( \underset{i \neq k}{\sum\limits_{i=1}^n} \alpha_i x_i \right) = f(x_k) \,. \label{f-aff-dep-2}
\end{align}
We have assumed that $f$ is injective, so Equation (\ref{f-aff-dep-2}) implies that
\begin{align}
\underset{i \neq k}{\sum\limits_{i=1}^n} \alpha_i x_i = x_k \,, \quad \underset{i \neq k}{\sum\limits_{i=1}^n} \alpha_i = 1 \,, \nonumber
\end{align}
so $x_1, \ldots, x_n \in A$ are affinely dependent.
\item (b) $\Rightarrow$ (a): Assume that $f: A \rightarrow B$ is an affine map that maps affinely independent points to affinely independent points. Let $x, y \in A$ be such that $f(x) = f(y)$. This means that $f(x)$ and $f(y)$ are affinely dependent (since each side of the equation $f(x) = f(y)$ can be seen as the trivial affine combination of one element). By the assumption that $f$ maps affinely independent points to affinely independent points, this means that $x$ and $y$ must be affinely dependent. The only affine combination is $x = y$, so $f$ is injective. \hfill \qedhere
\end{itemize}
\end{proof}

\subsection{Polytopes}
\label{polytope-section}

Now we investigate a special class of convex sets which are called \emph{polytopes}. As for the previous subsection, this is not a standard introduction to polytopes. Instead, this subsection is aimed at understanding and proving some particular properties of polytopes which will be important in the sections of Part \ref{result-part} of this thesis.

\begin{defi}
\label{polytope-def}
A compact convex subset $P$ of a finite-dimensional real vector space is a \textbf{polytope}\index{polytope} if $\ext(P)$ is a finite set. For a polytope $P$, an element of $\ext(P)$ is called a \textbf{vertex}\index{vertex} of $P$ (pl.: vertices).
\end{defi}

Note that the requirement that the set $P$ has to be compact makes a big difference: The nonnegative numbers in $\mathbb{R}$ or an interval of the form $[a, b)$ in $\mathbb{R}$ are both examples with finitely many (namely one) extreme points, but neither of them is a polytope since they are not compact.

It is assumed in the definition of a polytope that the vector space containing the polytope is finite-dimensional. We show in Appendix \ref{appendix-a} that this does not cause any loss of generality.

A polytope can equivalently be characterized as the convex hull of finitely many points. To see this, we make use of the following result of Carath\'eodory.

\begin{thm}[Carath\'eodory \cite{Caratheodory:1907p13250}, see {\cite[Chapter 2.3]{Gruenbaum}}]
\label{caratheodory}\index{Carath\'eodory, theorem by}
If $C$ is a compact subset of a finite-dimensional real vector space, then $\conv(C)$ is closed. In other words, for compact $C$ we have $\overline{\conv(C)} = \conv(C)$.
\end{thm}

\begin{prop}
\label{polytope-conv-finite}
A subset $P$ of a real vector space is a polytope if and only if it is the convex hull of finitely many points.
\end{prop}

\begin{proof}
A polytope $P$ is by definition a compact subset of a finite-dimensional vector space. This allows us to apply Theorem~\ref{minkowski} which implies that $P = \conv(\ext(P))$, where $\ext(P)$ is a finite set by the definition of a polytope. For the other direction, let $M$ be a finite set. Then $M$ is trivially compact, which by Theorem \ref{caratheodory} implies that $\conv(M)$ is closed. $M$ is finite, so $\conv(M)$ is also bounded. In finite-dimensional spaces, being closed and bounded is equivalent to being compact, so $\conv(M)$ is compact. The set $\conv(M) \backslash M$ does not contain any extreme points of $\conv(M)$ (as one can see from the definition of an extreme point), so $\ext(\conv(M)) \subset M$ which is a finite set. Thus, $\conv(M)$ is a polytope.
\end{proof}

Next, we want to introduce the notion of the dimension of a polytope. It is defined as the dimension of the affine hull.

\begin{defi}
\label{aff-indep-def}
We define the \textbf{dimension $\dim(\aff(M))$ of an affine hull}\index{dimension!of an affine hull} $\aff(M)$ of a subset $M$ of a finite-dimensional vector space $V$ as $n+1$, where $n$ is the maximal cardinality of a subset of $\aff(M)$ such that the subset is affinely independent. A subset of a vector space is \textbf{affinely independent}\index{affinely independent} if no element of the subset lies in the affine hull of the other elements of the subset.
\end{defi}

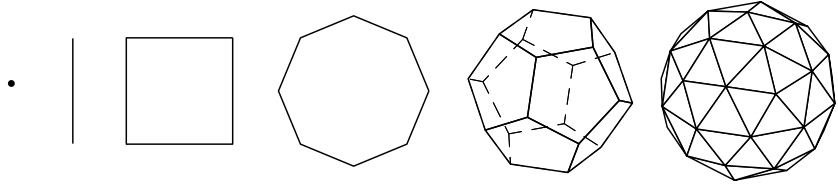
\begin{figure}[htb]
\centering

\begin{pspicture}[showgrid=false](0,-1)(0.7,1)
\psdot[dotsize=1pt 2](0,0.1)
\end{pspicture}
\begin{pspicture}[showgrid=false](0,-1)(1.3,1)
\psline[linewidth=0.7\pslinewidth](0,-0.7)(0,0.7)
\end{pspicture}
\begin{pspicture}[showgrid=false](0,-1)(1,1)
\PstSquare[PstPicture=false, linewidth=0.7\pslinewidth]
\end{pspicture}
\begin{pspicture}[showgrid=false, linewidth=0.7\pslinewidth](-1.2,-1)(1,1)
\PstPolygon[PolyNbSides=8, PstPicture=false]
\end{pspicture}
\begin{pspicture}[showgrid=false](-1.5,-1)(1,1) 
\psset{viewpoint=50 -20 30 rtp2xyz,Decran=70} 
\psSolid[object=geode, linewidth=0.7\pslinewidth,
action=draw,
dualreg, 
ngrid=5 0] 
\end{pspicture}
\begin{pspicture}(-1.5,-1)(1,1) 
\psset{viewpoint=50 -20 30 rtp2xyz,Decran=60} 
\psSolid[object=geode, linewidth=0.7\pslinewidth, 
action=draw**,
ngrid=5 1] 

\end{pspicture} 

\caption{From left to right, we have a 0-polytope (which is nothing but a point), a 1-polytope (in other words, a line), two 2-polytopes (polygons) and two 3-polytopes (polyhedra).}
\label{polytopes-1}
\end{figure}

With this definition at hand, we can characterize polytopes by the dimension of their affine hull. A few examples are shown in Figure \ref{polytopes-1}.

\begin{defi}
The \textbf{dimension $d$ of a polytope}\index{dimension!of a polytope} is the dimension of its affine hull, $d := \dim(\aff(P))$. The dimension of the empty polytope $\{ \}$ is defined to be $-1$. A $d$-dimensional polytope is called a \textbf{$d$-polytope}\index{polytope!$d$-polytope}. A $2$-polytope is a \textbf{polygon}\index{polygon}, a $3$-polytope is a \textbf{polyhedron}\index{polyhedron}.
\end{defi}

Polytopes can be given an intuitive geometrical picture: they are the intersection of finitely many closed half-spaces which are positioned in a way such that their intersection is bounded. In more technical terms, this reads as follows.

\begin{defi}
A subset $P$ of a finite-dimensional real vector space $V$ is called a \textbf{polyhedral set}\index{polyhedral set} provided that $P$ is the intersection of a finite family of closed half-spaces in $V$.
\end{defi}

\begin{thm}[{\cite[Chapter 3.1]{Gruenbaum}}]
\label{polytope-bounded-polyhedral}
A subset $P$ of a finite-dimensional real vector space $V$ is a polytope if and only if $P$ is a bounded polyhedral set.
\end{thm}

Now we are ready to give a good picture for the faces of a polytope. It turns out that a subset $F$ of a polytope $P$ is a proper face of $P$ if and only if it is the intersection of $P$ with an affine hyperplane\footnote{We say that a subset $M$ of a real vector space $V$ is an \emph{affine hyperplane} if there is a nonzero linear functional $f \in V^*$ and a $k \in \mathbb{R}$ such that $M = \{ v \in V \mid f(v) = k \}$.} which touches $P$ but which does not cut $P$. This result is established by the following two propositions.

\begin{prop}
\label{face-conv-ext}
If $F$ is a face of a compact convex subset $P$ of a finite-dimensional vector space $V$, then $F = \conv(\{v \in \ext(P) \mid v \in F \})$.
\end{prop}

\begin{proof}
Obviously, $\conv(\{ v \in \ext(P) \mid v \in F\} \subset F$ since $F$ is a convex set. For the other inclusion, let $w \in F$. By Theorem \ref{minkowski}, one has that $P = \conv(\ext(P))$. Let
\begin{align}
w = \sum\limits_{i = 1}^n \alpha_i v_i, \quad \alpha_i > 0, v_i \in \ext(P) \quad \forall i \in \{1, \ldots, n\} \label{conv-comb-ext-points}
\end{align}
be any convex combination of extreme points of $P$ with nonzero coefficients which gives $w$. By Proposition \ref{extreme-set-prop}, $v_1, \ldots, v_n \in F$. Hence, for every $w$ in $F$, it holds that every convex combination of extreme points of $P$ which yields $w$ is a convex combination of extreme points that are elements of $F$. This proves $F \subset \conv(\{v \in \ext(P) \mid v \in F\})$.
\end{proof}

\begin{prop}
\label{face-equivalence}
For a non-empty convex subset $F$ of a polytope $P$, the following are equivalent:
\begin{enumerate}[(a)]
\item $F$ is a proper face of $P$.
\item There is a closed half-space\footnote{We say that a subset $H$ of a real vector space $V$ is a \emph{closed half-space} if there is a nonzero linear functional $f \in V^*$ and a $k \in \mathbb{R}$ such that $H = \{ v \in V \mid f(v) \leq k \}$. The boundary of the half-space is given by $\partial H = \{ v \in V \mid f(v) = k\}$.} $H$ containing $P$ such that $F = P \cap \partial H$, where $\partial H$ is the affine hyperplane defined by the boundary of $H$.
\end{enumerate}
\end{prop}

\begin{proof}

We prove the implications (a) $\Rightarrow$ (b) and (b) $\Rightarrow$ (a) separately. \\
\begin{itemize}

\item (a) $\Rightarrow$ (b): 

Let $K = \conv(\{e \in \ext(P) \mid e \notin F \})$. We prove (a) $\Rightarrow$ (b) in three steps:
\begin{enumerate}[(i)]
\item At first, we show that $F \cap K = \emptyset$.
\item Then we show that $F \cap K = \emptyset$ implies the existence of a linear functional $f$ which takes a constant value $k$ on $F$ and satisfies $f(v) < k$ for all $v \in K$.
\item Finally, we show that $H = \{ v \in V \mid f(v) \leq k \}$ has the desired properties of (b).
\end{enumerate}
Before we prove the three steps, we make a few definitions. $P$ is a polytope, so it has finitely many, say $n$, extreme points. We define $\{e_1, \ldots, e_n\}$ to be the extreme points of $P$, i.e.
\begin{equation}
\label{ext-p-def}
\ext(P) = \{e_1, \ldots, e_n\}.
\end{equation}
Moreover, we define two index sets $I_F, I_K$ by
\begin{align}
&I_F = \{ i \in \{1, \ldots, n \} \mid e_i \in F \}, \label{i_f-def} \\
&I_K = \{ j \in \{1, \ldots, n \} \mid e_j \notin F \}. \label{i_k-def}
\end{align}
Now we prove each of the three steps.
\begin{enumerate}[(i)]

\item Recall that we have defined $K = \conv(\{e \in \ext(P) \mid e \notin F \})$. By Equation (\ref{i_k-def}), we have $K = \conv(\{e_i\}_{i \in I_K})$. Suppose that there is an $x \in K \cap F$. The vector $x$ is in $K$, so there is a convex combination
\begin{align}
x = \sum\limits_{j \in I_K} \alpha_j e_j, \quad \alpha_j \geq 0, \quad \sum\limits_{j \in I_K} \alpha_j = 1 \,. \nonumber
\end{align}
The vector $x$ is an element of the face $F$, so Proposition \ref{extreme-set-prop} implies that $e_j \in F$ for all $j$ with $\alpha_j > 0$. This leads to a contradiction since we have assumed that $e_j \notin F$ for all $j \in I_K$. This means that there cannot be an element $x \in F \cap K$, i.e. $F \cap K = \emptyset$.

\item $F$ is a \emph{proper} face, so Proposition \ref{face-conv-ext} implies that $K$ is nonempty. By Proposition \ref{polytope-conv-finite}, $K$ is a polytope since $\ext(P)$ is finite, so $K$ is compact and convex. The fact that $K = P \cap K$ implies
\begin{align}
\aff(F) \cap K = \aff(F) \cap (P \cap K) = (\aff(F) \cap P) \cap K \,. \label{useful-eq-1}
\end{align}
We know from Proposition \ref{face-cap-aff} that $\aff(F) \cap P = F$, so
\begin{align}
(\aff(F) \cap P) \cap K = F \cap K \overset{(i)}{=} \emptyset \,, \label{useful-eq-2}
\end{align}
where the last equality has been shown in the first step of the proof. Equations (\ref{useful-eq-1}) and (\ref{useful-eq-2}) imply that $\aff(F) \cap K = \emptyset$.

All in all, we have that $K$ is nonempty, compact and convex, $\aff(F)$ is closed and convex, and $\aff(F) \cap K = \emptyset$. This allows us to apply the separating hyperplane theorem. It says that there is a linear functional $f$ with
\begin{equation}
\label{sup-inf-equation}
\sup \{ f(v) \mid v \in K \} < \inf \{ f(v) \mid v \in \aff(F) \}.
\end{equation}
$\aff(F)$ is an affine hull on which the linear functional $f$ is lower bounded by $\sup\{f(v) \mid v \in K\}$. This implies that $f$ is constant on $\aff(F)$: If there were $y, z \in \aff(F)$ with $f(y) < f(z)$, then by choosing a large enough scalar $\alpha$ we would have that
\begin{align}
f(\underbrace{\alpha y + (1 - \alpha) z}_{\in \aff(F)}) = f(z) + \alpha(\underbrace{f(y) - f(z)}_{<0}) < \max \{f(v) \mid v \in K\} \,. \nonumber
\end{align}
Let $k := f(v)$ for some $v \in \aff(F)$ be the constant value that $f$ takes on $\aff(F)$. From (\ref{sup-inf-equation}) it follows that $f(v) < k$ for all $v \in K$.

\item Let $V$ denote the vector space containing the polytope $P$, let $H := \{ v \in V \mid f(v) \leq k \}$. Recall from (\ref{ext-p-def}) that $\ext(P) = \{e_1, \ldots, e_n\}$. Let $a$ be an arbitrary element of $P$. Then
\begin{equation}
\label{conv-a-comb}
a = \sum\limits_{i = 1}^n \alpha_i e_i \quad \text{for some numbers} \quad \alpha_i \geq 0, \quad \sum\limits_{i=1}^n \alpha_i = 1.
\end{equation}
From the Definitions (\ref{i_f-def}) and (\ref{i_k-def}), we see that $I_F \cup I_K = \{1, \ldots, n\}$. With these definitions, (\ref{conv-a-comb}) reads
\begin{align}
a = \sum\limits_{i \in I_K} \alpha_i e_i + \sum\limits_{j \in I_F} \alpha_j e_j \,, \nonumber
\end{align}
and therefore, by what we have shown in step (ii), we obtain
\begin{align}
f(a) = \sum\limits_{i \in I_K} \alpha_i \underbrace{f(e_i)}_{< k} + \sum\limits_{j \in I_F} \alpha_j \underbrace{f(e_j)}_{= k} \leq k \,, \nonumber
\end{align}
with equality if and only if $a \in F$ (since $\sum_{i=1}^n \alpha_i = 1$). This proves $P \subset H$ and $F = P \cap \partial H$, where $\partial H = \{ v \in V \mid f(v) = k \}$.

\end{enumerate}

\item (b) $\Rightarrow$ (a): Let the vector space containing $P$ be denoted by $V$. Let $H$ and $\partial H$ be sets of the form $H = \{ v \in V \mid f(v) \leq k \}$ and $\partial H = \{ v \in V \mid f(v) = k \}$ for some linear functional $f$ and some $k \in \mathbb{R}$. Let $F = P \cap \partial H$. Suppose there are $x, y \in P, 0 < \alpha < 1$ such that $\alpha x + (1 - \alpha) y \in F$. Then
\begin{align}
&f( \alpha x + (1-\alpha) y) = \alpha \underbrace{f(x)}_{\leq k} + (1-\alpha) \underbrace{f(y)}_{\leq k} = k \nonumber \\
&\Rightarrow f(x) = f(y) = k \quad \Rightarrow x, y \in F \,. \nonumber
\end{align}
Thus, $F$ is a proper face of $P$. \hfill \qedhere

\end{itemize}

\end{proof}

Note that the equivalence stated in Proposition \ref{face-equivalence} does not hold in the more general case of convex sets. In $\mathbb{R}^2$, let $C$ be the union of lower open half-space and the non-negative $x$-axis (which is a convex set), let $F$ consist of the origin. Then $F$ is a face, but $F$ is not of the form $F = C \cap \partial H$ for some closed half-space $H$. Proposition \ref{face-equivalence} has the following immediate consequence.

\begin{cor}
\label{face-of-polytope-polytope}
A face $F$ of a polytope $P$ is a polytope.
\end{cor}

\begin{proof}
The improper face $F = P$ is by assumption a polytope. If $F$ is a proper face, then it is the intersection of $P$ with an affine hyperplane $\partial H$. $P$ is compact and $\partial H$ is closed, so $F$ is compact. Proposition \ref{face-conv-ext} implies that $\ext(F)$ is finite, so $F$ is a polytope.
\end{proof}

Since we have defined the dimension of a polytope, Corollary \ref{face-of-polytope-polytope} suggests the definition of the dimension of a face.

\begin{defi}
\label{face-dimension-def}
The \textbf{dimension of a face $F$}\index{dimension!of a face} of a polytope is the dimension of $F$ as a polytope. We say that a face of dimension $d$ is a \textbf{$d$-face}\index{face!$d$-face}. If $F$ is a $(d-1)$-face of a $d$-polytope, then $F$ is called a \textbf{facet}\index{facet} of $P$.
\end{defi}

Now we have enough technical background to consider a very central class of polytopes which are called \emph{simplices}. They will be very important when we deal with generalized probabilistic theories. The theories with simplices as sets of states are precisely the classical theories, as we will explain in Example \ref{classical-theory-ex}.

\begin{ex}[Simplices]
\label{simplex-example}
A \textbf{$d$-simplex}\index{simplex}\index{simplex!$d$-simplex} is the convex hull of $d+1$ affinely independent points.

\begin{figure}[htb]
\centering

\begin{pspicture}[showgrid=false](0,-1)(1,1)
\psdot[dotsize=1pt 2](0,0.3)
\end{pspicture}
\begin{pspicture}[showgrid=false](0,-1)(2.5,1)
\psline[ linewidth=0.7\pslinewidth](0,-0.5)(0.75,1)
\end{pspicture}
\begin{pspicture}[showgrid=false](0,-1)(1.3,1)
\PstTriangle[PstPicture=false, linewidth=0.7\pslinewidth]
\end{pspicture}
\begin{pspicture}[showgrid=false](-1.5,-1)(1,1) 
\psset{viewpoint=10 20 30 rtp2xyz,Decran=70} 
\psSolid[Decran=4, linewidth=0.7\pslinewidth,
object=tetrahedron, 
r=3, 
action=draw]%
\end{pspicture}

\caption{From left to right, we have a 0-simplex (a point), a 1-simplex (a line), a 2-simplex (a triangle) and a 3-simplex (which is also called a tetrahedron).}
\label{simplex-figure}
\end{figure}
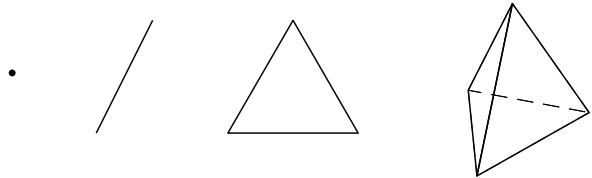

For every $d \in \mathbb{N}$, there is precisely one type of $d$-simplex.\footnote{Two $d$-simplices can be bijectively mapped to each other by an affine map. By what we have discussed at the end of Section~\ref{general-convex-section}, this means that any two $d$-simplices have the same convexity-structure.} From Proposition \ref{face-conv-ext}, we see that a face of a simplex is again a simplex since any subset of an affinely independent set of points is affinely independent. Given a $d$-simplex, we can easily construct a $(d+1)$-simplex. We simply have to add an affinely independent point and take the convex hull of this point and the $d$-simplex. For example, a tetrahedron can be constructed from a triangle by adding a point which is affinely independent of the triangle and taking the convex hull of the point and the triangle.
\hfill $\blacksquare$
\end{ex}

A characterizing property of simplices is the uniqueness of the convex combination of extreme points which gives an element of the simplex.

\begin{prop}
\label{simplex-unique-comb}
For a $d$-polytope $P$, the following are equivalent:
\begin{enumerate}[(a)]
\item $P$ is a simplex.
\item Every element $x \in P$ is a \emph{unique} convex combination of extreme points of $P$.
\end{enumerate}
\end{prop}

\begin{proof}
We prove the two directions separately. \\
\begin{itemize}

\item (a) $\Rightarrow$ (b):
Let $P$ be a $d$-simplex with extreme points $\ext(P) = \{ e_1, \ldots, e_{d+1} \}$ and let
\begin{align}
x = \sum\limits_{i=1}^{d+1} \alpha_i e_i = \sum\limits_{i=1}^{d+1} \beta_i e_i \label{two-conv-combs}
\end{align}
be two convex combinations of extreme points which yield $x$. Suppose that for any $k \in \{ 1, \ldots, d+1 \}$, we have that $\alpha_k \neq \beta_k$. Then, from (\ref{two-conv-combs}), we can construct an affine combination
\begin{align}
&e_k = \underset{i \neq k}{\sum\limits_{i=1}^{d+1}} \frac{\beta_i - \alpha_i}{\alpha_k - \beta_k} e_i \quad \text{with} \nonumber \\
&\underset{i \neq k}{\sum\limits_{i=1}^{d+1}} \frac{\beta_i - \alpha_i}{\alpha_k - \beta_k} = \frac{(1 - \beta_k) - (1 - \alpha_k)}{\alpha_k - \beta_k} = 1\,. \nonumber
\end{align}
But this is impossible since $P$ is assumed to be a simplex, for which (by the definition of a simplex) $\ext(P) = \{ e_1, \ldots, e_{d+1} \}$ is an affinely independent set. Therefore, $\alpha_k = \beta_k$ for every $k \in \{ 1, \ldots, d+1 \}$ since $k$ was arbitrary, so the convex combination of extreme points which gives $x$ is unique.

\item (b) $\Rightarrow$ (a): We have to show that $| \ext(P) | = d + 1$ and that $\ext(P)$ is affinely independent. Say that $\ext(P) = \{ e_1, \ldots, e_n \}$. Suppose there is a $k \in \{ 1, \ldots, n \}$ such that there exists an affine combination
\begin{align}
\underset{i \neq k}{\sum\limits_{i = 1}^n} \alpha_i e_i = e_k\,. \label{affine-assumption}
\end{align}
From this we can construct convex combinations:
\begin{align}
&I^+_k := \{ i \in \{1, \ldots, n\} \mid i \neq k, \alpha_i > 0\}\,, \nonumber \\
&I^-_k := \{ i \in \{1, \ldots, n\} \mid i \neq k, \alpha_i < 0\} \nonumber \\
\Rightarrow \quad &\sum\limits_{i \in I^+_k} \alpha_i e_i = e_k - \sum\limits_{j \in I^-_k} \alpha_j e_j \nonumber \\
\Leftrightarrow \quad &\sum\limits_{i \in I^+_k} \left( \frac{\alpha_i}{\sum\limits_{l \in I^+_k} \alpha_l} \right) e_i = \left( \frac{1}{\sum\limits_{l \in I^+_k} \alpha_l} \right) e_k + \sum\limits_{j \in I^-_k} \left( - \frac{\alpha_j}{\sum\limits_{l \in I^+_k} \alpha_l} \right) e_j\,. \label{two-convex-combs}
\end{align}
It is easily checked that both sides of equation (\ref{two-convex-combs}) are convex combinations. They obviously differ since the convex combination on the left side does not contain $e_k$ whereas the one on the right side does. This contradicts the assumption that every point in $P$ is a \emph{unique} convex combinations of elements of $\ext(P)$. Therefore, there cannot be a $k \in \{ 1, \ldots, n \}$ such that (\ref{affine-assumption}) holds which proves that $\ext(P)$ is an affinely independent set. The set $P$ is a $d$-polytope, so $\ext(P)$ must contain precisely $d+1$ affinely independent points. \hfill \qedhere

\end{itemize}
\end{proof}

Above, we have developed a half-space- and hyperplane-picture for polytopes and faces. Now, we prove a property of polytopes which will in turn allow us to prove a very important proposition for abstract state spaces in Section \ref{gpt-section}.

\begin{prop}
\label{halfspace-rep}
Let $P$ be a polytope. Assume that $P$ can be represented as the intersection of a \emph{given} finite set of closed half-spaces, $P = \bigcap_{i \in I} H_i$. Let $F$ be a face of $P$. Then there is an $l \in I$ such that $F$ is contained in the hyperplane defined by $H_l$, i.e. $F \subset \partial H_l$.
\end{prop}

\begin{proof}
Let the vector space $V$ containing $P$ be equipped with any inner product, turning $V$ into a Hilbert space. An affine hyperplane is a closed and convex subset of $V$, so according to the Hilbert Projection Theorem \ref{hilbert-projection},
\begin{equation}
\begin{array}{rlll}
d( \ \cdot \ , H_i): & V & \rightarrow & \mathbb{R} \nonumber \\
& x & \mapsto & \min\limits_{y \in \partial H_i} || x - y ||
\end{array}
\end{equation}
is well-defined for every $i \in I$. By Corollary \ref{face-of-polytope-polytope}, $F$ is a polytope, so $\ext(F)$ is a finite set. Let $n := | \ext(F) |$, say $\ext(F) = \{ e_1, \ldots, e_n \}$. Let
\begin{align}
x := \sum\limits_{i=1}^n \frac{1}{n} e_i \in F\,. \nonumber
\end{align}
The index set $I$ is finite, so $D_x := \min_{i \in I} d(x, \partial H_i)$ exists. Let $w \in V$ be a unit vector normal to $F$\footnote{Such a vector $w$ exists: According to Proposition \ref{face-equivalence}, there is a linear functional $f \in V^*$ defining a hyperplane $\partial \widetilde{H}$ which contains $F$. Then, the vector $w$ for which $f(\ \cdot \ ) = \langle \ \cdot \ , w \rangle$ (which exists by the Riesz Representation Theorem) is normal to $F$.}. Consider the line segment
\begin{align}
L := \{ \lambda (x + D_x w) + (1-\lambda) (x - D_x w) \mid \lambda \in [0,1] \}\,. \nonumber
\end{align}
The face $F$ has to be contained in every half-space $H_i$ (otherwise $F$ would not be contained in $P = \bigcap_{i \in I} H_i$). Moreover, we have that
\begin{align}
d(x, x + D_x w) &= || x - (x + D_x w) || = D_x || w || = D_x = \min_{i \in I} d(x, \partial H_i) \nonumber \\
&\leq d(x, \partial H_i) \quad \text{for all } i \in I\,. \nonumber
\end{align}
In words, we have just shown that for every half-space $H_i$, it holds that
\begin{itemize}
\item $x$ is in $H_i$ and
\item $x + D_x w$ is closer to $x$ than $\partial H_i$ is to $x$.
\end{itemize}
This implies that $(x + D_x w) \in H_i$ for every $i \in I$ and therefore $(x + D_x w) \in P = \bigcap_{i \in I} H_i$. Analogously, $(x - D_x w) \in P$.

The element of the line segment $L$ corresponding to $\lambda = \frac{1}{2}$ is contained in $F$. The set $F$ is a face of $P$, so by the definition of a face, the whole line segment $L$ is in $F$. The vector $w$ is normal to $F$, so $L$ can only be contained in $F$ if $D_x = 0$. By the definition of $D_x$, this implies that there is a hyperplane $H_l \in \{H_i\}_{i \in I}$ whose boundary $\partial H_l$ contains $x$. The fact that $x \in \partial H_l$ means that there is a linear functional $f \in V^*$ and a $k \in \mathbb{R}$ such that $H_l = \{ v \in V \mid f(v) \leq k \}$ and $f(x) = k$. It holds that $F \subset P \subset H_l$, so $f(e_j) \leq k$ for all $j \in \{ 1, \ldots, n \}$. Recall that $\{e_1, \ldots, e_n \} := \ext(F)$.
\begin{align}
&k = f(x) = \sum\limits_{j=1}^n \frac{1}{n} \underbrace{f(e_j)}_{\leq k} \quad \Rightarrow \quad f(e_j) = k \quad \forall j \in \{ 1, \ldots, n \} \nonumber \\
&\Rightarrow \quad e_j \in \partial H_l \quad \forall e_j \in \ext(F), \quad \partial H_l \text{ convex} \quad \Rightarrow \quad F \subset H_l\,. \tag*{\qedhere}
\end{align}
\end{proof}

Note that the statement of Proposition \ref{halfspace-rep} would be false if the set of closed half-spaces would not be assumed to be finite. This is shown in the following example.

\begin{ex}[The square as the intersection of closed half-spaces]
We show two different representations of the square $P = \{ (x, y) \in \mathbb{R}^2 \mid 0 \leq x, y \leq 1 \}$ as the intersection of closed half-spaces (c.f. Figure \ref{halfspace-figure}).

\begin{enumerate}[(a)]

\item In the first example, the square is represented as the intersection of four half-spaces:
\begin{align}
&H_4 = \{ (x, y) \in \mathbb{R}^2 \mid x \geq 0 \} \,, \nonumber \\
&H_3 = \{ (x, y) \in \mathbb{R}^2 \mid x \leq 1 \} \,, \nonumber \\
&H_2 = \{ (x, y) \in \mathbb{R}^2 \mid y \leq 1 \} \,, \nonumber \\
&H_0 = \{ (x, y) \in \mathbb{R}^2 \mid y \geq 0 \} \,, \nonumber \\
&I = \{ 4, 3, 2, 0\} \,, \nonumber \\
&P = \bigcap\limits_{i \in I} H_i \,.
\end{align}
In this case, Proposition \ref{halfspace-rep} applies. For each facet of the square (i.e. for each edge of the square), there is an $i \in I$ such that the boundary $\partial H_i$ of $H_i$ contains the face. For example, the bottom facet of the square is contained in $\partial H_0$.

\item In this example, the square is given by the intersection of the following \emph{infinite} family of half-spaces:
\begin{align}
&H_4 = \{ (x, y) \in \mathbb{R}^2 \mid x \geq 0 \} \,, \nonumber \\
&H_3 = \{ (x, y) \in \mathbb{R}^2 \mid x \leq 1 \} \,, \nonumber \\
&H_2 = \{ (x, y) \in \mathbb{R}^2 \mid y \leq 1 \} \,, \nonumber \\
&H_{1/n} = \left\{ (x, y) \in \mathbb{R}^2 \mid y \geq -\frac{1}{n} \right\} \quad \text{for every } n \in \mathbb{N}\,, \nonumber \\
&I' = \left\{ 4, 3, 2, 1, \frac{1}{2}, \frac{1}{3}, \frac{1}{4}, \ldots \right\} \,, \nonumber \\
&P = \bigcap\limits_{i \in I'} H_i \,. \nonumber
\end{align}
In this case, Proposition \ref{halfspace-rep} does not apply. Indeed, there is no $i \in I'$ such that the bottom facet of the square is contained in $\partial H_i$ since $0 \notin I'$.

\end{enumerate}

\begin{figure}[htb]
\centering

\begin{pspicture}[showgrid=false](-1,-1)(3,3.5)
\psframe[fillstyle=solid,fillcolor=lightgray](0,0)(2,2)
\psline[linestyle=dashed](-3,0)(5,0)
\psline[linestyle=dashed](-3,2)(5,2)
\psline[linestyle=dashed](2,-1)(2,3)
\psline[linestyle=dashed](0,-1)(0,3)
\uput[90](0,3){$\partial H_4$}
\uput[90](2,3){$\partial H_3$}
\uput[0](5,0){$\partial H_0$}
\uput[0](5,2){$\partial H_2$}
\uput[180](-3.5,3.3){(a)}
\psdot(0,0)
\psdot(2,0)
\psdot(0,2)
\psdot(2,2)
\uput[225](0,0){(0,0)}
\uput[315](2,0){(1,0)}
\uput[45](2,2){(1,1)}
\uput[135](0,2){(0,1)}
\uput[270](1,1.3){$P$}
\end{pspicture}

\begin{pspicture}[showgrid=false](-1,-2)(3,4.5)
\psframe[fillstyle=solid,fillcolor=lightgray](0,0)(2,2)
\psline[linestyle=dashed](-3,-2)(5,-2)
\psline[linestyle=dashed](-3,-1)(5,-1)
\psline[linestyle=dashed](-3,-0.666)(5,-0.666)
\psline[linestyle=dashed](-3,-0.5)(5,-0.5)
\psline[linestyle=dashed](-3,-0.4)(5,-0.4)
\psline[linestyle=dashed](-3,-0.333)(5,-0.333)
\psdot[dotsize=1.5pt](1,-0.25)
\psdot[dotsize=1.5pt](1,-0.15)
\psdot[dotsize=1.5pt](1,-0.05)
\psdot[dotsize=1.5pt](-2,-0.25)
\psdot[dotsize=1.5pt](-2,-0.15)
\psdot[dotsize=1.5pt](-2,-0.05)
\psdot[dotsize=1.5pt](4,-0.25)
\psdot[dotsize=1.5pt](4,-0.15)
\psdot[dotsize=1.5pt](4,-0.05)
\psdot[dotsize=1.5pt](5.5,-0.35)
\psdot[dotsize=1.5pt](5.5,-0.25)
\psdot[dotsize=1.5pt](5.5,-0.15)
\psdot[dotsize=1.5pt](5.5,-0.05)
\psline[linestyle=dashed](-3,2)(5,2)
\psline[linestyle=dashed](2,-2.3)(2,3)
\psline[linestyle=dashed](0,-2.3)(0,3)
\uput[180](-3.5,3.3){(b)}
\uput[90](0,3){$\partial H_4$}
\uput[90](2,3){$\partial H_3$}
\uput[0](5,2){$\partial H_2$}
\uput[0](5,-2){$\partial H_1$}
\uput[0](5,-1.1){$\partial H_{1/2}$}
\uput[0](5,-0.68){$\partial H_{1/3}$}
\psdot(0,0)
\psdot(2,0)
\psdot(0,2)
\psdot(2,2)
\uput[135](0,0){(0,0)}
\uput[45](2,0){(1,0)}
\uput[45](2,2){(1,1)}
\uput[135](0,2){(0,1)}
\uput[270](1,1.3){$P$}
\end{pspicture}

\caption{Example (a) shows the square as the intersection of four half-spaces. In this case, every facet of the square is contained in the boundary of one of the four half-spaces. In Example (b), however, there is no half-space whose boundary contains the bottom facet of the square. \hfill $\blacksquare$}
\label{halfspace-figure}
\end{figure}
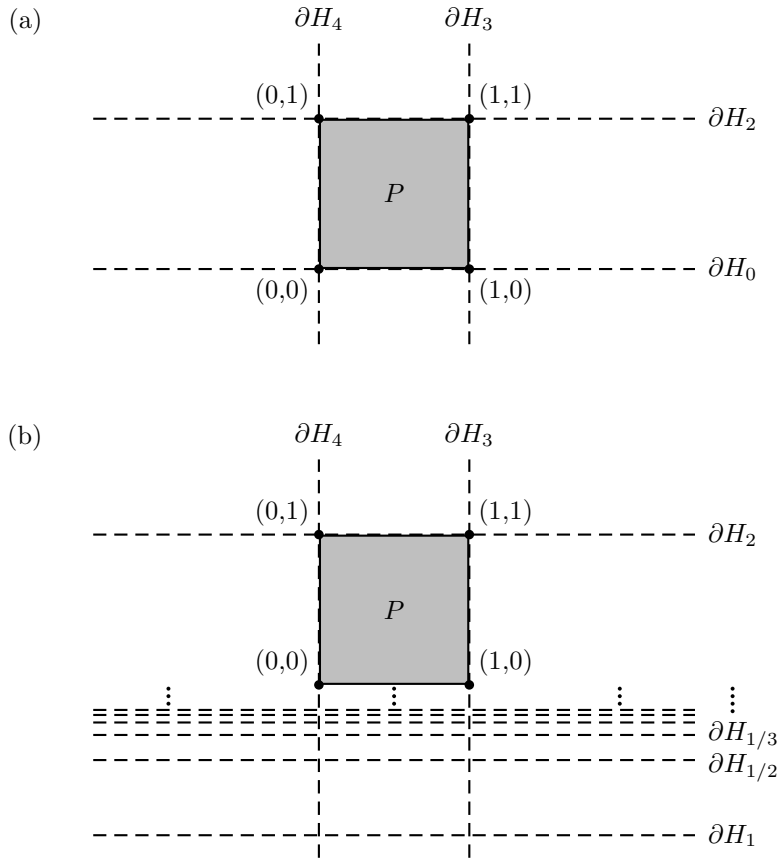

\end{ex}

Another important property of polytopes concerning their representation as the intersection of half-spaces is shown in the following theorem.

\begin{thm}[{\cite[Chapter 3.1]{Gruenbaum}}]
\label{polytope-facet-intersection}
Each polytope $P$ is the intersection of a finite family of closed half-spaces (c.f. Theorem \ref{polytope-bounded-polyhedral}). The smallest such family consists of those closed half-spaces containing $P$ whose boundaries are the affine hulls of the facets of $P$.
\end{thm}

Note that Theorem \ref{polytope-facet-intersection} does not imply Proposition \ref{halfspace-rep} (with the word ``face'' replaced by ``facet''): Theorem \ref{polytope-facet-intersection} only states the \emph{existence} of a family of half-spaces such that every facet is contained in one of the boundaries of the half-spaces, but it does not say that for a \emph{given} intersection, there must be a half-space with this property.

Let us get back to some intuitive properties of polytopes. We would expect that if $P$ is a $d$-polytope, then for every integer $0 \leq k \leq d$, $P$ has a $k$-face. This is indeed the case. More than that, the number of $k$-faces can be lower bounded by a positive number.

\begin{thm}[{\cite[Chapter 3.1]{Gruenbaum}}]
Let $P$ be a $d$-polytope, and for every integer $0 \leq k \leq d$, let $f_k(P)$ be the number of $k$-faces of $P$. Then
\begin{align}
f_k(P) \geq \binom{d+1}{k+1}\,. \nonumber
\end{align}
In particular, for every $0 \leq k \leq d$, $P$ has a $k$-face.
\end{thm}

With the aid of Theorem \ref{polytope-facet-intersection} and Proposition \ref{halfspace-rep}, we can prove another very intuitive and useful property of polytopes.

\begin{prop}
\label{face-sequence-prop}
Let $P$ be a polytope and let $F$ be a proper face of $P$. Then there exists a sequence
\begin{align}
\label{facet-sequence}
P = F_0 \supset F_1 \supset \ldots \supset F_k = F
\end{align}
of faces of $P$ such that $F_{i+1}$ is a facet of $F_i$ for every $i \in \{ 0, \ldots, k-1 \}$.
\end{prop}

\begin{proof}
By induction, it is sufficient to show that $F$ is a proper face of a facet of $P$. Let $\{H_i\}_{i \in I}$ be the finite family of half-spaces whose boundaries are the affine hulls of the facets of $P$. By Theorem \ref{polytope-facet-intersection}, $P = \bigcap_{i \in I} H_i$. By virtue of Proposition \ref{halfspace-rep}, there is an $l \in I$ such that $F \subset \partial H_l$. Proposition \ref{face-cap-aff} implies that $F_1 = \partial H_l \cap P$ is a facet of $P$. We know from Proposition \ref{face-equivalence} that there is a half-space $H$ containing $P$ such that $\partial H \cap P = F$. It follows that $F_1 \subset H$ and $F_1 \cap \partial H = F$, so $F$ is a face of $F_1$. This proves the existence of a sequence~(\ref{facet-sequence}) of \emph{subsets} such that $F_{i+1}$ is a facet of $F_i$ for every $i \in \{ 1, \ldots, k-1 \}$. By Proposition \ref{face-face-face}, those subsets are all faces of $P$.
\end{proof}

Before we dedicate ourselves to a few more properties of polytopes, this is a good point to introduce another example class of polytopes.

\begin{ex}[Pyramids]
\label{pyramid-example}
A \textbf{$d$-pyramid}\index{pyramid}\index{pyramid!$d$-pyramid} is a $d$-polytope $P$ such that there is some $(d-1)$-polytope $B$, called the \textbf{base}\index{base!of a pyramid} of the pyramid, such that $P = \conv(B \cup \{a\})$ for some point $a \in P$, which we call the \textbf{apex}\index{apex!of a pyramid} of the pyramid (c.f. Figure \ref{pyramids-figure}).

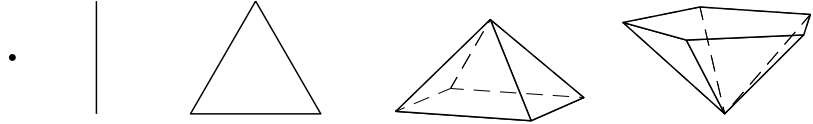
\begin{figure}[htb]
\centering

\begin{pspicture}[showgrid=false](0,-1)(1,1)
\psdot[dotsize=1pt 2](0,0.25)
\end{pspicture}
\begin{pspicture}[showgrid=false](0,-1)(1,1)
\psline[linewidth=0.7\pslinewidth](0,-0.5)(0,1)
\end{pspicture}
\begin{pspicture}[showgrid=false](-1,-1)(2,1)
\PstTriangle[PstPicture=false, linewidth=0.7\pslinewidth]
\end{pspicture}
\begin{pspicture}[showgrid=false](-1,0.5)(1.5,1.5)
\psset{viewpoint=40 16 8,Decran=100}
\psset{solidmemory}
\psSolid[object=new,linewidth=0.7\pslinewidth,
action=draw*,
name=A,
sommets= 
0 0 0 
0.420448 0.420448 0.5 
-0.420448 0.420448 0.5 
-0.420448 -0.420448 0.5 
0.420448 -0.420448 0.5 
0 0 1 
0 1.18921 1 
-1.18921 0 1 
0 -1.18921 1 
1.18921 0 1, 
faces={
[1 2 3 4]
[1 2 5]
[2 3 5]
[3 4 5]
[4 1 5]}]%
\end{pspicture}
\begin{pspicture}[showgrid=false](-1.5,-0.5)(1.5,1.5)
\psset{viewpoint=75 30 15,Decran=100}
\psset{solidmemory}
\psSolid[object=new,linewidth=0.7\pslinewidth,
action=draw*,
name=A,
fcol=7 (0.5 setfillopacity Red) 1 (0.5 setfillopacity Green) 10 (.5 setfillopacity Blue),
sommets= 
0 0 0 
0.343561 1.05737 1 
-0.899454 0.653491 1 
-0.899454 -0.653491 1 
0.343561 -1.05737 1 
1.11179 0 1, 
faces={
[0 2 1]
[0 3 2]
[0 4 3]
[0 5 4]
[0 1 5]
[1 2 3 4 5]}]%
\end{pspicture}

\caption{From left to right, we see a 0-, 1-, 2- and two 3-pyramids.}
\label{pyramids-figure}
\end{figure}

For $d=0,1,2$, there is only one type of $d$-pyramid simply because there is only one type of $(-1)$-, $0$- and $1$-polytope (which serves as a base for the $d$-pyramid). For $d \geq 3$, there are infinitely many different types of $d$-pyramids. If the base $B$ is a square and the apex $a$ is positioned centrally above $B$, we have the usual standard three-dimensional pyramid where this type of polytope gets its name from.
\hfill $\blacksquare$
\end{ex}

Now that we have just introduced pyramids, it is worth proving a lemma which will be useful in Section \ref{result-2-section}. To prove it, we make use of the following fact.

\begin{lemma}[{\cite[Chapter 3.1]{Gruenbaum}}]
\label{d-2-face-lemma}
If $P$ is a $d$-polytope, then each $(d-2)$-face $F$ of $P$ is contained in precisely two facets $F_1$ and $F_2$ of $P$, and $F = F_1 \cap F_2$.
\end{lemma}

This allows us to prove a fact about facets of pyramids that we naturally expect.

\begin{lemma}
\label{pyramid-facet-lemma}
Let $P$ be a pyramid with base $B$ and apex $a$, let $F$ be a facet of $B$. Then $\conv(F \cup \{a\})$ is a facet of $P$.
\end{lemma}

\begin{proof}
$F$ is a $(d-2)$-face of $P$. By Lemma \ref{d-2-face-lemma}, there are precisely two facets of $P$ containing $F$, one of which is $B$. Let the other facet of $P$ containing $F$ be denoted by $G$. We see from Proposition \ref{face-conv-ext} that $G$ contains more extreme points of $P$ than $F$. These additional extreme points have to be affinely independent of $F$ since $\aff(F) \cap P = F$ by Proposition \ref{face-cap-aff}. Consider one of these additional extreme points and let it be denoted by $p$. Then $\aff(G) = \aff(F \cup \{ p \})$ since $G$ is a $(d-1)$-polytope and $F$ is a $(d-2)$-polytope. If $p \in B$, then we have
\begin{align}
\aff(G) = \aff(F \cup \{p\}) = \aff(B) \,, \nonumber
\end{align}
which (by Proposition \ref{face-cap-aff}) implies
\begin{align}
\aff(G) \cap P = \aff(B) \cap P = B \nonumber
\end{align}
and therefore $G = B$. This contradicts our assumption that $G \neq B$, so $p$ cannot be contained in $B$. The only extreme point of $P$ which is not contained in $B$ is $a$, so $p = a$ and therefore $G = \conv(F \cup \{a\})$.
\end{proof}

\newpage

\section{Generalized probabilistic theories}
\label{gpt-section}

In this section, we introduce a framework of generalized probabilistic theories which generalizes classical theory and quantum theory to a more general setting. It is important to note that there is no standard framework for generalized probabilistic theories which is used overall. Instead, a few different frameworks have been considered. They mostly differ in the strength of their physical assumptions and the degree of the mathematical generality. The mathematical structure we are using in this thesis has been called the \emph{abstract state space}, see for example \cite{Barnum:2008p11512}, \cite{Barnum:2009p12289}, \cite{Barnum:2009p11466}, \cite{Barnum:2009p11048}. This section involves many mathematical definitions, but we will clarify their physical relevance by making examples concerning quantum theory, classical theory and some other special cases of generalized probabilistic theories.

We start with Section \ref{cones-and-ovs} where we give an introduction to cones and ordered vector spaces, which form the mathematical structure of abstract state spaces. In Section \ref{abs-st-sp-section}, we provide a derivation of the abstract state space formalism from physical assumptions. To our knowledge, such a derivation of the abstract state space formalism has not been published so far, so it is worth introducing the ideas behind the framework in this thesis. We will discuss a one-to-one correspondence between compact sets of states and abstract state spaces. Basically, abstract state spaces are the extension of the set of normalized states to the subnormalized states. At this point of our discussion, the state normalization is only a mathematical issue. Section \ref{meas-section} is dedicated to the definition of measurements on abstract state spaces and the investigation of their structure and properties. In Section \ref{norm-int-section}, we will give a physical interpretation of the state normalization. Section \ref{maximal-knowledge-section} is an attempt to give pure states a distinct physical interpretation. Finally, we will discuss transformations on abstract state spaces in Section \ref{trafo-section}.

\subsection{Cones and ordered vector spaces}
\label{cones-and-ovs}

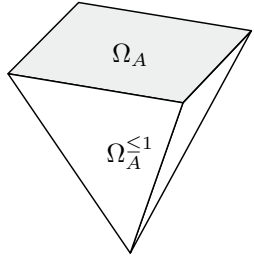
\begin{figure}[htb]
\centering

\begin{pspicture}[showgrid=false](-1.5,-1.2)(2,2)
\psset{viewpoint=40 16 20,Decran=140}
\psset{solidmemory}
\psSolid[object=new,
action=draw**, linewidth=0.7\pslinewidth,
name=A,
fcol=0 (0.15 setfillopacity Gray),
sommets= 
0 0 0 
0.420448 0.420448 0.5 
-0.420448 0.420448 0.5 
-0.420448 -0.420448 0.5 
0.420448 -0.420448 0.5 
0 0 -0.5 
0 1.18921 1 
-1.18921 0 1 
0 -1.18921 1 
1.18921 0 1, 
faces={
[1 2 3 4]
[5 2 1]
[5 3 2]
[5 4 3]
[5 1 4]}]%
\pstThreeDPut(0,0,1.5){$\Omega_A$}
\pstThreeDPut(0,0,0){$\Omega_A^{\leq 1}$}
\end{pspicture}

\label{cone-idea-figure}
\caption{This figure gives a geometric picture of the set of subnormalized states. Here, the set of normalized states $\Omega_A$ is assumed to be a square (gray). The set of subnormalized states $\Omega_A^{\leq 1}$ is given by all rescalings of $\Omega_A$ with scalars between 0 and 1.}

\end{figure}

We start this section with some intuition about cones in a physical theory. Suppose that the set of normalized states of a physical theory is given by a convex subset $\Omega_A$ of a vector space $A$ (we will make this assumption in Section \ref{abs-st-sp-section}). In quantum theory, for example, this is the set $\mathcal{S}(\mathcal{H})$ of density operators on a Hilbert space $\mathcal{H}$, and the normalization is given by the trace of the operator. As we will see in Sections \ref{norm-int-section} and \ref{trafo-section}, it is often useful not only to deal with normalized states but also with subnormalized states. In quantum theory, this means that it is useful to consider positive operators $\rho$ with $\tr(\rho) \in [0, 1]$ instead of density operators (with $\tr(\rho) = 1$) only. This extends the set of states under consideration from $\Omega_A$ to the set $\Omega_A^{\leq 1} = \{ \alpha \omega \in A \mid \alpha \in [0, 1], \omega \in \Omega_A\}$. Figure \ref{cone-idea-figure} shows this geometrically.

It is mathematically very convenient not only to consider rescalings with scalars between 0 and 1 but to consider a whole \emph{cone}. In this picture, a cone corresponds to rescalings of $\Omega_A$ with all positive scalars. As we will see below, this allows us to make use of mathematical structures like orders, dual orders, order intervals and more.

We start our definitions with the definition of a cone. Before we do this, it is convenient to introduce the following notation.

\begin{notation}
Throughout this thesis, we will use the following abbreviations. For any two subsets $M$ and $N$ of a real vector space and for any scalar $\alpha \in \mathbb{R}$, we denote
\begin{align}
&M + N := \{ m + n \mid m \in M, n \in N \} \,, \nonumber \\
&\alpha M := \{ \alpha m \mid m \in M \} \,. \nonumber
\end{align}
\end{notation}

\begin{defi}
\label{cone-def}
Let $V$ be a real vector space. A nonempty subset $K$ of $V$ is called a \textbf{cone}\index{cone} in $V$ if the following conditions are satisfied:\footnote{In the literature, a cone is sometimes defined by property (\ref{cone2}) alone. In this case, a cone satisfying (\ref{cone1}) is called a \emph{convex} cone, and a cone satisfying (\ref{cone3}) is called a \emph{salient} cone. We follow the definition in \cite{Aliprantis-Tourky} which coincides with our Definition \ref{cone-def}.}
\begin{align}
&\bullet \quad K + K = K, \label{cone1} \\
&\bullet \quad \alpha K = K \quad \forall \alpha \geq 0, \label{cone2} \\
&\bullet \quad K \cap (-K) = \{0\}. \label{cone3}
\end{align}
The \textbf{conical hull}\index{conical hull} of a subset $M$ of $V$ is given by
\begin{equation}
\cone(M) = \left\{ \sum\limits_{i=1}^n \alpha_i x_i \ \middle\vert \ n \in \{0, 1, 2, \ldots \}, x_i \in M, \alpha_i \geq 0 \right\} \,. \nonumber
\end{equation}
\end{defi}

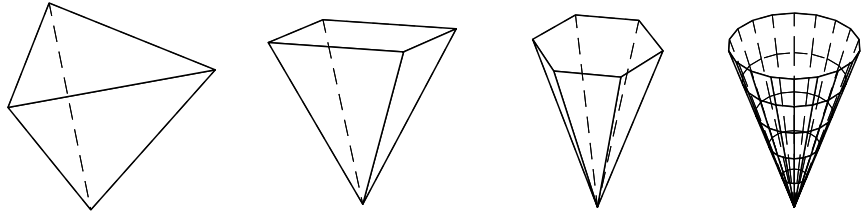
\begin{figure}[htb]
\centering

\begin{pspicture}[showgrid=false](-1.2,-1.5)(2,1.5)
\psset{viewpoint=50 -20 30 rtp2xyz,Decran=44} 
\psSolid[object=tetrahedron, linewidth=0.7\pslinewidth,
action=draw*,
RotY=180]
\end{pspicture}
\begin{pspicture}[showgrid=false](-1.5,-1.2)(2,1.5)
\psset{viewpoint=40 16 8,Decran=100}
\psset{solidmemory}
\psSolid[object=new,
action=draw*, linewidth=0.7\pslinewidth,
name=A,
sommets= 
0 0 0 
0.420448 0.420448 0.5 
-0.420448 0.420448 0.5 
-0.420448 -0.420448 0.5 
0.420448 -0.420448 0.5 
0 0 -0.5 
0 1.18921 1 
-1.18921 0 1 
0 -1.18921 1 
1.18921 0 1, 
faces={
[1 2 3 4]
[5 2 1]
[5 3 2]
[5 4 3]
[5 1 4]}]%
\end{pspicture}
\begin{pspicture}[showgrid=false](-1,-2.2)(1.5,0.5)
\psset{viewpoint=50 -20 30 rtp2xyz,Decran=22} 
\psSolid[object=cone, linewidth=0.7\pslinewidth,
mode=0,
action=draw*,
RotY=180]
\end{pspicture}
\begin{pspicture}[showgrid=false](-1,-2.2)(1,0.5)
\psset{viewpoint=50 -20 30 rtp2xyz,Decran=22} 
\psSolid[object=cone, linewidth=0.7\pslinewidth,
mode=4,
action=draw*,
RotY=180]
\end{pspicture}

\label{cone-figure-1}
\caption{A few examples of cones. For the sake of illustration, the cones have been truncated. They are actually infinitely high. The round cone on the very right is sometimes called the \emph{ice-cream cone} (for the obvious reason).}

\end{figure}

It is easy to verify that for any subset $M$ of a vector space $V$, the set $\cone(M)$ is a cone in $V$. Cones have an intuitive geometric picture. Figure \ref{cone-figure-1} shows examples of cones. Clearly, cones are convex sets. We will see below that there is a one-to-one correspondence between vector spaces with a cone and ordered vector spaces. The latter is given by the following two definitions.

\begin{defi}
A \textbf{partial order}\index{partial order} is a binary relation ``$\leq$'' over a set $P$ which is
\begin{align}
&\bullet \quad \text{reflexive:} && a \leq a \ \forall a \in P \,, \label{po1} \\
&\bullet \quad \text{antisymmetric:} && x \leq y \text{ and } y \leq x \quad \Rightarrow \quad x = y \,, \label{po2} \\
&\bullet \quad \text{transitive:} && x \leq y \text{ and } y \leq z \quad \Rightarrow \quad x \leq z \,. \label{po3}
\end{align}
\end{defi}

Sometimes, we will write $y \geq x$ for $x \leq y$.

\begin{defi}
\label{ovs}
A \textbf{partially ordered vector space}\index{partially ordered vector space|see{ordered vector space}} or \textbf{ordered vector space}\index{ordered vector space} is a real vector space $V$ and a partial order ``$\leq$'' over $V$ such that the following properties are satisfied:
\begin{align}
& \bullet \quad x \leq y \quad \Rightarrow \quad \alpha x \leq \alpha y &&\forall x, y \in V, \ \forall \alpha \geq 0 \,, \label{ovs1} \\
& \bullet \quad x \leq y \quad \Rightarrow \quad x + z \leq y + z &&\forall x, y, z \in V \,. \label{ovs2}
\end{align}
\end{defi}

This allows us to state a proposition which gives us a geometric picture for ordered vector spaces.

\begin{prop}
\label{cone-order-equiv}
Cones and ordered vector spaces obey the following correspondence:
\begin{enumerate}[(a)]
\item If $K$ is a cone in $V$, then $x \leq_K y :\Leftrightarrow y - x \in K$ defines a partial order on $V$ which turns $V$ into an ordered vector space.
\item If $(V, \leq)$ is an ordered vector space, then $V_+ := \{ v \in V \mid v \geq 0 \}$ defines a cone in $V$.
\end{enumerate}
\end{prop}

\begin{proof}

\begin{enumerate}[(a)]

\item We have to check the reflexivity, antisymmetry and transitivity of $\leq_K$ as well as the properties (\ref{ovs1}) and (\ref{ovs2}). Reflexivity is given if $x \leq_k x$, which means that $x - x = 0 \in K$. This is true since the zero vector is always an element of a cone by property (\ref{cone2}) or (\ref{cone3}). For antisymmetry, we need that $x - y \in K$ and $y - x = -(x-y) \in K$ implies $x = y$. This is true by property (\ref{cone3}) of a cone. Transitivity holds because property (\ref{cone1}) says that $y - x \in K$ and $z - y \in K$ implies $z - x = (z - y) + (y - x) \in K$. Property (\ref{ovs1}) follows directly from (\ref{cone2}) and (\ref{ovs2}) follows from $y - x \in K \Rightarrow (y + z) - (x + z) \in K$.

\item We have to check (\ref{cone1}) -- (\ref{cone3}). (\ref{cone1}) is immediately seen by noting that by (\ref{ovs2}), we have that $x \geq 0$ and $y \geq 0$ imply $x + y \geq 0$ and therefore $x + y \in K$. In a similar way, (\ref{cone2}) follows from (\ref{ovs1}). For (\ref{cone3}), suppose that $x \in K$ and $x \in -K$, i.e. $x \geq 0$ and $-x \geq 0$. Then by (\ref{ovs1}), we have that $2x \geq 0$. Now apply (\ref{ovs2}) to get $2x + (-x) \geq 0 + (-x)$, i.e. $x \geq -x$. Analogously, we get $-x \geq x$. By the antisymmetry of the order, we get $x = -x$ and therefore $x = 0$. We have inferred $x = 0$ from $x \in K$ and $x \in -K$ which shows $K \cap (-K) = \{0\}$.

\end{enumerate}
\end{proof}

This correspondence allows us to define the \emph{cone order} and the \emph{positive cone}.

\begin{defi}
\label{cone-order-def}
Let $(V, \leq)$ be an ordered vector space. The \textbf{positive cone}\index{cone!positive|see{positive cone}}\index{positive cone} $V_+$ of $V$ is given by
\begin{equation}
V_+ := \{ v \in V \mid v \geq 0 \} \,. \nonumber
\end{equation}
Conversely, let $V$ be real vector space and let $K$ be a cone in $V$. The \textbf{cone order}\index{cone order} ``$\leq_K$'' on $V$ induced by $K$ is given by
\begin{equation}
x \leq_K y :\Leftrightarrow y - x \in K \,. \nonumber
\end{equation}
If it is clear from the context by which cone the order is induced, the subscript $K$ is dropped and we write ``$\leq$'' instead of ``$\leq_K$''. Moreover, we write $x < y$ for $(x \leq y \text{ and } x \neq y)$, $x \geq y$ for $y \leq x$ and $x > y$ for ($x \geq y$ and $x \neq y$).
\end{defi}

Proposition \ref{cone-order-equiv} means that specifying an ordered vector space is equivalent to specifying a cone of a vector space, and we can refer to these two notions interchangeably.

There are many easily constructible examples of cones in $\mathbb{R}^n$, as suggested in Figure \ref{cone-figure-1}. Instead of explicitly writing down such a cone, we make a more abstract example.

\begin{ex}[The cone of positive operators on a Hilbert space]
\label{positive-operator-cone}
Let $\mathcal{H}$ be a finite-dimensional Hilbert space. The Hermitian operators on $\mathcal{H}$ form a real vector space $\Herm(\mathcal{H})$. Consider the subset of positive operators on $\mathcal{H}$. In comparison to the notation above, we have
\begin{align}
&V = \Herm(\mathcal{H}) \quad \text{for some finite-dimensional Hilbert space} \ \mathcal{H} \,, \nonumber \\
&V_+ = \{ T \in \Herm(\mathcal{H}) \mid \langle v | T | v \rangle \geq 0 \ \forall | v \rangle \in \mathcal{H} \} \,. \label{blelo}
\end{align}
It is easily verified that the positive operators fulfill the requirements (\ref{cone1}) -- (\ref{cone3}), i.e. that the positive operators indeed form a cone in $\mathcal{H}$. The name of \emph{positive} operators already suggests that this cone is induced by an order on $\Herm(\mathcal{H})$. The order is the usual operator order, given by
\begin{equation}
T \leq U \quad :\Leftrightarrow \quad \langle v | U - T | v \rangle \geq 0 \quad \forall | v \rangle \in \mathcal{H} \,. \label{blili}
\end{equation}
This example illustrates the close connection between ordered vector spaces and vector spaces with a cone. The conditions (\ref{blelo}) and (\ref{blili}) are very similar. It would be quite artificial to keep these two structures apart. \hfill $\blacksquare$
\end{ex}

The next definition that we make will be particularly important in the context of measurements on abstract state spaces in Section \ref{meas-section}.

\begin{defi}
For two elements $x, y \in V$ of a partially ordered set $V$ (in particular of an ordered vector space $V$), the set
\begin{equation}
[x, y] := \begin{cases} \{z \in V \mid x \leq z \leq y \} & \text{if } x \leq y \\ \emptyset & \text{if } x \nleq y \end{cases} \nonumber
\end{equation}
is called the \textbf{order interval}\index{order interval} from $x$ to $y$.
\end{defi}
\begin{figure}[htb]
\centering

\begin{pspicture}[showgrid=false](-1.7,-0.6)(2.3,2.4)
\psset{viewpoint=26 10 5,Decran=50}
\psset{solidmemory}
\psSolid[object=new, linewidth=0.7\pslinewidth,
action=draw*,
name=A,
sommets= 
0 0 0 
0.30729 0.945742 0.894427 
-0.804496 0.5845 0.894427 
-0.804496 -0.5845 0.894427 
0.30729 -0.945742 0.894427 
0.994412 0 0.894427 
-0.153645 -0.472871 0.552786 
0.402248 -0.29225 0.552786 
0.402248 0.29225 0.552786 
-0.153645 0.472871 0.552786 
-0.497206 0 0.552786 
0 0 1 
0.687121 2.11474 2 
-1.79891 1.30698 2 
-1.79891 -1.30698 2 
0.687121 -2.11474 2 
2.22357 0 2, 
faces={
[0 1 5]
[0 2 1]
[0 3 2]
[0 4 3]
[0 5 4]
[1 2 3 4 5]
}
]%
\uput[0](1.9,1){{\huge $\cap$}}
\uput[270](0,0){$\{ v \in V \mid v \geq x \}$}
\end{pspicture}
\begin{pspicture}[showgrid=false](-2.3,-2.6)(1.8,-0.3)
\psset{viewpoint=26 10 5,Decran=50}
\psset{solidmemory}
\psSolid[object=new, linewidth=0.7\pslinewidth,
action=draw*,
name=A,
sommets= 
0 0 0 
-0.30729 -0.945742 -0.894427 
0.804496 -0.5845 -0.894427 
0.804496 0.5845 -0.894427 
-0.30729 0.945742 -0.894427 
-0.994412 0 -0.894427 
0.153645 0.472871 -0.552786 
0.402248 -0.29225 0.552786 
0.402248 0.29225 0.552786 
-0.153645 0.472871 0.552786 
-0.497206 0 0.552786 
0 0 1 
0.687121 2.11474 2 
-1.79891 1.30698 2 
-1.79891 -1.30698 2 
0.687121 -2.11474 2 
2.22357 0 2, 
faces={
[0 1 2]
[0 2 3]
[0 3 4]
[0 4 5]
[0 5 1]
}]%
\uput[0](1.9,-1){{\huge =}}
\uput[270](0,-2){$\{ v \in V \mid v \leq y \}$}
\end{pspicture}
\begin{pspicture}[showgrid=false](-2.3,-0.6)(1.3,2.4)
\psset{viewpoint=26 10 5,Decran=57}
\psset{solidmemory}
\psSolid[object=new, linewidth=0.7\pslinewidth,
action=draw*,
name=A,
sommets= 
0 0 0 
0.152217 0.468477 0.414214 
-0.402248 0.29225 0.447214 
-0.402248 -0.29225 0.447214 
0.153645 -0.472871 0.447214 
0.497206 0 0.447214 
-0.153645 -0.472871 0.552786 
0.402248 -0.29225 0.552786 
0.402248 0.29225 0.552786 
-0.153645 0.472871 0.552786 
-0.497206 0 0.552786 
0 0 1 
0.343561 1.05737 1 
-0.899454 0.653491 1 
-0.899454 -0.653491 1 
0.343561 -1.05737 1 
1.11179 0 1, 
faces={
[0 1 8 5]
[0 5 7 4]
[0 4 6 3]
[0 3 10 2]
[0 2 9 1]
[6 4 7 11]
[7 5 8 11]
[8 1 9 11]
[9 2 10 11]
[10 3 6 11]
}]%
\uput[270](0,0){$[x, y]$}
\end{pspicture}

\caption{An order interval can be visualized as the intersection of an upward and a downward cone. For the sake of illustration, the cones have been truncated.}
\label{order-interval-figure}
\end{figure}
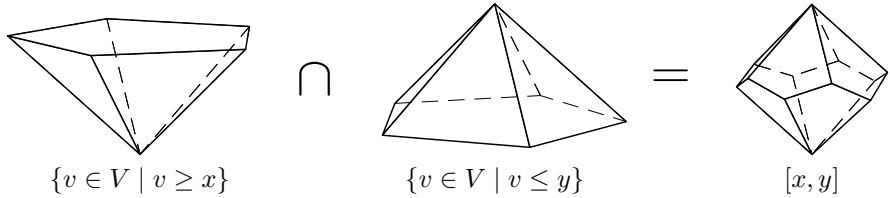

Another important concept in the context of measurements on abstract state spaces is the fact that a cone (an order) in a vector space $V$ induces a \emph{dual cone} (a \emph{dual order}) in the dual space $V^*$ of $V$. Before we define the dual cone (order), it is convenient to define what a positive linear functional on an ordered vector space is.

\begin{defi}
\label{positive-functional-def}
A linear functional $f \in V^*$ on an ordered vector space $V$ is \textbf{positive}\index{positive linear functional}\index{linear functional!positive|see{positive linear functional}} if $f(v) \geq 0 \ \forall v \in V_+$. The functional $f$ is called \textbf{strictly positive}\index{strictly positive linear functional}\index{linear functional!strictly positive|see{strictly positive linear functional}} if $f(v) > 0$ for all $v \in V_+ \backslash \{0\}$.
\end{defi}

\begin{defi}
\label{dual-cone-order}
Let $K$ be a cone in a real vector space $V$. The \textbf{dual cone}\index{dual cone} of $K$, denoted by $K^*$, is given by
\begin{equation}
\label{dual-cone-eq}
K^* := \{ f \in V^* \mid f(x) \geq 0 \ \forall x \in K \} \,. \nonumber
\end{equation}
Equivalently, the dual cone is the set of all positive linear functionals, where $K$ is regarded as the positive cone $V_+$. By Definition \ref{cone-order-def}, $K^*$ induces an order $\leq_{K^*}$ on $V^*$, which is called the \textbf{dual order}\index{dual order}.
\end{defi}

\begin{ex}[The dual cone of the positive operators and POVM elements]
\label{positive-operator-cone-2}
At this point, it is very instructive to reconsider the positive operators from Example \ref{positive-operator-cone}. A very natural way to look at the dual space of $\Herm(\mathcal{H})$ is via the Hilbert-Schmidt inner product
\begin{equation}
\langle T, U \rangle_\text{HS} := \tr(T^\dagger U) = \tr(TU) \,. \nonumber
\end{equation}
By the Riesz Representation Theorem, every vector $T$ in $\Herm(\mathcal{H})$ (i.e. every Hermitian operator $T$ on $\mathcal{H})$ induces a linear functional on $\Herm(\mathcal{H})$,
\begin{align}
\begin{array}{rccc}
f_T: & \Herm(\mathcal{H}) & \rightarrow & \mathbb{R} \\
& U & \mapsto & \tr(TU)
\end{array} \nonumber
\end{align}
This representation of linear functionals on $\Herm(\mathcal{H})$ by elements of $\Herm(\mathcal{H})$ naturally identifies $\Herm(\mathcal{H})$ with its dual space $(\Herm(\mathcal{H}))^*$. It does even more: It turns out that the linear functional $f_T$ is positive in the sense of Definition \ref{positive-functional-def} if and only if $T$ is a positive operator in $\Herm(\mathcal{H})$, i.e. $T \in V_+ = \{ T \in \Herm(\mathcal{H}) \mid \langle v | T | v \rangle \geq 0 \ \forall | v \rangle \in \mathcal{H} \}$. This means that the Hilbert Schmidt inner product, via the Riesz Representation Theorem, identifies the cone of positive operators $V_+ \subset \Herm(\mathcal{H})$ with its dual cone $V^*_+ \subset (\Herm(\mathcal{H}))^*$. By the equivalence of cones and orders, this also identifies the dual order with the order. This allows us to regard a positive operator as that what it is (a positive operator) as well as the linear functional associated with it. This expresses the fact that the cone of positive operators is \emph{strongly self-dual}\footnote{A positive cone $V_+$ is said to be \emph{strongly self-dual} if there is an invertible linear map $\phi: V_+^* \rightarrow V_+$ which is symmetric and positive, i.e. $f(\phi(e)) = e(\phi(f))$ for all $e, f \in V_+^*$ and $e(\phi(e)) \geq 0$ for all $e \in V^*$ \cite{Janotta:2011p10101}. We will not go into more detail concerning self-duality of cones.}.
To make this more clear, we consider the set of POVM elements on $\mathcal{H}$. As we will see in Section \ref{meas-section}, in the context of measurements, it is natural to consider order intervals of the dual cone. Here, the dual cone is $(\Herm(\mathcal{H}))^*$. If we denote the identity operator on $\mathcal{H}$ by $I$, then, as we will see below, the relevant order interval in quantum theory is the oder interval $[0, f_I]$ in $(\Herm(\mathcal{H}))^*$. We have that
\begin{equation}
[0, f_I] = \{ f \in (\Herm(\mathcal{H}))^* \mid 0 \leq f(T) \leq f_I(T) \ \forall T \in \Herm(\mathcal{H}) \} \,. \nonumber
\end{equation}
If we make use of the Hilbert Schmidt inner product and the Riesz Representation Theorem again, we can identify the order interval  $[0, f_I]$ in $(\Herm(\mathcal{H}))^*$ with the order interval $[0, I]$ in $\Herm(\mathcal{H})$. It reads as follows:
\begin{equation}
\label{povm}
[0, I] = \{ T \in \Herm(\mathcal{H}) \mid 0 \leq T \leq I \} \,. \nonumber
\end{equation}
The order interval $[0, I]$ is exactly the set of POVM elements on the Hilbert space $\mathcal{H}$. This way of making the functional behavior of the maps $f_T$ implicit by treating them as positive operators is very common and convenient. In fact, without this identification, we would have to treat POVMs as functionals rather than as operators. \hfill $\blacksquare$

\end{ex}

The next concept we are going to investigate is the notion of a \emph{base} of a cone.

\begin{defi}
Let $K$ be a cone in a vector space. A nonempty convex subset $\mathcal{B} \subset K \backslash \{0\}$ is said to be a \textbf{base}\footnote{This notion of a base has to be clearly distinguished from the base of a pyramid as introduced in Example \ref{pyramid-example}. In both cases, it is very common to call it a base, so we do not want to alter the terminology here. To make a distinction between the two, we denote the base of a pyramid by a normal $B$ and the base of a cone by a calligraphic $\mathcal{B}$.}\index{base!of a cone} for the cone $K$ if for each $x \in K \backslash \{0\}$ there exists $\lambda > 0$ and $b \in \mathcal{B}$ both uniquely determined such that $x = \lambda b$.
\end{defi}

It is important to notice that not every cone has a basis. For example,
\begin{align}
V_+ = \{ (x, y) \in \mathbb{R}^2 \mid y > 0 \} \cup \{ (x, y) \in \mathbb{R}^2 \mid y = 0, x \geq 0 \} \nonumber
\end{align}
is a cone in $\mathbb{R}^2$ but it has no base. The following theorem gives a precise characterization of the cones which allow for a base.

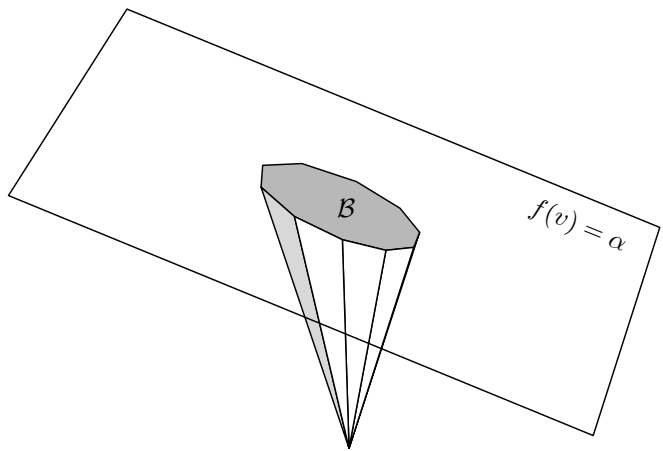
\begin{figure}[htb]
\centering

\begin{pspicture}[showgrid=false](-4.5,-3)(4.5,3.5) 
\psset{viewpoint=100 5 10 rtp2xyz,Decran=80, 
lightsrc=viewpoint,solidmemory,action=none} 
\psSolid[object=cone,linewidth=0.7\pslinewidth,RotY=180, mode=2,
a=4,name=my_octahedron,](0,0,3)
\psSolid[object=point, 
definition=solidcentreface, 
args=my_octahedron 1, 
name=G,] 
\psSolid[object=point, 
definition=mulv3d, 
args=G .8, 
name=H,] 
\psSolid[object=plan, linewidth=0.7\pslinewidth,
definition=solidface, 
args=my_octahedron 1, 
base=-4 4 -4 4, 
name=P,](H,,) 
\psSolid[object=load, linewidth=0.7\pslinewidth,
load=my_octahedron, 
plansepare={[0.2 0.4 1 -1]},
name=part] 
\psSolid[object=load,load=part1, 
action=draw**, linewidth=0.7\pslinewidth,
incolor={[rgb]{1 1 0.7}},] 
\psSolid[object=plan, linewidth=0.7\pslinewidth,
definition=equation, 
args={[0.2 0.4 1 -1]}, 
base=-5 5 -5 5,action=draw,name=awesome]
\uput[0]{-22}(2.2,0.8){$f(v)=\alpha$}
\pstThreeDPut(0.05,0,1){$\mathcal{B}$}
\end{pspicture} 

\caption{A visualization of Theorem \ref{base-strictly-positive}.}
\label{theorem-visualization}
\end{figure}

\begin{thm}[{\cite[Theorem 1.47]{Aliprantis-Tourky}}]
\label{base-strictly-positive}
A positive cone $V_+$ of an ordered vector space $V$ has a base if and only if $V$ admits a \emph{strictly} positive linear functional. More precisely, a subset $\mathcal{B}$ of $V_+$ is a base of $V_+$ if and only if there is a strictly positive linear functional $f \in V^*$ and a $\alpha > 0$ such that
\begin{equation}
\mathcal{B} = \{ v \in V_+ \mid f(v) = \alpha \} \quad \text{(see Figure \ref{theorem-visualization})} \,. \nonumber
\end{equation}
\end{thm}

\begin{ex}[Density operators as a base for the cone of positive operators]
\label{density-operator-base}
Once again, we consider the vector space $V = \Herm(\mathcal{H})$ of Hermitian operators on a finite-dimensional Hilbert space with the cone of positive operators (c.f. Examples \ref{positive-operator-cone} and \ref{positive-operator-cone-2}). The linear functional $f_I: T \mapsto \tr(T)$ is strictly positive, as one can easily see: Every positive operator has an eigenbasis and has only non-negative eigenvalues, so its trace is non-negative and vanishes if and only if it is the zero operator. According to Theorem \ref{base-strictly-positive}, this means that
\begin{equation}
\mathcal{S}(\mathcal{H}) = \{ \rho \in \Herm(\mathcal{H}) \mid \rho \text{ positive}, \tr(\rho) = 1 \} \,, \nonumber
\end{equation}
which is nothing but the set of density operators on $\mathcal{H}$, is a base for the cone of positive operators. \hfill $\blacksquare$
\end{ex}

In a finite-dimensional vector space, it is sometimes useful to make use of very basic topological properties of cone bases. Recall from Theorem \ref{unique-topo-thm} that in a finite-dimensional vector space, we do not have to specify a topology since in this case, there is a canonical topology compatible with the vector space structure. In Section \ref{abs-st-sp-section}, we will make use of the following fact.

\begin{thm}[{\cite[Chapter II.8]{Barvinok}}]
\label{compact-to-closed-thm}
Let $K$ be a cone in a finite-dimensional real vector space $V$ which has a compact base. Then $K$ is closed.
\end{thm}

Closed cones in finite-dimensional vector spaces are particularly neat because they show a certain duality property concerning their dual cone. The next proposition makes this statement more precise.

\begin{prop}
\label{dd-cone}
Let $V$ be a finite-dimensional real topological vector space, let $K$ be a closed cone in $V$. Then, the double dual cone $(K^*)^*$ in $(V^*)^* \cong V$ is identical to $K$.
\end{prop}

\begin{proof}
Recall that $V$ can be canonically identified with $(V^*)^*$ via $(v^*)^*(f) = f(v)$. Taking this into account, we have
\begin{align*}
K^* &= \{f \in V^* \mid f(v) \geq 0 \ \forall v \in K\} \\
(K^*)^* &= \{v \in V \mid f(v) \geq 0 \ \forall f \in K^*\}
\end{align*}
from which one can see that $K \subset (K^*)^*$. Thus, we have to show that $(K^*)^* \subset K$. Let $V$ be equipped with any norm (which necessarily induces the topology on $V$ since $V$ is finite-dimensional). Let $w \in V$, $w \notin K$. $K$ is closed and convex, so by virtue of the Hahn-Banach Theorem, there is a $g \in V^*$ with
\begin{equation}
\label{gwleqm}
g(w) < \inf \{ g(v) \mid v \in K \} =: k.
\end{equation}
We have that the zero vector $0$ is contained in $K$, so $k \leq 0$ since $g(0) = 0$. We cannot have $k < 0$: If $k < 0$, there is a $v \in K$ such that $g(v) < 0$. Multiplying $v$ with a large enough scalar $\alpha > 0$, we would have that $g(\alpha v) < g(w)$ while $\alpha v \in K$, which contradicts Inequality (\ref{gwleqm}). Therefore, $k = 0$ which implies that $g \in K^*$. We have shown that $w \notin K$ implies the existence of a linear functional $g$ with $g(w) < 0$, so $w \notin (K^*)^*$. This means that $(K^*)^* \subset K$ which completes the proof.
\end{proof}

Sometimes, it makes an important difference whether the vector space which contains a cone is chosen ``too big''. For example, one might consider the non-negative $y$-axis as a cone in $\mathbb{R}^2$. Another example is an ice-cream cone (c.f. Figure \ref{cone-figure-1}) as a cone in $\mathbb{R}^4$. In both cases, the linear span of the cone is a proper subspace of the vector space. In some contexts, one wants to exclude this case by requiring that the cone is \emph{generating}. A cone is generating if its linear span coincides with the vector space containing the cone. Noting that for a cone $K$, we have that $\spa(K) = K - K = \{ x - y \mid x, y \in K \}$, this gives the following definition.

\begin{defi}
A cone $K$ in a vector space $V$ is called \textbf{generating}\index{generating cone}\index{cone!generating|see{generating cone}} if $K - K = V$.
\end{defi}

The next notion we want to explain is the notion of an \emph{order unit}.

\begin{defi}
Let $V$ be an ordered vector space with positive cone $V_+$. A vector $e \in V_+$ is called an \textbf{order unit in $V$}\index{order unit}\index{order unit!in $V$} if for each $v \in V$ there exists some $\lambda > 0$ such that $v \leq \lambda e$. For the dual space, if $f \in V^*$ is an order unit in $V^*$ (with respect to the dual order), we might also say that $f$ is an \textbf{order unit on $V$}\index{order unit!on $V$}.
\end{defi}

In the context of abstract state spaces, where one considers closed and generating cones in finite-dimensional vector spaces (c.f. Section \ref{abs-st-sp-section}), the terms ``strictly positive linear functional'' and ``order unit'' are used synonymously. The following Theorem explains why one can do so.

\begin{thm}[{\cite[Theorem 3.5]{Aliprantis-Tourky}}]
\label{strictly-positive-order-unit}
For a closed and generating cone $V_+$ in a finite-dimensional vector space $V$ and for some $f \in V^*$, we have that $f$ is strictly positive if and only if $f$ is an order unit in $V^*$, i.e. an order unit on $V$.
\end{thm}

This concludes our mathematical introduction to cones and ordered vector spaces.

\subsection{The abstract state space}
\label{abs-st-sp-section}

In this subsection, we develop a particular kind of framework of generalized probabilistic theories which has also been called the \emph{abstract state space} formalism \cite{Barnum:2008p11512}, \cite{Barnum:2009p12289}, \cite{Barnum:2009p11466}, \cite{Barnum:2009p11048}. The introduction to this framework given here is not found in other references dealing with abstract state spaces. It reflects the view of the author of the present thesis and should not be regarded as a standard introduction. To our knowledge, such an introduction to abstract state spaces has not been published so far.

The framework relies on the following four central notions: \emph{probability}, \emph{system}, \emph{state} and \emph{measurement}. These notions will not be further specified here. Their meaning is assumed to be given. However, it is conceptionally important to notice that these notions do not have an independent meaning but only make sense in the context of each other. We will infer the framework from a number of assumptions. Those assumptions are part of the framework of an abstract state space. To distinguish these framework-based assumptions from the assumptions that we will make in Sections \ref{result-1-section} and \ref{result-2-section}, we call them ``Assumptions'', whereas we will refer to the assumptions in Sections \ref{result-1-section} and \ref{result-2-section} as ``Postulates''.

We start the derivation of an abstract state space by the specification of the \emph{set of normalized states}\index{set of normalized states}. As we have said above, we do not specify here what a state is but we assume that the meaning of this notion is given. However, we will explain what the term ``normalized'' stands for. We will explain at the end of the section why one can call them normalized from a mathematical point of view. In Section \ref{norm-int-section}, we will explain the physical interpretation of the state-normalization. For now, we might think of the set of normalized states as the set of those states of a system which are not conditioned on any event, whereas we will interpret the subnormalized states as states which are conditioned on a preceding random process.

The first assumption that we make is that the set of normalized states is a convex set. We have already motivated this assumption in Section \ref{randomization-section}. There we said that we want a set of states to be convex because we want to treat all probabilistic mixtures of states in a consistent way. In Section \ref{abstract-convexity-section}, we saw that from a very general and abstract point of view, this leads to the notion of a convex space. We mentioned that roughly speaking, convex spaces split into probabilistic and possibilistic spaces. Our concern are probabilistic theories, i.e. theories where the set of normalized states is a convex subset of a real vector space. According to the Theorem \ref{stone-thm} by Stone, this is equivalent to assuming that the set of normalized states is a cancellative convex space. This is our first assumption. In the literature, the cancellation property is usually not mentioned, but the set of states is assumed to be embedded in a vector space without further comments. We state the assumption of the cancellation property explicitly.

\begin{ass}
\label{canc-conv-ass}
For any system $A$, the \textbf{set of normalized states} is a cancellative convex space. In other words, the set of normalized states is a convex subset of a real vector space.
\end{ass}

The next assumption is very common not only in the framework of abstract state spaces but also in most (if not all) frameworks of generalized probabilistic theories that have been considered so far. It is of a purely technical nature, used to make the mathematics involved feasible. There is no immediate physical reason to make this assumption.

\begin{ass}
\label{finite-dim-ass}
The real vector space containing the convex subset of normalized states is finite-dimensional.
\end{ass}

With these two assumptions, we have a finite-dimensional real vector space at hand. Recall from Theorem \ref{unique-topo-thm} that in this case, we have a canonical topology on the vector space. This allows us to refer to topological properties of the space without explicitly specifying a topology. The next assumption that we make is of topological nature and reads as follows.

\begin{ass}
\label{compact-ass}
The set of normalized states is compact.
\end{ass}

From a mathematical point of view, this assumption facilitates a few technical issues. It also has the interesting physical \emph{consequence} that, by virtue of Theorem \ref{compact-to-closed-thm} and Proposition \ref{dd-cone}, it establishes some kind of duality between states and measurements. We will come back to this issue in Section \ref{meas-section}. However, the physical \emph{motivation} for this assumption is not completely undisputable. By Assumption \ref{finite-dim-ass}, we are in the finite-dimensional case, where the question of compactness divides into the questions of closedness and boundedness. We cannot give completely clear reasons for these two assumptions, but at least we want to say something about the closedness. Assume that a physical system can be prepared in a way such that certain statistics of measurements can be approximated arbitrarily well. In other words, assume that the state of a system can be prepared in states which are arbitrarily close to a certain ``state''. If one takes up the position that in this case, the approximated ``state'' should indeed be regarded as a state as well, then the assumption of closedness becomes natural. A more formal and mathematical argument would relate this to the fact that in a Hausdorff space, the set of all points which a series in a set can converge to is given by the closure of the set. We do not want to be dogmatic about this assumption. We make the assumption of compactness for technical reasons.

In the following, we motivate the abstract state space structure from Assumptions \ref{canc-conv-ass}, \ref{finite-dim-ass} and \ref{compact-ass}. To have an idea where this is going, we define what we mean by an abstract state space.

\begin{defi}
\label{abs-st-sp-def}
An \textbf{abstract state space}\index{abstract state space} is a tuple $(A, A_+, u_A)$, where $A$ is a finite-dimensional real topological vector space, $A_+$ is a closed and generating cone in $A$ and $u_A$ is an order unit in $A^*$ (i.e. an order unit on $A$).
\end{defi}

Here, in order to keep the introduction to the abstract state space formalism concise, we only give a rough picture of how a set of normalized states gives rise to an abstract state space at this point. For a precise mathematical proof of the equivalence of these two structures, we refer to Appendix \ref{appendix-b}. The proof is quite lengthy and is not necessary for an intuition for abstract state spaces, so we do not include it in this section here.

We start with a set of normalized states which satisfies Assumptions \ref{canc-conv-ass}, \ref{finite-dim-ass} and \ref{compact-ass}. Assumption \ref{canc-conv-ass} states that the set of normalized states is a cancellative convex space. In Section \ref{abstract-convexity-section}, we have seen that by Theorem \ref{stone-thm} (Stone), this means that we can assume that our set of normalized states is a convex subset of a real vector space. By Assumption \ref{finite-dim-ass}, this vector space is finite-dimensional. By Theorem \ref{unique-topo-thm}, we have a canonical topology on this vector space. Assumption \ref{compact-ass} states that in this topology, the set of normalized states is compact. We want to see that the set of normalized states can be seen as the base of a closed cone.

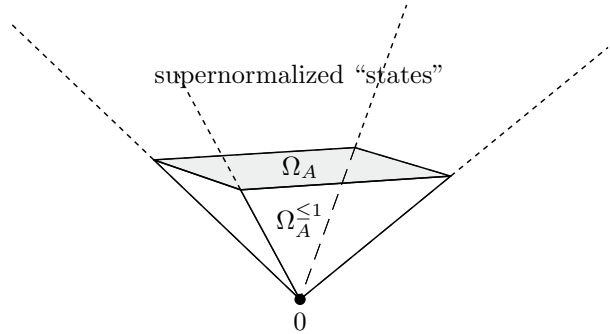
\begin{figure}[htb]
\centering

\begin{pspicture}[showgrid=false](-2,-0.5)(2,4.2)
\psset{viewpoint=26 10 5,Decran=50}
\psset{solidmemory}
\psSolid[object=new, linewidth=0.7\pslinewidth,
action=draw*,
name=B,
fcol=4 (.15 setfillopacity Gray),
sommets= 
0 0 0 
0.420448 0.420448 0.5 
-0.420448 0.420448 0.5 
-0.420448 -0.420448 0.5 
0.420448 -0.420448 0.5 
0 0 1 
0 1.18921 1 
-1.18921 0 1 
0 -1.18921 1 
1.18921 0 1, 
faces={
[0 7 6]
[0 8 7]
[0 9 8]
[0 6 9]
[6 7 8 9]
}]%
\psSolid[object=line, args=0 1.18921 1 0 2.37841 2, linestyle=dotted]
\psSolid[object=line, args=-1.18921 0 1 -2.37841 0 2, linestyle=dotted]
\psSolid[object=line, args=0 -1.18921 1 0 -2.37841 2, linestyle=dotted]
\psSolid[object=line, args=1.18921 0 1 2.37841 0 2, linestyle=dotted]
\psPoint(0,0,0.82){omega}
\psdots[dotsize=0](omega)
\uput[u](omega){$\Omega_A$}
\psPoint(0,0,0.38){omegal}
\psdots[dotsize=0](omegal)
\uput[u](omegal){$\Omega_A^{\leq 1}$}
\psPoint(0,0,1.5){omegas}
\psdots[dotsize=0](omegas)
\uput[u](omegas){supernormalized ``states''}
\psPoint(0,0,0){n}
\psdots[dotsize=0.15](n)
\uput[d](n){$0$}
\end{pspicture}

\caption{This figure visualizes how a compact convex set of normalized states forms the basis of a cone. The rescalings of $\Omega_A$ with factors between 0 and 1 form the set of subnormalized states $\Omega_A^{\leq 1}$. The rescalings with factors larger than 1 (the supernormalized ``states'') are not physical, but they are elements of the cone $A_+$.}
\label{cone-idea-figure-2}
\end{figure}

To see this, visualize a convex set $\Omega_A$ in a vector space $A$. Say that the affine hull $\aff(\Omega_A)$ of this set has dimension $d$. (For the sake of illustration, imagine that $\Omega_A$ is a square, i.e. $d=2$.) Assume that the vector space $A$ which contains $\Omega_A$ has dimension $d+1$. (In the case where $\Omega_A$ is a square, this means that $A$ is three-dimensional.) Imagine that $\Omega_A$ is placed ``somewhere above the origin'' of the vector space $A$ (in particular, $\Omega_A$ does not contain the origin). In this case, the set $\Omega_A$ generates a cone: The set $A_+ = \{ \alpha \omega \in A \mid \alpha \geq 0, \omega \in \Omega_A \}$ is a cone in $A$ (see figure \ref{cone-idea-figure-2}).

\begin{figure}[htb]
\centering

\begin{pspicture}[showgrid=false](-2,-0.5)(2,4.2)
\psset{viewpoint=26 10 5,Decran=50}
\psset{solidmemory}
\psSolid[object=new, linewidth=0.7\pslinewidth,
action=draw*,
name=B,
fcol=4 (.15 setfillopacity Gray),
sommets= 
0 0 0 
0.420448 0.420448 0.5 
-0.420448 0.420448 0.5 
-0.420448 -0.420448 0.5 
0.420448 -0.420448 0.5 
0 0 1 
0 1.18921 1 
-1.18921 0 1 
0 -1.18921 1 
1.18921 0 1, 
faces={
[0 7 6]
[0 8 7]
[0 9 8]
[0 6 9]
[6 7 8 9]
}]%
\psSolid[object=line, args=0 1.18921 1 0 2.37841 2, linestyle=dotted]
\psSolid[object=line, args=-1.18921 0 1 -2.37841 0 2, linestyle=dotted]
\psSolid[object=line, args=0 -1.18921 1 0 -2.37841 2, linestyle=dotted]
\psSolid[object=line, args=1.18921 0 1 2.37841 0 2, linestyle=dotted]
\psPoint(0,0,1.7){omegas}
\psdots[dotsize=0](omegas)
\uput[u](omegas){$A_+$}
\psPoint(0,0,0.82){omega}
\psdots[dotsize=0](omega)
\uput[u](omega){$\Omega_A$}
\psPoint(4.7,0,1){f}
\psdots[dotsize=0](f)
\uput[u](f){$u_A(\omega) = 1$}
\psSolid[object=plan, linewidth=0.7\pslinewidth,
definition=equation, 
args={[0 0 1 -1]}, 
base=-3 3 -2 2,action=draw,name=awesome]
\end{pspicture}

\caption{This figure visualizes the role of the order unit $u_A$ in an abstract state space.}
\label{cone-idea-figure-3}
\end{figure}
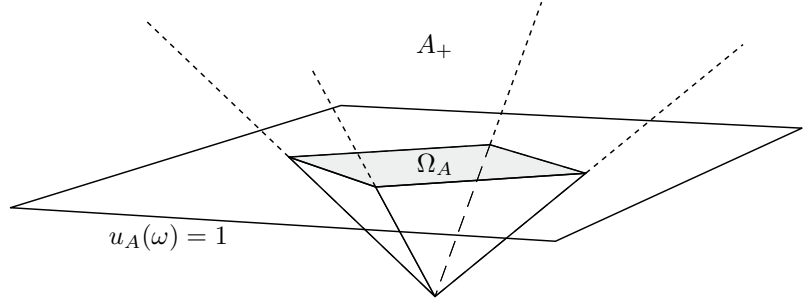

The cone $A_+$ is generating since we have assumed that the vector space is of only one dimension higher than the set of normalized states. The set of normalized states $\Omega_A$ is a base of the cone $A_+$. This cone is closed by virtue of Theorem \ref{compact-to-closed-thm} since $\Omega_A$ is compact. By Theorem \ref{base-strictly-positive}, there is a strictly positive linear functional, which we call $u_A$, such that $\Omega_A = \{ \omega \in A_+ \mid u_A(\omega) = 1 \}$ (see Figure \ref{cone-idea-figure-3}). In Theorem \ref{strictly-positive-order-unit}, we have seen that we can equivalently say that $u_A$ is an order unit in $A^*$, or an order unit on $A$.

This gives us an abstract state space $(A, A_+, u_A)$. We have only given a rough picture here. We have not \emph{proved} our claims. Moreover, we have not said whether the abstract state space $(A, A_+, u_A)$ constructed from the set $\Omega_A$ is unique. Conversely, we have not answered the question whether every abstract state space in turn gives rise to a compact convex set of normalized states $\Omega_A$. Roughly speaking, it turns out that both questions can be answered in the affirmative. There is a one-to-one correspondence between compact convex subsets of finite-dimensional vector spaces and abstract state spaces. For more details, we refer to Appendix \ref{appendix-b}.

In the following, when we talk about generalized probabilistic theories, we will always work in the abstract state space formalism. We make the following definitions.

\begin{defi}
For an abstract state space $(A, A_+, u_A)$, we define the \textbf{set of normalized states}\index{set of normalized states} by
\begin{align}
\Omega_A := \{ \omega \in A_+ \mid u_A(\omega) = 1 \} \,. \nonumber
\end{align}
In analogy to quantum theory, the extreme points of $\Omega_A$ are called \textbf{pure states}\index{pure state}\index{state!pure|see{pure state}}. We define the \textbf{set of subnormalized states}\index{set of subnormalized states} $\Omega_A^{\leq 1}$ by
\begin{align}
\Omega_A^{\leq 1} := \{ \omega \in A_+ \mid u_A(\omega) \leq 1 \} \,. \nonumber
\end{align}
\end{defi}

The interpretation of $\Omega_A^{\leq 1}$ will become clear in Section \ref{norm-int-section}. For an abstract state space $(A, A_+, u_A)$, we will often refer to the abstract state space merely by $A$ instead of the whole tuple $(A, A_+, u_A)$.

\begin{ex}[Quantum theory]
We already know all structures that describe quantum theory as an abstract state space from the Examples \ref{set-of-density-operators}, \ref{positive-operator-cone}, \ref{positive-operator-cone-2} and \ref{density-operator-base}:
\begin{align}
&\bullet \quad A = \Herm(\mathcal{H}) \text{ for some finite-dimensional Hilbert space } \mathcal{H} \,, \nonumber \\
&\bullet \quad A_+ = \{T \in \Herm(\mathcal{H}) \mid T \geq 0 \}, \text{ the set of positive operators on } \mathcal{H} \,, \nonumber \\
&\bullet \quad u_A = f_I: T \mapsto \tr(T) \,, \nonumber \\
&\bullet \quad \Omega_A = \mathcal{S}(\mathcal{H}) \,, \text{ the set of density operators on } \mathcal{H} \,. \tag*{$\blacksquare$}
\end{align}
\end{ex}

So far, we have used the term of a set of \emph{normalized} states without explaining the reason for that. Now, we argue mathematically about this notion. Given the positive cone $A_+$ of an abstract state space $A$, the set $\Omega_A$ is completely characterized by the order unit $u_A$ on $A$. It is not a priori clear why a linear functional should give rise to a set which is called \emph{normalized}, since one would expect that this is associated with a norm. It turns out, in fact, that given $A_+ \subset A$ and $u_A$ (and therefore $\Omega_A$), there is a canonical choice of a norm on $A$. We can get this norm by defining a norm $|| \cdot ||_{A^*}$ on $A^*$ induced by $\Omega_A$ and then take the dual norm $|| \cdot ||_A$ on $A$. The norm $|| \cdot ||_{A^*}$ is defined by
\begin{equation}
\label{norm-on-a-star}
|| f ||_{A^*} = \sup\limits_{\omega \in \Omega_A} | f(\omega) | \,.
\end{equation}

\begin{prop}
If $A_+ \subset A$ is a generating cone, $u_A$ is an order unit on $A$ and $\Omega_A = \{ \omega \in A_+ \mid u_A(\omega) = 1 \}$, then $|| f ||_{A^*} = \sup\limits_{\omega \in \Omega_A} | f(\omega) |$ defines a norm on $A^*$.
\end{prop}

\begin{proof}
We have to verify that $|| \cdot ||_{A^*}$ satisfies (a) positive homogeneity, (b) the triangle inequality and (c) definiteness, which are all very easy to check.
\begin{align}
&\text{(a)}: &&|| \alpha f ||_{A^*} = \sup\limits_{\omega \in \Omega_A} | \alpha f(\omega) | = | \alpha | \sup\limits_{\omega \in \Omega_A} | f(\omega) | = | \alpha | \ || f ||_{A^*} \nonumber \\
&\text{(b)}: &&|| f + g ||_{A^*} = \sup\limits_{\omega \in \Omega_A} | f(\omega) + g(\omega) | \leq \sup\limits_{\omega \in \Omega_A} | f(\omega) | + | g(\omega) | \nonumber \\
& &&\leq \sup\limits_{\omega \in \Omega_A} | f(\omega) | + \sup\limits_{\omega' \in \Omega_A} | g(\omega') | = || f ||_{A^*} + || g ||_{A^*} \nonumber \\
&\text{(c)}: &&|| f || = 0 \ \Rightarrow \ \Omega_A \subset \text{ker}(f) \ \Rightarrow \ \spa(\Omega_A) \subset \text{ker}(f) \nonumber \\
& &&\Rightarrow \ \text{ker}(f) = A \ (\text{since } \spa(\Omega_A) = A) \ \Rightarrow \ f = 0 \tag*{\qedhere}
\end{align}
\end{proof}

We will make use of the norm $|| \ \cdot \ ||_{A^*}$ in the proof of Proposition \ref{e_a-polytope} below.

\begin{defi}
For a normed vector space $V$ with norm $|| \cdot ||_V$, the \textbf{dual norm}\index{dual norm} $|| \cdot ||_{V^*}$ on $V^*$ is given by
\begin{equation}
\label{dual-norm}
|| f ||_{V^*} := \sup \{ |f(v)| \mid v \in V, || v ||_V \leq 1 \}.
\end{equation}
\end{defi}

It is not difficult to check that this gives indeed a norm. Substituting $V = A^*$ in this definition, we obtain a norm $|| \cdot ||_A$ on $A$. Comparing (\ref{dual-norm}) with $\Omega_A = \{ \omega \in A_+ \mid u_A(\omega) = 1 \}$, we see that
\begin{align}
|| \omega ||_A = u_A(\omega) \quad \forall \omega \in A_+\,. \nonumber
\end{align}
This explains, from a mathematical point of view, how it is justified to call $\Omega_A$ the set of \emph{normalized} states. We will see in Section \ref{norm-int-section} how the states in $\Omega_A$ are interpreted as opposed to subnormalized states in $\Omega_A^{\leq 1}$.

\subsection{Measurements on abstract state spaces}
\label{meas-section}

We have seen in Section \ref{abs-st-sp-section} that the specification of a compact convex set of normalized states gives rise to the structure of an abstract state space. This structure basically adds the state normalization to the framework. In this section, we will see that the abstract state space in turn gives rise to the mathematical structure of measurements.

Suppose that a measurement $\mathcal{M}$ is performed on a system. This measurement has some finite set of possible outcomes $\mathcal{I}_\mathcal{M} = \{ 1, \ldots, n \}$. In the abstract state space formalism, $\mathcal{M}$ is represented by a set of functions $\mathcal{M} = \{ e_1, \ldots, e_n \}$, where $e_k$ is associated with the measurement outcome $k \in \mathcal{I}_\mathcal{M}$. These functions are to be interpreted as follows. Suppose the system, prior to the measurement, is in the initial state $\omega$. If the measurement is performed on the system, then $e_k(\omega)$ is the probability that the measurement outcome is $k$. To allow a physical interpretation, these functions have to satisfy four properties. In the following, we discuss these four consistency properties.

The first property links to the idea that we have explained in Section \ref{randomization-section}. Suppose that the system undergoes a random process $\mathcal{R}$. This random process has the possible outcomes 1 and 2 which have the probabilities $p_1$ and $p_2$. If the outcome is 1, the state of the system after the random process is $\omega_1$, in the other case it is $\omega_2$. If we do not know the outcome of the random process, we have argued in Section \ref{randomization-section} that we would describe the state after the random process by the probabilistic mixture $\widetilde \omega = p_1 \omega_1 + p_2 \omega_2$. Assume that after the random process, we measure the system with respect to $\mathcal{M} = \{ e_1, \ldots, e_n \}$. The probability for the measurement outcome $k$ is given by $e_k(\widetilde \omega) = e_k( p_1 \omega_1 + p_2 \omega_2 )$. Recapitulate this situation. We do not know the outcome of the random process $\mathcal{R}$ which brings us to take the probabilistic mixture of the state. Then we apply $e_k$ to get the probability for the measurement outcome $k$. From a physical point of view, however, there is no reason why we should not calculate the probability for outcome $k$ by first calculating it given that we know the outcome of $\mathcal{R}$ and then take the probabilistic mixture of the probabilities $e_k(\omega_1)$ and $e_k(\omega_2)$. In other words, it is not physically determined ``where to mix''. Therefore, we expect that
\begin{align}
\label{conv-lin-equation}
e_k(p_1 \omega_1 + p_2 \omega_2) = p_1 e_k(\omega_1) + p_2 e_k(\omega) \,.
\end{align}
In Section \ref{randomization-section}, we have seen that in the case of quantum theory, the linearity of the expression
\begin{align}
p_l = \sum\limits_k p_k p_{l|k} = \sum\limits_k p_k \tr(Q_l \rho_k) \nonumber
\end{align}
(in other words the linearity of the function $[\rho_k \mapsto \tr(Q_l \rho_k)]$) allows us to regard probabilistic mixtures of states to be equivalent to probabilistic mixtures of probabilities. But in a generalized probabilistic theory, this is not a priori given. This means that in order to treat probabilistic mixtures in a consistent way, we have to \emph{assume} that Equation (\ref{conv-lin-equation}) holds, i.e. that the functions $e_k$ are convex-linear.

The physically reasonable domain of the functions $e_k$ is the set of subnormalized states $\Omega_A^{\leq 1}$ of an abstract state space $A$. (For the mathematical convenience, we will treat them as functions $e_k: A \rightarrow \mathbb{R}$ below.) The second property that we require from the functions $e_k$ is that they map the zero vector $0 \in A$ to zero. The reason for that is that (as we will see in Section \ref{norm-int-section}) we interpret the state $0 \in A$ as the state of a system which is conditioned on an impossible event. The joint probability of an impossible event and some other event must necessarily vanish. The first two properties that we have discussed imply that the functions $e_k$ are linear functionals (as we will see below).

If we want to interpret the values $e_k(\omega)$ as probabilities, then we have to require that
\begin{align}
0 \leq e_k(\omega) \leq 1 \quad \forall \omega \in \Omega_A^{\leq 1} \,. \nonumber
\end{align}
Equivalently, we can require that (given the $e_k$ are linear)
\begin{align}
0 \leq e_k(\omega) \leq 1 \quad \forall \omega \in \Omega_A \,. \nonumber
\end{align}
This is the third property that we demand from the functions $e_k$. Finally, it is very natural to assume that if we perform a measurement, it is certain that we get \emph{some} outcome. This leads to the fourth consistency property:
\begin{align}
\sum\limits_{k \in \mathcal{I}_\mathcal{M}} e_k(\omega) = 1 \quad \forall \omega \in \Omega_A \,. \nonumber
\end{align}

These four consistency requirements are \emph{necessary} for a physical interpretation of the functions $\{e_k\}_{k \in \mathcal{I}_\mathcal{M}}$. However, it is not clear why these requirements should be \emph{sufficient} in the sense that any set of functions $\{e_k\}_{k \in \mathcal{I}_\mathcal{M}}$ which satisfies the four conditions should correspond to a physical measurement. In the abstract state space formalism, it is assumed that any set of functions $\{e_k\}_{k \in \mathcal{I}_\mathcal{M}}$ satisfying the above mathematical requirements corresponds to a physical measurement. It should be pointed out that this causes a loss of generality of the framework. One might think of physical theories where not all mathematically defined measurements are possible, and such theories are not encompassed by the abstract state space formalism.

We state our assumption explicitly.

\begin{ass}
\label{meas-ass}
Any finite set $\mathcal{M} = \{ e_k \}_{k \in \mathcal{I}_\mathcal{M}}$ of functions $e_k: \Omega_A^{\leq 1} \rightarrow \mathbb{R}$ for which
\begin{align}
&\bullet \quad \text{each }e_k \text{ is convex-linear on } \Omega_A^{\leq 1} \,, \label{meas1} \\
&\bullet \quad e_k(0) = 0 \quad \forall k \in \mathcal{I}_\mathcal{M} \,, \label{meas2} \\
&\bullet \quad 0 \leq e_k(\omega) \leq 1 \quad \forall \omega \in \Omega_A, \forall k \in \mathcal{I}_\mathcal{M} \,, \label{meas3} \\
&\bullet \quad \sum\limits_{k \in \mathcal{I}_\mathcal{M}} e_k(\omega) = 1 \quad \forall \omega \in \Omega_A \label{meas4}
\end{align}
corresponds to a physical measurement. For an initial state $\omega \in \Omega_A$ prior to the measurement, the value $e_k(\omega)$ is the probability that a measurement with respect to $\mathcal{M} = \{ e_k \}_{k \in \mathcal{I}_\mathcal{M}}$ gives the outcome $k \in \mathcal{I}_\mathcal{M}$.
\end{ass}

\begin{prop}
Let $A$ be an abstract state space, let $\mathcal{M} = \{ e_k \}_{k \in \mathcal{I}_\mathcal{M}}$ be a set of functions $e_k: \Omega_A^{\leq 1} \rightarrow \mathbb{R}$. Then $\mathcal{M} = \{ e_k \}_{k \in \mathcal{I}_\mathcal{M}}$ satisfies properties (\ref{meas1}), (\ref{meas2}) and (\ref{meas3}) if and only if every function $e_k$ extends to a linear functional $e_k \in A^*$ which lies in the order interval between the zero functional and $u_A$ in the dual cone, i.e. $e_k \in [0, u_A] \subset A^*$. Property (\ref{meas4}) is satisfied if and only if $\sum_{k \in \mathcal{I}_\mathcal{M}} e_k = u_A$.
\end{prop}

\begin{proof}
We know from Theorem \ref{convex-linear-affine} that every convex-linear function is a linear function plus a translation. This implies that every convex-linear function which leaves the origin invariant is linear. Therefore, (\ref{meas1}) and (\ref{meas2}) imply that the $e_k$ extend to linear functionals, so we can say $e_k \in A^*$. The inequality $0 \leq e_k(\omega) \leq 1$ for all $\omega \in \Omega_A$ can be rewritten as $0 \leq e_k(\omega) \leq u_A(\omega)$ for all $\omega \in \Omega_A$. The set $\Omega_A$ is a basis of $A_+$, so $\{\alpha \omega \mid \alpha \geq 0, \omega \in \Omega_A \} = A_+$. Thus, by the linearity of the functionals $e_k$, the condition (\ref{meas3}) extends to the whole cone, i.e. $0 \leq e_k(\omega) \leq u_A(\omega)$ for all $\omega \in A_+$. This means that (\ref{meas3}) implies that the $e_k$ lie in $[0, u_A] := \{ f \in A^* \mid 0 \leq f \leq u_A \} \subset A^*$ in the dual order. Conversely, it is readily verified that every element of $[0, u_A]$ satisfies the properties (\ref{meas1}), (\ref{meas2}) and (\ref{meas3}). By an analogous argumentation, (\ref{meas4}) is equivalent to $\sum_{k \in \mathcal{I}_\mathcal{M}} e_k = u_A$.
\end{proof}

Functions that satisfy the properties (\ref{meas1}), (\ref{meas2}) and (\ref{meas3}) are commonly called \emph{effects}. The above leads us to the following definition.

\begin{defi}
\label{meas-def}
For an abstract state space $A$, we define the \textbf{set of effects}\index{set of effects} by the order interval $E_A := [0, u_A] = \{ f \in A^* \mid 0 \leq f \leq u_A \}$ in the dual order. An element $e \in E_A$ is called an \textbf{effect}\index{effect}. A \textbf{measurement}\index{measurement} is a set $\mathcal{M} = \{ e_1, \ldots, e_n \}$ of effects such that $\sum_{k=1}^n e_k = u_A$.
\end{defi}

It is easy to see that the set of effects $E_A$ is a convex set. It is the order interval from $0$ to $u_A$ and therefore it is the intersection of the positive (upward) cone $A_+^* = \{ f \in A^* \mid f \geq 0 \}$ in $A^*$ and the downward cone $\{ f \in A^* \mid f \leq u_A\}$ in $A^*$ (c.f. Figure \ref{order-interval-figure}). Thus, $E_A$ is the intersection of two convex sets and therefore convex.

There is something interesting to notice at this point. The cone $A^*_+$, which we could roughly call the ``effect cone'', is the dual cone of the ``state cone'' $A_+$. The state cone $A_+$ is a closed cone in a finite-dimensional space. By Proposition \ref{dd-cone}, the double dual cone $(A_+^*)^*$, i.e. the dual cone of the dual cone, is identical to the cone $A_+$. This establishes some kind of duality between states and measurements. Instead of specifying the triple $(A, A_+, u_A)$, one could just as well define an abstract state space by the triple $(A^*, A_+^*, u_A)$. In other words, one could specify an abstract state space by the definition of the measurements instead of by the definition of the states. This might be an attractive idea for people who take up the position that measurements are ``more operational'' than states.

Below, we will often focus on measurements that consist of effects which are extreme points of $E_A$. It is convenient to give these effects a special name.

\begin{defi}
An extreme point $e$ of the set of effects $E_A$ is called a \textbf{pure} effect. A measurement $\mathcal{M} = \{ e_1, \ldots, e_n \}$ is called a \textbf{pure measurement} if all effects $e_1, \ldots, e_n$ of the measurement are pure.
\end{defi}

\begin{ex}[POVMs in quantum theory]
\label{povm-in-qt}
We know from Example \ref{positive-operator-cone-2} that the dual cone $A_+^*$ of the cone $A_+$ of positive operators can be identified with $A_+$. We have also seen that $[0, f_I] \simeq [0, I]$ is the set of POVM elements. From the above definition, we get that a measurement in quantum theory is given by a set $\{ P_i \}_{i=1}^n$ of positive operators such that $\sum_{i=1}^n P_i = I$, as expected. There is something important to notice. The set of POVMs contains (orthogonal)\footnote{We are restricted to the set of Hermitian operators, in which an operator $P$ is a projector if and only if it is an orthogonal projector.} projectors. As we will see in Section \ref{pure-meas-section}, the projectors are precisely the pure effects in quantum theory. However, not every POVM element is a projector. Every positive operator $P$ with eigenvalues smaller or equal to one is a POVM element, and whenever one of the eigenvalues $\alpha$ satisfies $0 < \alpha < 1$, $P$ is not a projector. We will discuss in Section \ref{pure-meas-section} that there are important differences between projectors and other positive operators in the discussion of post-measurement states.
\end{ex}

\begin{ex}[The polygon models \cite{Janotta:2011p10101}]
\label{polygon-models}
The polygon models form a whole class of abstract state spaces. They have been studied in the context of nonlocality, but they also provide interesting examples in our context. In Section \ref{rep-subsp-section}, we will consider polygon models as counterexamples to properties of abstract state spaces that we assume to be physical.

For every $n \in \{3, 4, 5, \ldots\}$, there is a polygon model, which is defined by
\begin{align}
&\bullet \quad A = \mathbbm{R}^3, \nonumber \\
&\bullet \quad A_+ = \cone(\Omega_A^n), \text{ where} \nonumber \\
&\bullet \quad \Omega_A^n = \conv (\{ \omega_1, \ldots, \omega_n \}) \subset \mathbbm{R}^3, \text{ where} \nonumber \\
&\qquad \qquad \omega_i = \left(\begin{array}{c}r_n \cos \left( \frac{2 \pi i}{n} \right) \\r_n \sin \left( \frac{2 \pi i}{n} \right) \\1\end{array}\right) \in \mathbbm{R}^3 \,, \label{omega_i} \\
&\qquad \qquad r_n = \sqrt{\frac{1}{\cos(\frac{\pi}{n})}}\,. \nonumber
\end{align}
This induces the following set of effects:

\begin{align}
&\text{If $n$ is even,} \nonumber \\
&\bullet \quad E_A^n = \conv \{0, e_1, \ldots, e_n, u_A \}, \text{ where} \nonumber \\
&\qquad \quad u_A = \left(\begin{array}{c}0 \\0 \\1\end{array}\right), \quad e_i = \frac{1}{2} \left(\begin{array}{c}r_n \cos \left( \frac{(2i - 1) \pi}{n} \right) \\r_n \sin \left( \frac{(2i - 1) \pi}{n} \right) \\1\end{array}\right) \in (\mathbbm{R}^3)^* \cong \mathbbm{R}^3. \nonumber \\
&\text{If $n$ is odd,} \nonumber \\
&\bullet \quad E_A^n = \conv \{0, e_1, \ldots, e_n, u_A - e_1, \ldots, u_A - e_n, u_A \}, \text{ where} \nonumber \\
&\qquad \quad u_A = \left(\begin{array}{c}0 \\0 \\1\end{array}\right), \quad e_i = \frac{1}{1+r_n^2} \left(\begin{array}{c}r_n \cos \left( \frac{2 \pi i}{n} \right) \\r_n \sin \left( \frac{2\pi i}{n} \right) \\1\end{array}\right) \in (\mathbbm{R}^3)^* \cong \mathbbm{R}^3\,. \label{e_i-odd}
\end{align}

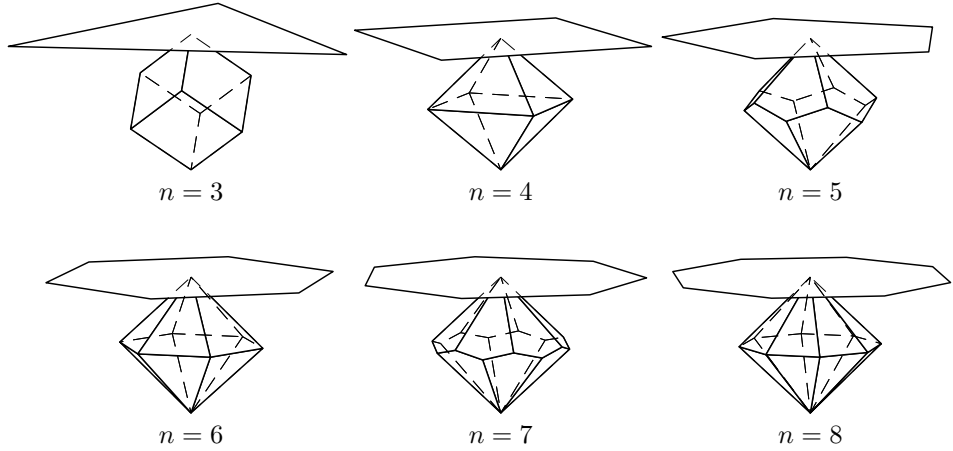
\begin{figure}[htb]
\centering

\begin{pspicture}[showgrid=false](-2,-0.5)(2,2.7)
\psset{viewpoint=20 50 10,Decran=100}
\psset{solidmemory}
\psSolid[object=new, linewidth=0.7\pslinewidth,
action=draw*,
name=C,
sommets=
0 0 0 
-0.707107 1.22474 1 
-0.707107 -1.22474 1 
1.41421 0 1 
-0.235702 0.408248 0.333333 
-0.235702 -0.408248 0.333333 
0.471405 0 0.333333 
0.235702 -0.408248 0.666667 
0.235702 0.408248 0.666667 
-0.471405 0 0.666667 
0 0 1, 
faces={
[0 4 8 6]
[0 6 7 5]
[0 5 9 4]
[7 6 8 10]
[8 4 9 10]
[9 5 7 10]
[1 2 3]
}]%
\uput[270](0,0){$n=3$}
\end{pspicture}
\begin{pspicture}[showgrid=false](-2,-0.5)(2,2.7)
\psset{viewpoint=26 10 5,Decran=50}
\psset{solidmemory}
\psSolid[object=new, linewidth=0.7\pslinewidth,
action=draw*,
name=B,
sommets= 
0 0 0 
0.420448 0.420448 0.5 
-0.420448 0.420448 0.5 
-0.420448 -0.420448 0.5 
0.420448 -0.420448 0.5 
0 0 1 
0 1.18921 1 
-1.18921 0 1 
0 -1.18921 1 
1.18921 0 1, 
faces={
[1 4 0]
[0 2 1]
[0 3 2]
[0 4 3]
[1 2 5]
[2 3 5]
[3 4 5]
[4 1 5]
[6 7 8 9]}]%
\uput[270](0,0){$n=4$}
\end{pspicture}
\begin{pspicture}[showgrid=false](-2,-0.5)(1.5,2.7)
\psset{viewpoint=26 10 5,Decran=50}
\psset{solidmemory}
\psSolid[object=new, linewidth=0.7\pslinewidth,
action=draw*,
name=A,
sommets= 
0 0 0 
0.152217 0.468477 0.414214 
-0.402248 0.29225 0.447214 
-0.402248 -0.29225 0.447214 
0.153645 -0.472871 0.447214 
0.497206 0 0.447214 
-0.153645 -0.472871 0.552786 
0.402248 -0.29225 0.552786 
0.402248 0.29225 0.552786 
-0.153645 0.472871 0.552786 
-0.497206 0 0.552786 
0 0 1 
0.343561 1.05737 1 
-0.899454 0.653491 1 
-0.899454 -0.653491 1 
0.343561 -1.05737 1 
1.11179 0 1, 
faces={
[0 1 8 5]
[0 5 7 4]
[0 4 6 3]
[0 3 10 2]
[0 2 9 1]
[6 4 7 11]
[7 5 8 11]
[8 1 9 11]
[9 2 10 11]
[10 3 6 11]
[12 13 14 15 16]}]%
\uput[270](0,0){$n=5$}
\end{pspicture}

\begin{pspicture}[showgrid=false](-2,-0.5)(2,2.7)
\psset{viewpoint=26 7 5,Decran=50}
\psset{solidmemory}
\psSolid[object=new, linewidth=0.7\pslinewidth,
action=draw*,
name=A,
sommets=
0 0 0 
0.537285 0.930605 1 
-0.537285 0.930605 1 
-1.07457 0 1 
-0.537285 -0.930605 1 
0.537285 -0.930605 1 
1.07457 0 1 
0.465302 0.268642 0.5 
0 0.537285 0.5 
-0.465302 0.268642 0.5 
-0.465302 -0.268642 0.5 
0 -0.537285 0.5 
0.465302 -0.268642 0.5 
0 0 1, 
faces={
[0 8 7]
[0 9 8]
[0 10 9]
[0 11 10]
[0 12 11]
[0 7 12]
[7 8 13]
[8 9 13]
[9 10 13]
[10 11 13]
[11 12 13]
[12 7 13]
[1 2 3 4 5 6]
}]
\uput[270](0,0){$n=6$}
\end{pspicture}
\begin{pspicture}[showgrid=false](-2,-0.5)(2,2.7)
\psset{viewpoint=26 7 5,Decran=50}
\psset{solidmemory}
\psSolid[object=new, linewidth=0.7\pslinewidth,
action=draw*,
name=A,
sommets=
0 0 0 
0.656862 0.82368 1 
-0.234432 1.02711 1 
-0.949194 0.457108 1 
-0.949194 -0.457108 1 
-0.234432 -1.02711 1 
0.656862 -0.82368 1 
1.05353 0 1 
0.311322 0.390385 0.473952 
-0.111109 0.486802 0.473952 
-0.449873 0.216647 0.473952 
-0.449873 -0.216647 0.473952 
-0.111109 -0.486802 0.473952 
0.311322 -0.390385 0.473952 
0.499321 0 0.473952 
-0.311322 -0.390385 0.526048 
0.111109 -0.486802 0.526048 
0.449873 -0.216647 0.526048 
0.449873 0.216647 0.526048 
0.111109 0.486802 0.526048 
-0.311322 0.390385 0.526048 
-0.499321 0 0.526048 
0 0 1, 
faces={
[0 13 16 12]
[0 12 15 11]
[0 11 21 10]
[0 10 20 9]
[0 9 19 8]
[0 8 18 14]
[0 14 17 13]
[16 13 17 22]
[17 14 18 22]
[18 8 19 22]
[19 9 20 22]
[20 10 21 22]
[21 11 15 22]
[15 12 16 22]
[1 2 3 4 5 6 7]
}]
\uput[270](0,0){$n=7$}
\end{pspicture}
\begin{pspicture}[showgrid=false](-2,-0.5)(1.5,2.7)
\psset{viewpoint=26 7 5,Decran=50}
\psset{solidmemory}
\psSolid[object=new, linewidth=0.7\pslinewidth,
action=draw*,
name=A,
sommets=
0 0 0 
0.73566 0.73566 1 
0 1.04038 1 
-0.73566 0.73566 1 
-1.04038 0 1 
-0.73566 -0.73566 1 
0 -1.04038 1 
0.73566 -0.73566 1 
1.04038 0 1 
0.480593 0.199068 0.5 
0.199068 0.480593 0.5 
-0.199068 0.480593 0.5 
-0.480593 0.199068 0.5 
-0.480593 -0.199068 0.5 
-0.199068 -0.480593 0.5 
0.199068 -0.480593 0.5 
0.480593 -0.199068 0.5 
0 0 1, 
faces={
[0 10 9]
[0 11 10]
[0 12 11]
[0 13 12]
[0 14 13]
[0 15 14]
[0 16 15]
[0 9 16]
[9 10 17]
[10 11 17]
[11 12 17]
[12 13 17]
[13 14 17]
[14 15 17]
[15 16 17]
[16 9 17]
[1 2 3 4 5 6 7 8]
}]
\uput[270](0,0){$n=8$}
\end{pspicture}

\caption{The polygon models for $n=3$ to $n=8$. The surface (polygon) at the top of each model represents the set of normalized states $\Omega_A^n$, while the crystal-like shape below the polygons represents the set of effects $E_A^n$. }
\label{3d-polygons}
\end{figure}

The fact that the polygon models are defined in $\mathbb{R}^3$ makes them particularly neat because they can be visualized (c.f. Figure \ref{3d-polygons}). It is easy to see that each set of effects $E_A$ is the intersection of an upward and a downward cone (c.f. Figure \ref{order-interval-figure}). The polygon model corresponding to $n=3$ is precisely a classical system with three pure states. We will examine classical systems in more detail in Example \ref{classical-theory-ex} below. The $n=4$ polygon model corresponds to a so-called \emph{gbit} (this stands for ``generalized bit''). A gbit represents the local state space of a frequently discussed bipartite model which is called the \emph{PR-box} or \emph{nonlocal box} \cite{Popescu:1994p5991}. 
\hfill $\blacksquare$
\end{ex}

\begin{ex}[Classical theory]
\label{classical-theory-ex}

This example is very central, both in probabilistic theories in general and in this thesis. We say that an abstract state space $A$ is a \textbf{classical theory}\index{classical theory} if the set of normalized states $\Omega_A$ is a simplex\index{simplex}. The reason why such an abstract state space is called classical is that this allows us to interpret a state as a classical probability distribution. To see this, recall from Proposition \ref{simplex-unique-comb} that each point in a simplex is a unique convex combination of its extreme points:
\begin{align}
&\Omega_A \text{ simplex,} \quad \ext(P) = \{ \omega_1, \ldots, \omega_n \}, \quad \omega \in \Omega_A \nonumber \\
&\Rightarrow \quad \exists \text{ unique } \ (p_1, \ldots, p_n) \text{ such that } \omega = \sum\limits_{i=1}^n p_i \omega_i, \quad p_i > 0, \quad \sum\limits_{i=1}^n p_1 = 1 \,. \nonumber
\end{align}
If we interpret $\omega_1, \ldots, \omega_n$ as mutually exclusive properties of the physical system, then we can interpret $(p_1, \ldots, p_n)$ as a probability distribution over these properties. The characterizing properties of a classical system are the fact that a state represents a unique probability distribution over its pure states and that these pure states can be perfectly distinguished, as we will discuss in the following.
\hfill $\blacksquare$
\end{ex}

In operational terms, we say that some states $\omega_1, \ldots, \omega_n$ are perfectly distinguishable if we can perform a measurement whose result allows us to determine in which of the states $\omega_1, \ldots, \omega_n$ the system was prior to the measurement (given that it was in one of these states). In a rigorous form, perfect distinguishability reads as follows.

\begin{defi}
\label{perf-dist-st-def}
Let $A$ be an abstract state space. We say that states $\omega_1, \ldots, \omega_n \in \Omega_A$ are \textbf{perfectly distinguishable}\index{perfectly distinguishable states}\index{states!perfectly distinguishable|see{perfectly distinguishable states}} if there is a measurement $\{e_1, \ldots, e_n\} \subset E_A$ on $A$ such that
\begin{align}
e_i(\omega_j) = \delta_{ij} \quad \forall i, j \in \{ 1, \ldots, n \} \,. \nonumber
\end{align}
\end{defi}

The following proposition can also be found in \cite[Lemma 24]{Muller:2011wl}.

\begin{prop}
\label{perf-dist-classical}
Let $A$ be a $d$-dimensional abstract state space. The following statements are equivalent.
\begin{enumerate}[(a)]
\item There are $d$ perfectly distinguishable states $\omega_1, \ldots, \omega_d \in \Omega_A$.
\item $\Omega_A$ is a $(d-1)$-simplex with $\ext(\Omega_A) = \omega_1, \ldots, \omega_d$, i.e. $A$ is a classical theory.
\end{enumerate}
\end{prop}

\begin{proof}
We prove the two implications separately.
\begin{itemize}

\item (a) $\Rightarrow$ (b): 
If the states $\omega_1, \ldots, \omega_d \in \Omega_A$ are perfectly distinguishable, then there are effects $f_1, \ldots, f_d \in E_A$ such that $f_i(\omega_j) = \delta_{ij}$ for all $i, j \in \{ 1, \ldots, d \}$. If the states $\omega_1, \ldots, \omega_d$ were linearly dependent in $A$, i.e. if there was an $i \in \{ 1, \ldots, d \}$ such that
\begin{align}
\omega_i = \underset{k \neq i}{\sum\limits_{k=1}^d} \alpha_k \omega_k \quad \text{with } \alpha_k \in \mathbb{R}\,, \label{lin-dep-contra}
\end{align}
then
\begin{align}
\underbrace{f_i(\omega_i)}_{1} \neq \underset{k \neq i}{\sum\limits_{k=1}^d} \alpha_k \underbrace{f_i(\omega_k)}_{0}\,, \nonumber
\end{align}
would lead to a contradiction to (\ref{lin-dep-contra}), so the states $\omega_1, \ldots, \omega_d$ are linearly independent vectors in $A$. Thus, they form a basis for $A$. For every $\omega \in \Omega_A$, we have that
\begin{align}
&\omega = \sum\limits_{k=1}^d \beta_k \omega_k \quad \text{for unique numbers } \beta_k \in \mathbb{R} \,, \label{is-conv-comb1}\\
&u_A(\omega) = \sum\limits_{k=1}^d \beta_k = 1\,, \label{is-conv-comb2} \\
&0 \leq f_i(\omega) = \sum\limits_{k=1}^d \beta_k f_i(\omega_k) = \beta_i \,. \label{is-conv-comb3}
\end{align}
Equations (\ref{is-conv-comb1}), (\ref{is-conv-comb2}) and (\ref{is-conv-comb3}) imply that any state $\omega \in \Omega_A$ is a unique convex combination of the $d$ states $\omega_1, \ldots, \omega_d$. This implies that $\Omega_A \subset \conv(\{ \omega_1, \ldots, \omega_d \}) \Rightarrow \Omega_A = \conv(\{\omega_1, \ldots, \omega_n\}) \Rightarrow \ext(\Omega_A) \subset \{ \omega_1, \ldots, \omega_d\}$. We have that $\aff(\Omega_A) = \aff(\ext(\Omega_A))$ is $(d-1)$-dimensional, so $\ext(\Omega_A)$ contains at least $d$ elements. Therefore, $\ext(\Omega_A) = \{ \omega_1, \ldots, \omega_d\}$. We have proved that every element of $\Omega_A$ is a unique convex combination of its $d$ extreme points. By Proposition \ref{simplex-unique-comb}, this implies that $\Omega_A$ is a $(d-1)$-simplex with $\ext(\Omega_A) = \{\omega_1, \ldots, \omega_d\}$.

\item (b) $\Rightarrow$ (a): If $\Omega_A$ is a $(d-1)$-simplex, then $\ext(\Omega_A) = \{ \omega_1, \ldots, \omega_d \}$ is an affinely independent set of vectors such that $\conv(\ext(\Omega_A)) = \Omega_A$. Suppose that $\omega_1, \ldots, \omega_d$ are linearly dependent in $A$, i.e.
\begin{align}
\sum\limits_{i=1}^d \alpha_i \omega_i = 0 \quad \text{for some real numbers } \alpha_i \,. \nonumber
\end{align}
Then we would have that $0 \in \aff(\Omega_A)$ because
\begin{align}
\sum\limits_{i=1}^d \frac{\alpha_i}{\left( \sum_{j=1}^d \alpha_j \right)} \omega_i = 0 \nonumber
\end{align}
would be an affine combination of elements in $\Omega_A$. But the zero-vector cannot be an element of $\aff(\Omega_A)$ because $\aff(\Omega_A) = \{ \omega \in A \mid u_A(\omega) = 1 \}$ but $u_A(0) = 0$. Thus, $\omega_1, \ldots, \omega_d$ are linearly independent vectors in the $d$-dimensional vector space $A$. This means that they form a basis. Let $f_1, \ldots, f_d \in A^*$ be the dual basis with respect to $\omega_1, \ldots, \omega_d$, i.e.
\begin{align}
f_i(\omega_j) = \delta_{ij} \,. \label{dual-basis}
\end{align}
The only thing we are left to show is that $f_1, \ldots, f_d \in E_A = \{ f \in A^* \mid 0 \leq f(\omega) \leq 1 \ \forall \omega \in \Omega_A \}$. This follows from $\Omega_A = \conv(\omega_1, \ldots, \omega_d)$ and (\ref{dual-basis}):
\begin{align}
&\omega \in \Omega_A \nonumber \\
&\Rightarrow \quad f_k(\omega) = f_k \underbrace{\left( \sum\limits_{i=1}^d \beta_i \omega_i \right)}_{\text{convex sum}} = \sum\limits_{i=1}^d \beta_i \underbrace{f_k(\omega_i)}_{\in [0,1]} \in [0,1] \quad \forall k \in \{1, \ldots, d\} \nonumber \\
&\Rightarrow \quad f_1, \ldots, f_d \in E_A \,. \tag*{\qedhere}
\end{align}

\end{itemize}
\end{proof}

\begin{ex}[Polytopic theories]
\label{polytopic-theory-ex}
This is the class of theories for which we will derive the results in Part \ref{result-part} of this thesis. We say that an abstract state space $A$ is a \textbf{polytopic theory} if the set $\Omega_A$ is a polytope. We have already seen examples of polytopic theories: The polygon models (Example \ref{polygon-models}) are all polytopic theories since a polygon is a polytope. Every classical theory (Example \ref{classical-theory-ex}) has a simplex as the set of normalized states, so it as a polytopic theory as well. Besides these two classes, one might think of any other polytope serving as the set of normalized states, e.g. a cube, a pyramid or any higher-dimensional polytope. \hfill $\blacksquare$
\end{ex}

Now that we know how measurements are defined, we want to investigate some of their properties. At first, we have a closer look at at some properties of pure effects, before we study a few properties of the set of states $E_A$ in the case where the set of normalized states $\Omega_A$ is a polytope.

Given a pure effect $f \in E_A$, there is always a ``complementary'' effect $\overline f$ such that $\{ f, \overline f \}$ is a pure measurement, as shown by the following proposition.

\begin{prop}
\label{u-f-pure}
Let $A$ be an abstract state space. If $f \in E_A$ is pure, then $\overline f := u_A - f \in E_A$ is pure.
\end{prop}

\begin{proof}
Let $\omega \in \Omega_A$.
\begin{align}
(u - f)(\omega) = 1 - f(\omega) \in [0, 1] \quad \Rightarrow \quad u-f \in E_A \,. \nonumber
\end{align}
Let $g, h \in E_A$ such that $\lambda g + (1-\lambda) h = u-f$.
\begin{align}
\Rightarrow f &= u-\lambda g - (1-\lambda)h \nonumber \\
&= \lambda u + (1-\lambda)u - \lambda g - (1-\lambda) h \nonumber \\
&= \lambda(\underbrace{u-g}_{\in E_A}) + (1-\lambda)(\underbrace{u-h}_{\in E_A}) \nonumber
\end{align}
The effect $f$ is pure, so $u-g = u-h = f$ and therefore $g = h = u-f$.
\end{proof}

We will refer a few times to this kind of complementary effect, so it is practical to give it this name.

\begin{defi}
For a pure effect $f \in E_A$, the effect $\overline f := u_A - f$ is called the \textbf{complementary effect}\index{complementary effect}\index{effect!complementary|see{complementary effect}} to $f$.
\end{defi}

Another very central property of a pure effect $f$ is that $f$ is naturally associated with a face $F_f$ of the set of normalized states. This is established in the following.

\begin{prop}
\label{pure-have-1}
Let $A$ be an abstract state space, let $f \neq 0$ be a pure effect on $A$. Then there exists a state $\omega \in \Omega_A$ such that $f(w) = 1$.
\end{prop}

\begin{proof}
The effect $f$ is nonzero and $\Omega_A$ is not contained in a linear hyperplane\footnote{We say that a subset $M$ of a finite-dimensional vector space $V$ is a \emph{linear hyperplane} if there is a nonzero linear functional $g \in V^*$ such that $M = \{ v \in V \mid g(v) = 0 \}$.} of $V$, so there are states on which $f$ is positive. The effect $f$ is a continuous real-valued function on the compact set $\Omega_A$, so there is a maximum of $f$ on $\Omega_A$. Let $m := \max_{\sigma \in \Omega_A}f(\sigma)$. We have said that $m > 0$, and by the definition of an effect, we have that $0 < m \leq 1$. Thus, $0 < \frac{m}{2-m} \leq 1$. Note that $\frac{1}{m} f \in E_A$ and $\frac{1}{2} f \in E_A$. We take the convex combination
\begin{align}
f = \frac{m}{2-m} \left( \frac{1}{m} f \right) + \left(1 - \frac{m}{2-m} \right) \left(\frac{1}{2} f \right) \,. \nonumber
\end{align}
But $f$ is an extreme point of $E_A$, so $\frac{1}{m}f = f \Rightarrow m = 1$.
\end{proof}

\begin{prop}
\label{max-value-face}
If $f \in V^*$ is a nonzero linear functional on a compact convex set $C$, then $F^m_f := \{x \in C \mid f(x) = m\}$, where $m = \max\limits_{x \in C}f(x)$, is a face of $C$.
\end{prop}

\begin{proof}
The functional $f$ is a continuous real-valued function on the compact set $C$ and therefore has a maximum on $C$ which we call $m$. Let $v, w \in C$, $0 < \alpha < 1$. Suppose that $\alpha v + (1-\alpha) w \in F^m_f$, i.e.
\begin{equation}
\label{f-face-equation}
\alpha \underbrace{f(v)}_{\leq m} + (1-\alpha)\underbrace{f(w)}_{\leq m} = m.
\end{equation}
Equation (\ref{f-face-equation}) is clearly satisfied if $f(v) = f(w) = m$. If $f(v) < m$ (or $f(w) < m$), then, in order to satisfy equation (\ref{f-face-equation}), $f(w) > m$ (or $f(v) > m$, respectively), which contradicts the fact that $m$ is the maximum of $f$ on $C$. Thus, $f(v) = f(w) = 1$, i.e. $v, w \in F^m_f$, which implies that $F^m_f$ is a face of $C$ (c.f. Definition \ref{face-def}).
\end{proof}

Analogously, Proposition \ref{max-value-face} holds in the case where $m$ is the minimal value of $f$ on $C$, but we will not make use of this fact, so $F^m_f$ is considered with respect to the \emph{maximal} value $m$ of $f$ on $C$.

\begin{cor}
\label{nonempty-face}
Combining Propositions \ref{pure-have-1} and \ref{max-value-face}, one has that for every nonzero pure effect $f$, $F_f := F^1_f$ is a nonempty face of $\Omega$.
\end{cor}

\begin{defi}
\label{opposite-face-def}
For a pure effect $f \in E_A$, the \textbf{face $F_f$ associated with $f$}\index{face!associated with an effect $f$} is defined by $F_f := F_f^1 = \{ \omega \in \Omega_A \mid f(\omega) = 1 \}$. The \textbf{opposite face}\index{face!opposite}\index{opposite face|see{face!opposite}} is given by $\overline F_f := F_{\overline f} = \{ \omega \in \Omega_A \mid \overline f (\omega) = 1 \} = \{ \omega \in \Omega_A \mid f(\omega) = 0 \}$.
\end{defi}

\begin{ex}[Associated faces and opposite faces in the polygon models]
\label{associated-faces-ex}
We reconsider the polygon models which we have seen in Example \ref{polygon-models}. In the case where the polygon has an even number of vertices, the pure effects and the associated faces are of a different character than in the case where the number of vertices is odd. We consider the two cases separately.
\begin{enumerate}[(a)]

\item $n$ is even: In this case, we have the $n$ pure effects $e_1, \ldots, e_n \in E_A^n$, the unit effect $u_A \in E_A$ and the zero effect $0 \in E_A^n$, so alltogether, there are $n+2$ pure effects. Let $\widehat E_A^n = E_A^n \backslash \{u_A, 0\} = \{e_1, \ldots, e_n\}$.
For each effect $e_k \in \widehat E_A^n$, the face associated with $e_k$ is a facet of the polygon, i.e. $F_{e_k}$ is an edge of the polygon. Moreover, the complementary effect $\overline{e_k} = u_A - e_k$ always coincides with some other effect $e_l \in \widehat E_A^n$. Therefore, the face $\overline{F_{e_k}} = F_{\overline{e_k}}$ opposite to $F_{e_k}$ is an edge as well.

As an example, consider the the square, which is the polygon model corresponding to $n=4$ (see Figure \ref{square-figure}). For the pure effect $e_3 \in \widehat E_A^4$, the associated face $F_{e_3}$ is an edge. The effect $\overline{e_3}$ complementary to $e_3$ is $\overline{e_3} = e_1$, and the face $\overline{F_{e_3}}$ opposite to $F_{e_3}$ is the edge $\overline{F_{e_3}} = F_{e_1}$.
\begin{figure}[htb]
\centering

\begin{pspicture}[showgrid=false](-2,0)(4,3.5)
\psset{viewpoint=26 10 5,Decran=70}
\psset{solidmemory}
\psSolid[object=new,linewidth=0.5\pslinewidth,
action=draw*,
name=B,
sommets= 
0 0 0 
0.420448 0.420448 0.5 
-0.420448 0.420448 0.5 
-0.420448 -0.420448 0.5 
0.420448 -0.420448 0.5 
0 0 1 
0 1.18921 1 
-1.18921 0 1 
0 -1.18921 1 
1.18921 0 1, 
faces={
[1 4 0]
[0 2 1]
[0 3 2]
[0 4 3]
[1 2 5]
[2 3 5]
[3 4 5]
[4 1 5]
[6 7 8 9]}]%
\psPoint(-0.420448, -0.420448, 0.5){e3}
\psdots[dotsize=0.15](e3)
\uput[ur](e3){$e_3$}
\psSolid[object=line, linewidth=2\pslinewidth,args=-1.18921 0 1 0 -1.18921 1]
\psSolid[object=line, linewidth=2\pslinewidth,args=-1.18921 0 1.01 0 -1.18921 1.01]
\psPoint(-0.594604, -0.594604, 1){h}
\uput[u](h){$F_{e_3}$}
\end{pspicture}
\begin{pspicture}[showgrid=false](-2,0)(2,2.7)
\psset{viewpoint=26 10 5,Decran=70}
\psset{solidmemory}
\psSolid[object=new,linewidth=0.5\pslinewidth,
action=draw*,
name=B,
sommets= 
0 0 0 
0.420448 0.420448 0.5 
-0.420448 0.420448 0.5 
-0.420448 -0.420448 0.5 
0.420448 -0.420448 0.5 
0 0 1 
0 1.18921 1 
-1.18921 0 1 
0 -1.18921 1 
1.18921 0 1, 
faces={
[1 4 0]
[0 2 1]
[0 3 2]
[0 4 3]
[1 2 5]
[2 3 5]
[3 4 5]
[4 1 5]
[6 7 8 9]}]%
\psPoint(0.420448, 0.420448, 0.5){e1}
\psdots[dotsize=0.15](e1)
\uput[d](e1){$\overline{e_3} = e_1$}
\psSolid[object=line, linewidth=2\pslinewidth,args=0 1.18921 1 1.18921 0 1]
\psSolid[object=line, linewidth=2\pslinewidth,args=0 1.18921 1.01 1.18921 0 1.01]
\psPoint(0.727673, 0.528686, 1){g}
\uput[u](g){$\overline{F_{e_3}} = F_{e_1}$}
\end{pspicture}

\caption{This figure shows the $n=4$ polygon model (the square) with the face $F_{e_3}$ associated with the pure effect $e_3$ and the opposite face $\overline{F_{e_3}}$ (which is associated with the complementary effect $\overline{e_3}$).}
\label{square-figure}
\end{figure}
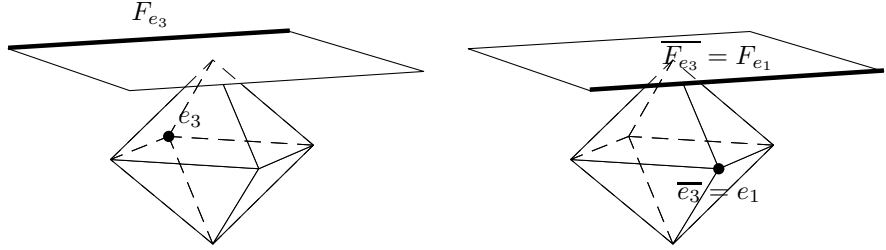

\item $n$ is odd: Here we also have pure effects $e_1, \ldots, e_n, u_A, 0 \in E_A^n$, but in this case, these are not all the pure effects. For each of the pure effects $e_k \in \{e_1, \ldots, e_n\}$, the complementary effect $\overline{e_k} = u_A - e_k$ (which, by Proposition \ref{u-f-pure} is a pure effect) does \emph{not} coincide with some other effect in $\{e_1, \ldots, e_n\}$. Alltogether, this makes a total of $2n + 2$ pure effects. The face $F_{e_k}$ associated with some effect $e_k \in \{e_1, \ldots, e_n\}$ consists of only one state, namely $\omega_k$. On the other hand, the opposite face $\overline{F_{e_k}}$ is an edge of the polygon.

As an example, consider the regular pentagon, which is the polygon model corresponding to $n = 5$ (see Figure \ref{pentagon-figure}). The face $F_{e_3}$ associated with the pure effect $e_3 \in E_A^5$ consists of only the state $\omega_3$. The opposite face $\overline{F_{e_3}}$, however, is an edge of the pentagon.
\begin{figure}[htb]
\centering

\begin{pspicture}[showgrid=false](-2,-0.5)(4,3.5)
\psset{viewpoint=26 10 5,Decran=70}
\psset{solidmemory}
\psSolid[object=new,linewidth=0.5\pslinewidth,
action=draw*,
name=A,
sommets= 
0 0 0 
0.152217 0.468477 0.414214 
-0.402248 0.29225 0.447214 
-0.402248 -0.29225 0.447214 
0.153645 -0.472871 0.447214 
0.497206 0 0.447214 
-0.153645 -0.472871 0.552786 
0.402248 -0.29225 0.552786 
0.402248 0.29225 0.552786 
-0.153645 0.472871 0.552786 
-0.497206 0 0.552786 
0 0 1 
0.343561 1.05737 1 
-0.899454 0.653491 1 
-0.899454 -0.653491 1 
0.343561 -1.05737 1 
1.11179 0 1, 
faces={
[0 1 8 5]
[0 5 7 4]
[0 4 6 3]
[0 3 10 2]
[0 2 9 1]
[6 4 7 11]
[7 5 8 11]
[8 1 9 11]
[9 2 10 11]
[10 3 6 11]
[12 13 14 15 16]}]%
\psPoint(-0.402248, -0.29225, 0.447214){e3}
\psdots[dotsize=0.15](e3)
\uput[u](e3){$e_3$}
\psPoint(-0.899454, -0.653491, 1){w3}
\psdots[dotsize=0.15](w3)
\uput[ur](w3){$F_{e_3} = \{ \omega_3 \}$}
\end{pspicture}
\begin{pspicture}[showgrid=false](-2,-0.5)(1.5,3.5)
\psset{viewpoint=26 10 5,Decran=70}
\psset{solidmemory}
\psSolid[object=new,linewidth=0.5\pslinewidth,
action=draw*,
name=A,
sommets= 
0 0 0 
0.152217 0.468477 0.414214 
-0.402248 0.29225 0.447214 
-0.402248 -0.29225 0.447214 
0.153645 -0.472871 0.447214 
0.497206 0 0.447214 
-0.153645 -0.472871 0.552786 
0.402248 -0.29225 0.552786 
0.402248 0.29225 0.552786 
-0.153645 0.472871 0.552786 
-0.497206 0 0.552786 
0 0 1 
0.343561 1.05737 1 
-0.899454 0.653491 1 
-0.899454 -0.653491 1 
0.343561 -1.05737 1 
1.11179 0 1, 
faces={
[0 1 8 5]
[0 5 7 4]
[0 4 6 3]
[0 3 10 2]
[0 2 9 1]
[6 4 7 11]
[7 5 8 11]
[8 1 9 11]
[9 2 10 11]
[10 3 6 11]
[12 13 14 15 16]}]%
\psPoint(0.402248, 0.29225, 0.552786){e3bar}
\psdots[dotsize=0.15](e3bar)
\uput[d](e3bar){$\overline{e_3}$}
\psSolid[object=line, linewidth=2\pslinewidth,args=0.343561 1.05737 1 1.11179 0 1]
\psSolid[object=line, linewidth=2\pslinewidth,args=0.343561 1.05737 1.01 1.11179 0 1.01]
\psPoint(0.727673, 0.528686, 1){f}
\uput[u](f){$\overline{F_{e_3}}$}
\end{pspicture}

\caption{This figure shows the $n=5$ polygon model (the regular pentagon) with the face $F_{e_3} = \{\omega_3\}$ associated with the pure effect $e_3$ and the opposite face $\overline{F_{e_3}}$ (which is associated with the complementary effect $\overline{e_3}$). \hfill $\blacksquare$}
\label{pentagon-figure}
\end{figure}
\end{enumerate}

\end{ex}

Now we want to see what properties $E_A$ has in the case where $\Omega_A$ is a polytope. The first thing we prove is that in this case, $E_A$ is a polytope as well.

\begin{prop}
\label{e_a-polytope}
Let $A$ be an abstract state space such that $\Omega_A$ is a polytope. Then $E_A$ is a polytope. In particular, $E_A$ has only finitely many pure effects.
\end{prop}

\begin{proof}
Let $A^*$ be equipped with the norm (\ref{norm-on-a-star}) that we have introduced in Section \ref{abs-st-sp-section},
\begin{align}
|| f ||_{A^*} = \sup\limits_{\omega \in \Omega_A} | f(\omega) | \,. \nonumber
\end{align}
$E_A$ is obviously contained in the unit ball in $A^*$ with respect to this norm. This implies that $E_A$ is bounded. By the definition of $E_A$, we have that
\begin{align}
E_A = \{f \in A^* \mid 0 \leq f(\omega) \leq 1 \ \forall \omega \in \Omega_A \}. \nonumber
\end{align}
We can rewrite this as
\begin{align}
E_A = \{f \in A^* \mid f(\omega) \geq 0 \ \forall \omega \in \Omega_A \} \cap \{f \in A^* \mid f(\omega) \leq 1 \ \forall \omega \in \Omega_A\}\,. \nonumber
\end{align}
If the two inequalities $f(\omega_i) \geq 0$ and $f(\omega_i) \leq 1$ are satisfied for a family $\{\omega_i\}_i$ of states in $\Omega_A$, then they are also satisfied for all states in the convex hull $\conv(\{\omega_i\}_i)$ of the family. $\Omega_A$ is compact (by the definition of a polytope), so by Theorem \ref{minkowski}, it is the convex hull of its extreme points. Therefore,
\begin{align}
E_A = \{f \in A^* \mid f(\omega) \geq 0 \ \forall \omega \in \Omega_A \text{ pure} \} \cap \{f \in A^* \mid f(\omega) \leq 1 \ \forall \omega \in \Omega_A \text{ pure}\}\,. \nonumber
\end{align}
$\Omega_A$ is a polytope, so it has finitely many extreme points. Thus, the above equation implies that $E_A$ is the intersection of a finite family of closed half-spaces and therefore a polyhedral set. By virtue of Theorem \ref{polytope-bounded-polyhedral}, $E_A$ is a polytope.
\end{proof}

Above, we have established that for every pure effect $f \in E_A$, there is a non-empty associated face $F_f$ given by $f(\omega) = 1 \Leftrightarrow \omega \in F_f$ for all $\omega \in \Omega_A$. An interesting question is whether the converse is true: If $F$ is a face of $\Omega_A$, is there a pure effect $f \in E_A$ such that $F = F_f$? It turns out that this is not the case. A counterexample\label{counterexample-mention}: If $F$ is a vertex of the $n=4$ polygon-model, then there are effects $f$ such that $F = \{ \omega \in \Omega_A \mid f(\omega) = 1\}$, but none of these effects is pure. However, there is a weaker version of the statement which \emph{is} true.

\begin{prop}
\label{rest-possible}
Let $A$ be an abstract state space such that $\Omega_A$ is a polytope. If $F \subset \Omega_A$ is a facet of $\Omega_A$, then there exists a pure effect $g \in E_A$ such that $g(\omega) = 1 \Leftrightarrow \omega \in F$, i.e. such that $F = F_g$.
\end{prop}

\begin{proof}
Let
\begin{align}
\chi_A = \{\omega \in A \mid u_A(\omega)=1\} \,, \quad \Omega_A = \{ \omega \in A_+ \mid u_A(\omega) = 1\} \,. \nonumber
\end{align}
Recall from Proposition \ref{dd-cone} that for a closed cone $K$, one has that $(K^*)^* = K$. The cone $A_+$ is closed by the definition of an abstract state space. This implies
\begin{align}
A_+ = \{ \omega \in A \mid f(\omega) \geq 0 \ \forall f \in A_+^* \} \,. \label{blolu}
\end{align}
This allows us to write
\begin{align}
\Omega_A &= \{ \omega \in A_+ \mid u_A(\omega) = 1\} \overset{(\ref{blolu})}{=} \{ \omega \in A \mid u_A(\omega) = 1, f(\omega) \geq 0 \ \forall f \in A_+^* \} \nonumber \\
&= \{ \omega \in \chi_A \mid f(\omega) \geq 0 \ \forall f \in A_+^* \} \,. \label{blala}
\end{align}
For every $f \in A_+^*$, there is an $\alpha > 0$ such that $\alpha f \in E_A$.\footnote{This can be seen as follows. A positive linear functional $f \in A_+^*$ is a nonnegative continuous function on the compact set $\Omega_A$. Thus, it has a nonnegative maximum $m := \max_{\omega \in \Omega_A} f(\omega)$. If $m=0$, then $f \in E_A$. If $m > 0$, then $\alpha f \in E_A$ with $\alpha = \frac{1}{m} > 0$.} For any $\omega \in \chi_A$, one has that $f(\omega) \geq 0$ if and only if $\alpha f(\omega) \geq 0$. This allows us to rewrite (\ref{blala}):
\begin{align}
\Omega_A = \{ \omega \in \chi_A \mid f(\omega) \geq 0 \ \forall f \in E_A \} \,.
\end{align}
Note that
\begin{align}
\label{bijection}
f \mapsto u_A - f \text{ is a bijection on } E_A \,.
\end{align}
Therefore,
\begin{align}
&f(\omega) \geq 0 \Leftrightarrow (u-f)(\omega) \leq 1 \label{u-f} \\
\Rightarrow \quad \Omega_A &=  \{ \omega \in \chi_A \mid f(\omega) \geq 0 \ \forall f \in E_A\} \nonumber \\
&= \{ \omega \in \chi_A \mid f(\omega) \leq 1 \ \forall f \in E_A \} \text{ by (\ref{bijection}) and (\ref{u-f}).} \nonumber
\end{align}
Note that if $f_i(\omega) \leq 1$ is satisfied for a family $\{f_i\}_i$ of functionals, then it is satisfied for all functionals in the family's convex hull, $f(\omega) \leq 1$ for all $f \in \conv(\{f_i\}_i)$. By virtue of Theorem \ref{minkowski}, $\Omega_A$ is the convex hull of its extreme points, so
\begin{align}
\Omega_A = \{ \omega \in \chi_A \mid f(\omega) \leq 1 \ \forall f \in E_A \text{ pure} \} \,. \nonumber
\end{align}
For $f=0$ and $f=u_A$, we have that $\{ \omega \in \chi_A \mid f(\omega) \leq 1\}$ is all of $\chi_A$, so
\begin{align}
&\Omega_A = \{ \omega \in \chi_A \mid f(\omega) \leq 1 \ \forall f \in \widehat{E_A}\}, \text{ where} \label{what-omega-is} \\
&\widehat{E_A} = \{ f \in E_A \mid f \text{ pure}, \ f \neq 0, \ f \neq u_A \} \,. \nonumber
\end{align}
By Proposition \ref{e_a-polytope}, $\widehat{E_A}$ is finite. Therefore, we see from (\ref{what-omega-is}) that $\Omega_A$ is the intersection of finitely many half-spaces, where the boundary of each half-space is the set of points in $\chi_A$ at which a pure effect has value 1. Lemma \ref{halfspace-rep} implies that there is a pure effect $g$ such that $g(\omega) = 1$ for all $\omega \in F$.

It remains to show that $g(\omega) < 1$ if $\omega \in \Omega_A, \omega \notin F$. Let $\sigma \in \Omega_A, \sigma \notin F$. The point $\sigma$ is not an element of $\aff(F)$, but $\aff(F)$ is equal to $\{ \omega \in \chi_A \mid g(\omega) = 1 \}$ since $F$ is a facet (a maximal proper face). Thus, $\sigma$ cannot be an element of $\Omega_A$, so we have a contradiction.
\end{proof}

\subsection{The physical interpretation of the state normalization}
\label{norm-int-section}

So far, the only subset of the abstract state space $A$ we have talked about in a physical context is the cone-base $\Omega_A$, which we also call the \emph{set of normalized states}. In Section \ref{abs-st-sp-section}, we have argued mathematically why this notion makes sense. Now we will give a physical interpretation of the ``norm'' $u_A(\omega)$ of a state $\omega \in \Omega_A^{\leq 1}$. In the following, when we talk about states, we mean elements of the set of subnormalized states $\Omega_A^{\leq 1} = \{ \omega \in A_+ \mid u_A(\omega) \leq 1 \}$, and when we say \emph{normalized states}, we mean elements of $\Omega_A = \{ \omega \in A_+ \mid u_A(\omega) = 1 \}$.

Suppose that a random process $\mathcal{R}$ with the possible outcomes $\mathcal{I}_\mathcal{R} = \{1, \ldots, n\}$ takes place. Assume that the state of a system is prepared in a state $\omega_k \in \Omega_A$ which depends on the outcome $k \in \mathcal{I}_\mathcal{R}$. In Section \ref{randomization-section}, we argued that if we are ignorant about the outcome of the random process, our description of the system is given by the normalized state
\begin{equation}
\label{mixture}
\widetilde \omega = \sum\limits_{k = 1}^n p_k \omega_k \,,
\end{equation}
where $p_k$ is the probability for the outcome $k \in \mathcal{I}_\mathcal{R}$. This is the probabilistic mixture of the states $\{ \omega_k \}_{k \in \mathcal{I}_\mathcal{R}} \subset \Omega_A$. If we \emph{know} that the outcome of the random process is $m \in \mathcal{I}_\mathcal{R}$, i.e. \emph{conditioned on} the outcome $m$, we describe the system by the normalized state $\omega_m$. These two situations (being ignorant about the outcome and knowing the outcome) thus lead to the states (\ref{mixture}) and $\omega_m$, both of which are meaningful. 

Besides these two treatments of states (which both deal with normalized states), there is a third one which is meaningful in the presence of random processes (which deals with unnormalized states). Consider the following:
\begin{itemize}
\item Suppose that a random process prepares the system in the state $\omega_k$ with probability $p_k$.
\item Subsequent to the random process, a measurement $\mathcal{M}$ on system $A$ is performed.
\end{itemize}
In Section \ref{meas-section}, we have seen that measurement outcomes $k \in \mathcal{I}_\mathcal{M}$ are associated with linear functionals $e_k \in \{e_k\}_{k \in \mathcal{I}_\mathcal{M}}$. If $\omega \in \Omega_A$ is a normalized state (prior to the measurement) and $l \in \mathcal{I}_\mathcal{M}$ is a possible outcome of the measurement $\mathcal{M}$, then $e_l(\omega)$ gives the probability that the outcome of the measurement is $l$. In this context, however, it makes sense to consider a subnormalized state $\sigma = p_k \omega_k$ (instead of a normalized state $\omega$). Notice that $p_k \omega_k \notin \Omega_A$ whenever $p_k < 1$. One can interpret this state as follows.

For $e_l \in \mathcal{M}$, we have that
\begin{align}
e_l(p_k \omega_k) = p_k \cdot e_l(\omega_k) \,. \nonumber
\end{align}
The value $p_k$ is the probability that the outcome of the random process is $k$. The value $e_l(\omega_k)$ is the probability that the subsequent measurement gives the outcome $l$, \emph{conditioned on} the fact that the outcome of the random process was $k$. Thus, $e_l(p_k \omega) = e_l(\sigma)$ is the joint probability for the event that the outcome of the random process is $k$ and the outcome of the subsequent measurement is $l$. Thus, subnormalized states give us descriptions of joint probabilities. Moreover, from the subnormalized state $\sigma$, we can read out the probability $p_k$ separately by applying the order unit $u_A$ because $u_A(\sigma) = u_A(p_k \omega_k) = p_k$.

This gives an interpretation for states $\omega$ with $0 \leq u_A(\omega) < 1$. Therefore, the set of physically meaningful states is given by the set of subnormalized states
\begin{equation}
\Omega_A^{\leq 1} = \{ \omega \in A_+ \mid u_A(\omega) \leq 1 \} \,. \nonumber
\end{equation}

\subsection{Pure states and maximal knowledge}
\label{maximal-knowledge-section}

Suppose that a physical system is in some state $\omega \in \Omega_A$. Assume that we want to perform a pure\footnote{We will see in Section \ref{pure-meas-section} how pure measurements are distinguished from other measurements.} measurement $\mathcal{M} = \{ e_k \}_{k \in \mathcal{I}_\mathcal{M}}$ on the system. In general, we cannot predict the outcome $k \in \mathcal{I}_\mathcal{M}$ of the measurement with certainty. Our prediction of the outcome is given by a probability distribution, and this distribution might have probabilities that are neither zero nor one. In a classical theory (c.f. Example \ref{classical-theory-ex}), there is a natural way to interpret this situation. When the outcome of a measurement cannot be predicted with certainty, then classically this can be interpreted as the circumstance that we do not \emph{know} enough about the state of the system to make a definite prediction.\footnote{We do not want to be dogmatic about this \emph{Bayesian} interpretation of probability. The goal of this section is to explain that pure states have a physical interpretation that distinguishes them from mixed states. Such an interpretation is also possible from other viewpoints regarding probability. We, however, describe the physical distinction of pure states from mixed states in the Bayesian picture.} For example, we might assume that the physical system is a die. Suppose we are sitting on a table. We drop the die on the floor and it rolls under the table where we cannot see it. How many pips does the die show?

We describe the set of normalized states of this system by a 6-simplex. Each vertex $\omega_k$ of the simplex, $k = 1, \ldots, 6$, corresponds to a definite number of pips. As long as we do not have a look under the table to find out how many pips the die shows, we describe the state of the die by a mixed state. If the die is unbiased, we describe its state by the probabilistic mixture $\omega = \sum_{k=1}^6 \frac{1}{6} \omega_k$ since we have no idea which number the die shows. If the die is biased, e.g. with an additional weight on the face with one pip, then it might be more likely that the die shows one pip, $\widetilde \omega = \frac{1}{2} \omega_1 + \sum_{k=2}^6 \frac{1}{10} \omega_k$. We perform a pure measurement on this system by having a look under the table to find out how many pips the die shows. In both cases ($\omega$ and $\widetilde \omega$), we cannot predict the outcome with certainty. It is natural to say that we cannot predict the outcome because we do not \emph{know} enough about the die under the table. If the die under the table were in a \emph{pure} state, we \emph{could} predict the outcome with certainty. We would have \emph{maximal knowledge} about the system.

For a quantum system, e.g. a spin-$\frac{1}{2}$ particle, the situation is different. Even if the system is in a pure state, for example in the up-state $| \uparrow_z \rangle$ with respect to the $z$-axis, there are pure measurements for which we cannot predict the outcome with certainty. For instance, if we were to predict the outcome of a spin-measurement with respect to the $x$-axis, we would assign the probability $\frac{1}{2}$ for both of the outcomes ``up'' and ``down''. But in this case, unlike the case of a die, the fact that we cannot predict the outcome with certainty can \emph{not} be interpreted as the fact that we do not know enough about the state of the system. In the case of a die, the state $\widetilde \omega = \frac{1}{2} \omega_1 + \sum_{k=2}^6 \frac{1}{10} \omega_k$ can be interpreted as ``with probability $\frac{1}{2}$, the system is in the pure state $\omega_1$, with probability $\frac{1}{10}$, the system is in the pure state $\omega_2$'' and so on. In this situation, the fact that we cannot predict the outcome with certainty can be interpreted as being due to the circumstance that we have incomplete knowledge about the state of the system, since we can represent the state as a probabilistic mixture of other states. In the quantum case, where the state is given by the \emph{pure} state $| \uparrow_x \rangle$, this interpretation does not apply. The state of the system is pure, so it cannot be represented as a (nontrivial) probabilistic mixture of other states. We can say that we have \emph{maximal knowledge} about the state of the system. The fact that we still cannot predict the outcome with certainty (although we have maximal knowledge the state) might therefore be interpreted as an inherent property of the theory, rather than being due to our ignorance about the state.

This interpretation of pure states as states of maximal knowledge sheds new light on Theorem \ref{minkowski}. This theorem by Minkowski states that a compact convex subset of a finite-dimensional vector space is the convex hull of its extreme points. In other words, a compact convex set in a finite-dimensional vector space is fully characterized by its extreme points. In Section \ref{abs-st-sp-section}, we have assumed that the set of normalized states is a compact convex set in a finite-dimensional space, and the extreme points $\ext(\Omega_A)$ are precisely the pure states. In summary, we might therefore say the following. 

\begin{quote}
\emph{Pure states are states of maximal knowledge, and the set of states is fully characterized by these states of maximal knowledge}.
\end{quote}
In other words:
\begin{quote}
\emph{Every state is a probabilistic mixture of states of maximal knowledge} (in the case of pure states, this mixture is trivial).
\end{quote}

\subsection{Transformations on abstract state spaces}
\label{trafo-section}

In this section, we investigate the concept of transformations on abstract state spaces. Roughly speaking, a transformation is a map from the set of subnormalized states $\Omega_A^{\leq 1}$ to itself which maps an initial state to a final state.

The way transformations are treated here is different from how transformations are normally defined. The reason is that in this thesis, we never consider multi-partite systems. We always consider systems of only one constituent. Thus, we do not specify how systems are combined to form bi-partite or multi-partite systems. This makes a difference in the definition of transformations. In the treatment of multi-partite systems, one has to require a consistency property of transformations which is called \emph{complete positivity}. The definition of complete positivity depends on the specific way in which multiple systems are combined in a physical theory. As we do not specify a rule for how to combine systems, we cannot give a definition of complete positivity. Instead, we only require the weaker property of \emph{positivity} (we will explain this property below). Nonetheless, this will not lead to problems. Since positivity is weaker than complete positivity (no matter how systems are combined), the class of positive transformations is larger than the class of completely positive transformations. Thus, any result that is inferred for positive transformations also holds for completely positive transformations.

Transformations naturally arise in two contexts: dynamics and measurements. In this thesis, we do not deal with dynamics, so we forget about this aspect in the following and focus on measurements. When we describe measurements, there are two aspects involved. If we only want to describe one single measurement, we only need to care about one of the two aspects. This aspect is the probability distribution of the measurement outcomes, i.e. the measurement statistic. But if we want to describe consecutive measurements, we also need another aspect. We need to care about how a measurement influences the measurement statistic of a subsequent measurement. If we perform two consecutive measurements on a system, it might be the case that the statistic of the second measurement depends on the outcome of the first measurement. For a full description of how the statistic of the second measurement might be influenced, we need to specify the state of the system after the first measurement. This specification is made by a transformation which maps the initial state to the post-measurement state.

The first of the two aspects that we have just described is fully covered by describing a measurement by a set of effects. To meet the second aspect, we need to treat a measurement as a set of transformations. Below, we will call such a set an \emph{operation}.

To discuss the second aspect of a measurement in more detail, suppose that we perform a measurement $\mathcal{M} = \{ e_1, \ldots, e_n \}$ with outcomes $\mathcal{I}_\mathcal{M} = \{ 1, \ldots, n \}$ on a system. Assume that subsequent to this measurement, we make some other measurement $\mathcal{N}$ with outcomes $\mathcal{I}_\mathcal{N}$. If we have a full description of this situation, then this means that we can infer the measurement statistic of the second measurement for all possible choices of $\mathcal{N}$. This in turn means that we need to have a description of the state after the first measurement, since a state is exactly the mathematical object that gives us the measurement statistic for every possible measurement $\mathcal{N}$. Two states of a system are different if and only if there is some measurement for which the two states induce different statistics. This means that a full description of consecutive measurements necessarily involves the description of post-measurement states.

Above, we mentioned that the statistics of a second measurement can depend on the outcome of the first measurement. In other words, the post-measurement state can depend on the outcome of the measurement. For the moment, we describe the transition from the initial state (prior to the measurement) to the post-measurement state by a map $t_k: \Omega_A \rightarrow \Omega_A$. This map $t_k$ depends on the outcome $k \in \mathcal{I}_\mathcal{M}$ of the $\mathcal{M}$-measurement. It takes an initial state and maps it to the post-measurement state for the case that the outcome of the measurement is $k$. Thus, if we want to describe both aspects of a measurement $\mathcal{M}$ --- the outcome statistic of $\mathcal{M}$ and the influence on the statistic of any subsequent measurement $\mathcal{N}$ --- we might achieve this by a set of tuples $\{ (e_k, t_k) \}_{k \in \mathcal{I}_\mathcal{M}}$, where the first element of each tuple is an effect $e_k$ and the second element is a map $t_k: \Omega_A \rightarrow \Omega_A$.

So far, we have only talked about normalized initial and post-measurement states $\omega \in \Omega_A$. We have not made use of the fact that an abstract state space provides the structure to deal with the state normalization. If we use this extra structure, we can combine the effect $e_k$ and the map $t_k: \Omega_A \rightarrow \Omega_A$ to form a \emph{transformation} $\widetilde T_k$ which encompasses both aspects of a measurement. We achieve this by defining $\widetilde T_k(\omega) = e_k(\omega) t_k(\omega)$. This gives a map $\widetilde T_k: \Omega_A \rightarrow \Omega_A^{\leq 1}$. From this transformation $\widetilde T_k$, we can infer both the measurement statistic and the post-measurement state:
\begin{align}
&e_k(\omega) = (u_A \circ \widetilde T_k)(\omega) \,, \nonumber \\
&t_k(\omega) = \frac{\widetilde T_k(\omega)}{(u_A \circ \widetilde T_k)(\omega)} \,. \nonumber
\end{align}
To allow concatenations of transformations, we want to extend the map $\widetilde T_k: \Omega_A \rightarrow \Omega_A^{\leq 1}$ to a map $\Omega_A^{\leq 1} \rightarrow \Omega_A^{\leq 1}$. For reasons of mathematical convenience, we extend it to a map $T_k: \Omega_A \rightarrow \Omega_A$. For a physical interpretation of the transformation $T_k$, we have to require four properties of $T_k$.

The first two properties are the convex-linearity of the transformation and that it leaves the origin invariant. These two properties are completely analogously to the first two properties that we demanded for effects in Section \ref{meas-section}, so we only recall shortly the reasons for these conditions. The convex-linearity expresses the fact that there is no physical specification of whether we should take probabilistic mixtures of states or of probabilities. Thus, we regard them as identical and reach the requirement of the convex-linearity. We require $T_k(0) = 0$ because the zero-state is the state conditioned on an impossible event. These two properties together imply (as in the case of effects in Section \ref{meas-section}) the \emph{linearity} of the transformation.

The third and the fourth property arise from the requirement that if we restrict $T_k$ to the set of subnormalized states $\Omega_A^{\leq 1}$, then we should get a map that maps to the subnormalized states $\Omega_A^{\leq 1}$. This requirement splits up into the third and fourth property. The third property is the \emph{positivity} of $T_k$.

\begin{defi}
\label{positive-map-def}
A map $\phi: A \rightarrow B$ between ordered vector spaces $A$ and $B$ is \textbf{positive}\index{positive map} if $\phi(A_+) \subset B_+$.
\end{defi}

The fourth property is that $T_k$ does not increase the norm of the state, i.e. $u_A(T_k(\omega)) \leq 1$ for all $\omega \in \Omega_A^{\leq 1}$, or equivalently $u_A(T_k(\omega)) \leq 1$ for all $\omega \in \Omega_A$. These four properties lead us to the definition of a transformation. To make this definition in more generality, we define a transformation as a map between possibly different abstract state spaces $A$ and $B$.

\begin{defi}
\label{trafo-def}
A \textbf{transformation}\index{transformation} $T: A \rightarrow B$ between abstract state spaces $A$ and $B$ is a map which fulfills the following conditions:
\begin{align}
&\bullet \quad T \text{ is linear.} \label{trafo1} \\
&\bullet \quad T \text{ is positive.} \label{trafo2} \\
&\bullet \quad T \text{ does not increase the norm, i.e. } u_B(T(\omega)) \leq 1 \text{ for all } \omega \in \Omega_A. \label{trafo3}
\end{align}
Given that $T$ is linear, conditions (\ref{trafo2}) and (\ref{trafo3}) can be summarized as $T(\Omega_A) \subset T(\Omega_A^{\leq 1})$.
The \textbf{effect induced by the transformation}\index{effect!induced by a transformation} is given by $e_T = u_B \circ T$.
\end{defi}

We have described how transformations arise in the context of a measurement. More generally, we can think of transformations as arising from any sort of random process. Suppose there is a random process, and in the course of the random process, a transformation takes place which depends on the outcome of the random process. We call this an \emph{operation}.

\begin{defi}
An \textbf{operation}\index{operation} $\mathcal{O} = \{ T_1, \ldots, T_k \}$ between abstract state spaces $A$ and $B$ is a family of transformations $T_k: A \rightarrow B$ such that the effects induced by the transformations sum up to the unit effect:
\begin{align}
\sum\limits_{k=1}^n (u_B \circ T_k)(\omega) = \sum\limits_{k=1}^n e_{T_k}(\omega) = 1 \quad \forall \omega \in \Omega_A \,. \label{ind-eff-sum-1}
\end{align}
It has the interpretation that if $\omega$ is the initial state of the system prior to the operation, then with probability $e_{T_k}(\omega) = (u_B \circ T_k)(\omega)$, the operation transforms the state according to $T_k$:
\begin{align}
\omega \mapsto \frac{T_k(\omega)}{(u_B \circ T_k)(\omega)} \,. \nonumber
\end{align}
\end{defi}

An operation is a stronger formulation than a measurement, in the sense that every operation induces a measurement.

\begin{prop}
Every operation induces a measurement.
\end{prop}

\begin{proof}
At first, we check that for every transformation $T_k \in \mathcal{O}$, the effect $e_{T_k}$ induced by $T_k$ is indeed an effect. The function $e_{T_k}$ a linear functional by (\ref{trafo1}). We have that $0 \leq f(\omega) \leq 1$ for all $\omega \in \Omega_A$ by (\ref{trafo2}) and (\ref{trafo3}). Finally, the effects sum up to one because of Equation (\ref{ind-eff-sum-1}).
\end{proof}

For now, we close the discussion of transformations at this point. We will come back to the issue of transformations in Section \ref{non-dist-section}, where we will finish this part of the framework. We will make the assumption that the post-measurement states of pure measurements are given by transformations which induce the measurement. The reason why we restrict to pure measurements will become clear in Section \ref{pure-meas-section}.

\newpage

\part{The application of the techniques and the results}
\label{result-part}

In Part \ref{introduction-part}, we have learned about a particular framework for generalized probabilistic theories called the \emph{abstract state space} formalism. We have seen that probabilistic mixtures give rise to convex sets of states in Section \ref{randomization-section}. In Section \ref{gpt-section}, we have seen how convex sets fit into the abstract state space formalism. Section \ref{general-convex-section} was dedicated to the study of some properties of convex sets in general, before we investigated polytopes as a special type of convex sets in Section \ref{polytope-section}. Polytopes will be of particular importance in Part \ref{result-part}. When we introduce some physical principles and state them as ``Postulates'', we do this in the full generality of the abstract state space formalism. But the results that we derive from these postulates are based on the restriction to polytopic theories (i.e. theories where the set $\Omega_A$ of normalized states is a polytope). In other words, we derive the results for theories with only finitely many pure states.

Admittedly, the restriction to polytopic sets of states is artificial and has no physical justification. While classical theory (with a simplex as the set of states) is a polytopic theory, quantum theory is not. A quantum system with a two-dimensional Hilbert space, for example, has a set of normalized states which is convex-isomorphic to a closed unit ball (the Bloch sphere), so it has continuously many pure states. Hence it is worth saying a few words about how quantum theory is related to the results of Part \ref{result-part}.

The idea of the postulates is to
\begin{enumerate}[(a)]
\item identify physical principles that are satisfied by quantum theory and
\item determine generalized probabilistic theories which violate these postulates.
\end{enumerate}
Referring to what we have said in the introduction of this thesis, this can be regarded as a step towards the higher goal of \emph{inferring quantum theory from physical principles}. This goal would be achieved if we would find physical principles that (a) are satisfied by quantum theory but which (b') rule out all probabilistic theories except for quantum theory. As (b') seems to be difficult to achieve, it might already be a step forward to identify physical principles that only rule out some class of probabilistic theories (but not all except for quantum theory). This is the concern of the present thesis. In this sense, the restriction to polytopes is justified, since we manage to rule out a class of probabilistic theories by postulating physical principles that are satisfied by quantum theory.

We want to point out that the depth of the insight into quantum theory that we gain by (a) and (b) significantly depends on two conditions on the principles:
\begin{enumerate}
\item The principles should, as much as possible, be of a physical nature rather than of a mathematical nature. The more this physical aspect has an \emph{operational} interpretation (rather than being a hardly accessible, very abstract idea), the less mysterious is our picture of quantum theory.
\item Within a certain minimal strength of deduction, the principles should (appear to) be as weak as possible. Strong physical principles should be \emph{inferred} from a few weak principles rather than being assumed from the beginning.
\end{enumerate}
We approach the idea that we have just described in two different ways by considering two different approaches to postulate physical principles.

We discuss the first approach in Section \ref{result-1-section}. There we consider three principles (or postulates). Postulate \ref{rep-postulate} is called \emph{repeatability}. It demands that if we perform a pure measurement twice in a row, then we get the same outcome both times. This requirement constrains the set of possible post-measurement states to a certain subset of $\Omega_A$. Postulate \ref{subsp-postulate} states that the set of all states satisfying the repeatability condition have a certain subspace structure, so we call it the \emph{subspace principle}. Postulate \ref{discr-postulate} is what we call the \emph{state discrimination principle}. Suppose that two sets $\Lambda_1$ and $\Lambda_2$ of states can be perfectly distinguished by a measurement. Assume that in addition, two subsets $\Lambda_3, \Lambda_4 \subset \Lambda_2$ can be perfectly distinguished from each other. Postulate \ref{discr-postulate} claims that in this case, the sets $\Lambda_1, \Lambda_3$ and $\Lambda_4$ can be perfectly distinguished. We will see that every polytopic theory satisfying Postulates \ref{rep-postulate}, \ref{subsp-postulate} and \ref{discr-postulate} is a classical theory, i.e. a theory where the set of normalized states is a simplex.

The second approach is the main result of this thesis and is presented in Section \ref{result-2-section}. It achieves the same (inferring classical theory from polytopic theory) by only postulating one simple and plausible physical principle. This principle states that if we know the result of a measurement in advance with certainty, then we can perform this measurement without disturbing the statistics of any other measurement. In other words, a measurement that does not provide any information does not disturb the state. We will argue that such a measurement can be seen as the readout of classical information.

\newpage

\section{Repeatability, subspaces and a state discrimination principle}
\label{result-1-section}

In this section, we consider three physical principles, which we will state as ``Postulates'', and study their consequences. Although from a technical point of view, it is not necessary to introduce post-measurement states, the interpretation of two of the postulates relies on post-measurement states to some extend. This forces us to be careful. In Section \ref{pure-meas-section}, we will discuss that Definition \ref{meas-def} of a measurement encompasses a class of measurements which is too general for a consistent treatment of post-measurement states. We explain why it is necessary to restrict to pure measurements when we talk about post-measurement states.

In Section \ref{rep-subsp-section}, we introduce the principle of \emph{repeatability} and the \emph{subspace principle}. Section \ref{discrimination-section} is dedicated to the \emph{state discrimination principle}. Finally, we will show in Section \ref{result-1-result} that a polytopic theory which satisfies these three principles is precisely a classical theory, i.e. a theory where the set of states is a simplex.

\subsection{Post-measurement states and pure measurements}
\label{pure-meas-section}

In the following sections, we will be concerned with post-measurement states. Whatever we assume about post-measurement states in generalized probabilistic theories, when we apply it to quantum theory, it should not contradict the known laws of quantum theory. It is not our goal to disprove quantum theory but to understand what makes it special. To respect this, we have to be careful that we do not make statements about a too large class of measurements. In the following, we state this more precisely.

In Section \ref{trafo-section}, we have explained that in order to describe consecutive measurements in a probabilistic theory, we need a rule for assigning post-measurement states. This is a rule which, given an initial state $\omega$ and an effect $e$, gives the post-measurement state $\omega_\text{post}$ for the case where we perform a measurement on a system in the state $\omega$ and obtain the outcome associated with the effect $e$. This is an assignment $(\omega, e) \mapsto \omega_\text{post}(\omega, e)$. We call this an \emph{update rule} for short.

In quantum theory, the effects are given by POVM elements. We have an update rule for the case where the POVM element $P$ is a projector, i.e. $P^2 = P$. This update rule is called the \emph{von Neumann-L\"uders projection}. It makes the assignment
\begin{align}
(\rho, P) \mapsto \rho_\text{post}(\rho, P) = \frac{P \rho P}{\tr(P \rho)} \,. \label{von-neumann-lueders}
\end{align}
But as we have mentioned in Example \ref{povm-in-qt}, projectors are not the only effects in quantum theory. One might ask whether the von Neumann-L\"uders projection can be generalized. POVMs are not intended to make statements about post-measurement states but only about the statistics of measurement outcomes in a single-shot measurement. However, we might forget about the actual purpose of POVMs for a moment and ask whether the the von Neumann-L\"uders projection can be extended to arbitrary POVM elements. It turns out that this cannot be achieved in a consistent way.

To see this, we consider two different situations which lead to the same POVMs but to different post-measurement states. The following example achieves this by two different global projective measurements on a larger system.

\begin{ex}[Projective measurements on a larger system]
\label{meas-on-larger}

Let $\mathcal{H}_A \cong \mathbb{C}^2$ be the Hilbert space of a system $A$, let $\mathcal{F} = \{F_1, F_2, F_3\}$ be the POVM on $\mathcal{H}_A$ given by
\begin{align}
F_1 = \left(\begin{array}{cc}\frac{2}{3} & 0 \\0 & 0\end{array}\right), \quad F_2 = \left(\begin{array}{cc}\frac{1}{6} & \frac{1}{2\sqrt{3}} \\\frac{1}{2\sqrt{3}} & \frac{1}{2}\end{array}\right), \quad F_3 = \left(\begin{array}{cc}\frac{1}{6} & -\frac{1}{2\sqrt{3}} \\-\frac{1}{2\sqrt{3}} & \frac{1}{2}\end{array}\right). \nonumber
\end{align}
This POVM can be seen as being induced by a projective measurement on a larger system which contains $A$ as a subsystem. To see this, let $\mathcal{H}_B \cong \mathbb{C}^2$ be the Hilbert space of an ancilla system $B$ and let
\begin{align}
\rho_B = \left(\begin{array}{cc}1 & 0 \\0 & 0\end{array}\right) \in \mathcal{S}(\mathcal{H}_B) \nonumber
\end{align}
be the state of system $B$. In the Kronecker product matrix representation, let $\mathcal{G} = \{G_1, G_2, G_3, G_4\}$ be the projective POVM on $\mathcal{H}_A \otimes \mathcal{H}_B$ given by
\begin{align}
&G_1 = \left(\begin{array}{cccc}\frac{2}{3} & \frac{\sqrt{2}}{3} & 0 & 0 \\\frac{\sqrt{2}}{3} & \frac{1}{3} & 0 & 0 \\0 & 0 & 0 & 0 \\0 & 0 & 0 & 0\end{array}\right),
&&G_2 = \left(\begin{array}{cccc}\frac{1}{6} & -\frac{1}{3\sqrt{2}} & \frac{1}{2\sqrt{3}} & 0 \\-\frac{1}{3\sqrt{2}} & \frac{1}{3} & -\frac{1}{\sqrt{6}} & 0 \\\frac{1}{2\sqrt{3}} & -\frac{1}{\sqrt{6}} & \frac{1}{2} & 0 \\0 & 0 & 0 & 0\end{array}\right), \nonumber \\
&G_3 = \left(\begin{array}{cccc}\frac{1}{6} & -\frac{1}{3\sqrt{2}} & -\frac{1}{2\sqrt{3}} & 0 \\-\frac{1}{3\sqrt{2}} & \frac{1}{3} & \frac{1}{\sqrt{6}} & 0 \\-\frac{1}{2\sqrt{3}} & \frac{1}{\sqrt{6}} & \frac{1}{2} & 0 \\0 & 0 & 0 & 0\end{array}\right),
&&G_4 = \left(\begin{array}{cccc}0 & 0 & 0 & 0 \\0 & 0 & 0 & 0 \\0 & 0 & 0 & 0 \\0 & 0 & 0 & 1\end{array}\right). \nonumber
\end{align}
It is easily calculated that for any state $\rho_A \in \mathcal{S}(\mathcal{H}_A)$, we have that
\begin{align}
\tr(F_i \rho_A) = \tr(G_i \rho_A \otimes \rho_B) \ \forall i \in \{1,2,3\}, \quad \tr(G_4 \rho_A \otimes \rho_B) = 0 \,. \nonumber
\end{align}
In the same way, instead of $\mathcal{G}$, we could have chosen the projective POVM $\tilde{\mathcal{G}} = \{\tilde G_1, \tilde G_2, \tilde G_3, \tilde G_4\}$ on $\mathcal{H}_A \otimes \mathcal{H}_B$ given by
\begin{align}
&\tilde G_1 = \left(\begin{array}{cccc}\frac{2}{3} & 0 & 0 & \frac{\sqrt{2}}{3} \\0 & 0 & 0 & 0 \\0 & 0 & 0 & 0 \\\frac{\sqrt{2}}{3} & 0 & 0 & \frac{1}{3}\end{array}\right),
&&\tilde G_2 = \left(\begin{array}{cccc}\frac{1}{6} & 0 & \frac{1}{2\sqrt{3}} & -\frac{1}{3\sqrt{2}} \\0 & 0 & 0 & 0 \\\frac{1}{2\sqrt{3}} & 0 & \frac{1}{2} & -\frac{1}{\sqrt{6}} \\-\frac{1}{3\sqrt{2}} & 0 & -\frac{1}{\sqrt{6}} & \frac{1}{3}\end{array}\right), \nonumber \\
&\tilde G_3 = \left(\begin{array}{cccc}\frac{1}{6} & 0 & -\frac{1}{2\sqrt{3}} & -\frac{1}{3\sqrt{2}} \\0 & 0 & 0 & 0 \\-\frac{1}{2\sqrt{3}} & 0 & \frac{1}{2} & \frac{1}{\sqrt{6}} \\-\frac{1}{3\sqrt{2}} & 0 & \frac{1}{\sqrt{6}} & \frac{1}{3}\end{array}\right),
&&\tilde G_4 = \left(\begin{array}{cccc}0 & 0 & 0 & 0 \\0 & 1 & 0 & 0 \\0 & 0 & 0 & 0 \\0 & 0 & 0 & 0\end{array}\right).  \nonumber
\end{align}
This would induce $\mathcal{F}$ as well:
\begin{align}
\tr(F_i \rho_A) = \tr(\tilde G_i \rho_A \otimes \rho_B) \ \forall i \in \{1,2,3\}, \quad \tr(\tilde G_4 \rho_A \otimes \rho_B) = 0 \nonumber
\end{align}
for any state $\rho_A \in \mathcal{S}(\mathcal{H}_A)$.

If we calculate the post-measurement state $\rho_{AB,\mathcal{G}}'$ associated with $\mathcal{G}$ and trace out the ancilla system $B$ to get the post-measurement $\rho'_{A,\mathcal{G}}$ of system $A$, we get a different state than if we do the same for $\tilde{\mathcal{G}}$, $\rho'_{A,\mathcal{\tilde{\mathcal{G}}}} \neq \rho'_{A,\mathcal{G}}$. For example,
\begin{align}
&\tr_B \left( \frac{G_1 \rho_A \otimes \rho_B G_1}{\tr(G_1 \rho_A \otimes \rho_B)} \right) = \left(\begin{array}{cc}1 & 0 \\0 & 0\end{array}\right) \quad \text{(independent of $\rho_A$), but} \nonumber \\
&\tr_B \left( \frac{\tilde G_1 \rho_A \otimes \rho_B \tilde G_1}{\tr(\tilde G_1 \rho_A \otimes \rho_B)} \right) = \left(\begin{array}{cc}\frac{2}{3} & 0 \\0 & \frac{1}{3}\end{array}\right) \quad \text{(independent of $\rho_A$ as well)} \,. \nonumber
\end{align}
This shows that from the POVM $\mathcal{F}$ alone, there cannot be a consistent update rule. \hfill $\blacksquare$

\end{ex}

Example \ref{meas-on-larger} shows that the update rule for generalized probabilistic theories should only make statements about post-measurement states for a class of effects which, in the case of quantum theory, reduces to orthogonal projectors. In generalized probabilistic theories, we no longer have the notion of projectors as effects, since effects are elements of a more abstract convex set $E_A$. Therefore, we have to find a criterion formulated in the language of convex sets. It turns out that projectors are precisely the pure effects of quantum theory.

\begin{prop}
\label{extreme_projector}
Let $(A, A_+, u_A)$ be a finite-dimensional quantum theory, i.e. let
\begin{itemize}
\item $\mathcal{H}$ be an $n$-dimensional Hilbert space, let
\item $\mathcal{S}(\mathcal{H})$ be the set of density operators, on $\mathcal{H}$, let
\item $A = \Herm(\mathcal{H})$ be the set of all Hermitian operators on $\mathcal{H}$, let
\item $A_+$ be the cone of all positive operators on $\mathcal{H}$, let
\item $I$ be the identity operator on $\mathcal{H}$, let
\item $E_A = [0,I] = \{T \in \Herm(\mathcal{H}) \mid 0 \leq T \leq I\} \subset A_+$.
\end{itemize}
Then $T$ is a pure effect (i.e. an extreme point of $E_A$) if and only if $T$ is a projector.\footnote{When we say projector, we always include the zero operator $0$ and the identity operator $I$.}
\end{prop}

\begin{proof}
At first, we prove that any element of $[0,I]$ can be written as a convex combination of (orthogonal) projectors. Write $T$ in its eigendecomposition:
\begin{equation*}
T = \sum\limits_{i = 1}^n \lambda_i \mid e_i \rangle \langle e_i \mid = \sum\limits_{i=1}^n \lambda_i P_i \,,
\end{equation*}
where $\lambda_1 \geq \lambda_2 \geq \ldots \lambda_n$. The operator $T$ can be written as
\begin{align}
T &= \lambda_1 P_1 + \lambda_2 P_2 + \ldots + \lambda_n P_n \nonumber \\
&= (\lambda_1 - \lambda_2)P_1 + (\lambda_2 - \lambda_3)P_1 + \ldots + (\lambda_{n-1} - \lambda_n)P_1 + \lambda_n P_1 \nonumber \\
&\qquad +(\lambda_2 - \lambda_3)P_2 + \ldots + (\lambda_{n-1} - \lambda _ n)P_2 + \lambda_n P_2 \nonumber \\
&\qquad + \ldots \nonumber \\
&\qquad + (\lambda_{n-1} - \lambda_{n})P_{n-1} + \lambda_n P_{n-1} \nonumber \\
&\qquad + \lambda_n P_n \nonumber \\
&= (\lambda_1 - \lambda_2) P_1 + (\lambda_2 - \lambda_3)(P_1 + P_2) + \ldots \nonumber \\
&\qquad + (\lambda_{n-1} - \lambda_n)(P_1 + \ldots + P_{n-1}) + \lambda_n (P_1 + \ldots + P_n) \nonumber \\
&= \sum\limits_{i=1}^n p_i \Pi_i \,, \quad p_i = \lambda_i - \lambda_{i+1} \geq 0 \quad (\lambda_{n+1} = 0) \,, \quad \Pi_i = \sum\limits_{k=1}^i P_k \,. \nonumber
\end{align}
Define $p_0 = 1-\lambda_1 \geq 0$, so that
\begin{equation*}
T = p_0 0 + \sum\limits_{i=1}^n p_i \Pi_i \quad \text{and} \quad \sum\limits_{i=0}^n p_i = 1 \,.
\end{equation*}
This proves that every POVM-element is a convex combination of projectors.
Next prove that a projector cannot be written as a proper convex combination of elements of $[0,I]$. Let $\Pi$ be a projector and suppose that
\begin{equation*}
\Pi = \lambda E + (1-\lambda) F \,, \quad 0 < \lambda < 1 \,, \quad E, F \in [0,I] \,.
\end{equation*}
For any $| \psi \rangle \in \text{ker}(\Pi)$, we have
\begin{align}
0 &= \langle \psi | \Pi | \psi \rangle = \lambda \underbrace{\langle \psi | E | \psi \rangle}_{\geq 0} + (1-\lambda) \underbrace{\langle \psi | F | \psi \rangle}_{\geq 0} \geq 0 \nonumber \\
&\Rightarrow \langle \psi | E | \psi \rangle = \langle \psi | F | \psi \rangle = 0 \,. \label{ip0}
\end{align}
Let $C$ be an operator satisfying $C^* C = E$, and let $D$ satisfy $D^* D = F$. Then (\ref{ip0}) implies
\begin{equation*}
\langle C \psi | C \psi \rangle = || C \psi ||^2 = 0 \quad \Rightarrow \quad C | \psi \rangle = 0 \quad \Rightarrow \quad E | \psi \rangle = 0 \quad \forall | \psi \rangle \in \text{ker}(\Pi) \,.
\end{equation*}
Likewise, we have that
\begin{equation*}
F | \psi \rangle = 0 \quad \forall | \psi \rangle \in \text{ker}(\Pi) \,.
\end{equation*}
Thus,
\begin{equation*}
\text{supp}(F), \text{supp}(E) \subset \text{supp}(\Pi) \,.
\end{equation*}
Let $| \phi \rangle \in \text{supp}(\Pi)$.
\begin{align}
1 &= \langle \phi | \Pi | \phi \rangle = \lambda \underbrace{\langle \phi | E | \phi \rangle}_{\leq 1} + (1-\lambda) \underbrace{\langle \phi | F | \phi \rangle}_{\leq 1} \leq 1 \nonumber \\
&\Rightarrow \langle \phi | E | \phi \rangle = \langle \phi | F | \phi \rangle = 1 \nonumber \\
&\Rightarrow E = F = \Pi \,. \nonumber \tag*{\qedhere}
\end{align}
\end{proof}

Proposition \ref{extreme_projector} provides us with the criterion that allows us to distinguish between projectors and other effects in quantum theory in the language of abstract state spaces: projectors are pure effects. By what we have discussed above, this means that we make the following restriction:
\begin{quote}\emph{Whenever we make statements about post-measurement states or consecutive measurements, we restrict ourselves to pure effects}.\end{quote}

\subsection{Repeatability and subspaces}
\label{rep-subsp-section}

The first principle that we want to postulate is the principle of \emph{repeatability}. This is a statement about consecutive measurements and therefore about post-measurement states. By the discussion of Section \ref{pure-meas-section}, this means that we restrict the statement of the principle to the case of \emph{pure} effects.

\begin{post}[Repeatability]
\label{rep-postulate}
If we perform a pure measurement twice in a row, then we get the same outcome both times. In other words, if we perform a pure measurement $\mathcal{M} = \{ e_1, \ldots, e_n \} \subset E_A$ and get an outcome associated with $e_i \in \mathcal{M}$, then the post-measurement state $\omega_\text{post}$ satisfies $e_i(\omega_\text{post}) = 1$.
\end{post}

We can link this postulate to a result that we have inferred in Section \ref{meas-section}. Corollary \ref{nonempty-face} states that for a pure effect $e \in E_A$, the set of all states $\omega \in \Omega_A$ satisfying $e(\omega) = 1$ is a face of $\Omega_A$. We called it the \emph{face associated with $e$}. In the context of Postulate \ref{rep-postulate}, we will also call it the \emph{face of possible post-measurement states}.

From a purely technical point of view, we will not explicitly make use of this postulate. Instead, it will be implicitly contained in Postulate \ref{subsp-postulate}. In other words, Postulate \ref{rep-postulate} will be the motivation for Postulate \ref{subsp-postulate}. In order to make clear what physical ideas are behind our assumptions, we state Postulate~\ref{rep-postulate} explicitly, although this is technically not necessary.

It is very natural to postulate repeatability. From the von-Neumann L\"uders projection (\ref{von-neumann-lueders}), it is easy to see that quantum theory satisfies Postulate \ref{subsp-postulate}:
\begin{align}
\omega_\text{post} = \frac{P \rho P}{\tr(P \rho)} \quad \Rightarrow \quad f_P(\omega_\text{post}) = \tr \left( P \frac{P \rho P}{\tr(P \rho)} \right) = \frac{\tr(P \rho)}{\tr(P \rho)} = 1 \,. \nonumber
\end{align}

In the following, we introduce a concept of subspaces. Roughly speaking, a subspace is a subset of the set of states which can be treated as a set of states in its own right, just as if it would be associated with a different, ``smaller'' kind of system. From a purely mathematical point of view, a subset of a set of states only needs to satisfy very little in order to induce the structure of an abstract state space. It only needs to be a compact convex subset, as is shown in the following proposition.

\begin{prop}
\label{ind-abs-prop}
For an abstract state space $(A, A_+, u_A)$ and a compact convex subset $\Omega_S \subset \Omega_A$, the triple $(S, S_+, u_S)$, where $S = \spa(\Omega_S)$, $S_+ = \cone(\Omega_S)$ (recall Definition \ref{cone-def}) and $u_S = u_A|_S$, is an abstract state space in the sense of Definition \ref{abs-st-sp-def}.
\end{prop}

\begin{proof}
According to Definition \ref{abs-st-sp-def}, we have to check that (a) $S$ is a finite-dimensional real topological vector space, (b) $S_+$ is a closed and generating cone in $S$ and (c) $u_S$ is an order unit in $S^*$. (a) is trivial. For (b), recall Theorem \ref{compact-to-closed-thm} to see that $S_+$ is closed (trivially, $S_+$ is a cone). $S_+$ is generating since $S$ is chosen to be the span of the basis $\Omega_S$. For (c), note that $u_S$ is the restriction of a strictly positive map and therefore is itself a strictly positive map. By Theorem \ref{strictly-positive-order-unit}, $u_S$ is an order unit in $S^*$.
\end{proof}

To simplify the terminology in the following discussion, we define this kind of abstract state space, which is induced by the choice of a compact convex subset of a set of states as an \emph{induced abstract state space}.

\begin{defi}[Induced abstract state space]
\label{ind-abs-def}
For an abstract state space $(A, A_+, u_A)$ and a compact convex subset $\Omega_S \subset \Omega_A$, the triple $(S, S_+, u_S)$, where $S = \spa(\Omega_S)$, $S_+ = \cone(\Omega_S)$ and $u_S = u_A|_S$, is referred to as the \textbf{abstract state space $S$ induced by $\Omega_S \subset \Omega_A$}\index{induced abstract state space}\index{abstract state space!induced by a subset|see{induced abstract state space}}. In this case, we say that the subset $\Omega_S$ \textbf{induces an abstract state space}.
\end{defi}

From a mathematical point of view, Proposition \ref{ind-abs-prop} is a correct statement, and there is no reason to forbid Definition \ref{ind-abs-def}. From a physical point of view, however, it is not sensible to consider abstract state spaces induced by arbitrary compact convex subsets $\Omega_S \subset \Omega_A$ as being physical. To see this, we first develop an intuitive picture of subspaces, before we specify \emph{physical} properties that we regard as necessary conditions for a ``subspace'' to be satisfied. Then we translate them into the mathematical language of abstract state spaces. Finally, we will give some (physical and unphysical) examples.

A subspace of a system can be thought of as a subset of the set of states that arises from the fact that we are ignorant about some properties that the system can have. In other words, a subspace arises in situations where we are unaware about the existence of some states of the system. As an example, think of a source $Q$ which emits physical systems which seem to be of some particular type. We describe this type of system by an abstract state space $(S, S_+, u_S)$. It might be that our description $(S, S_+, u_S)$ of the system only describes the actual physical system partially. This might be caused by the fact that the source $Q$ is built in a way which constrains the state of the system. 

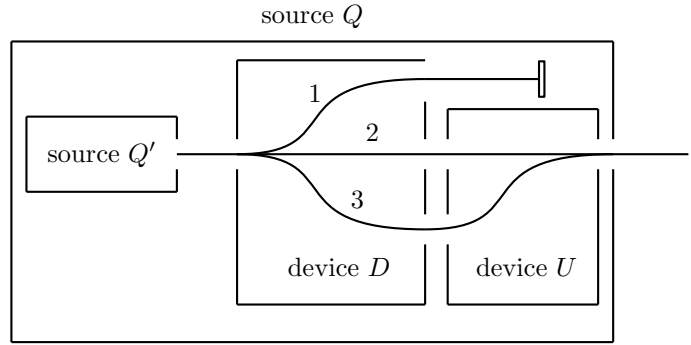
\begin{figure}[htb]
\centering

\begin{pspicture}[showgrid=false](-3,-1)(8,3.5)
\psline(-2,-1)(-2,3)
\psline(-2,3)(6,3)
\psline(-2,-1)(6,-1)
\psline(-1.8,1)(-1.8,2)
\psline(-1.8,1)(0.2,1)
\psline(-1.8,2)(0.2,2)
\psline(0.2,2)(0.2,1.7)
\psline(0.2,1)(0.2,1.3)
\psline(0.2,1.5)(7,1.5) 
\psbezier[showpoints=false]{}(1,1.5)(2.5,1.5)(1.5,2.5)(3.5,2.5)
\psbezier[showpoints=false]{}(1,1.5)(2.5,1.5)(1.5,0.5)(3.5,0.5)
\psline(1,-0.5)(3.5,-0.5)
\psline(1,2.75)(3.5,2.75)
\psline(1,-0.5)(1,1.3)
\psline(1,1.7)(1,2.75)
\psline(3.5,-0.5)(3.5,0.3)
\psline(3.5,0.7)(3.5,1.3)
\psline(3.5,1.7)(3.5,2.2)
\psline(6,-1)(6,1.3)
\psline(6,3)(6,1.7)
\psline(3.5,2.5)(5,2.5)
\psframe(5,2.25)(5.1,2.75)
\psbezier[showpoints=false]{}(3.5,0.5)(5,0.5)(4,1.5)(6,1.5)
\uput[0](-1.7,1.5){source $Q'$}
\uput[0](1.5,0){device $D$}
\uput[0](4,0){device $U$}
\uput[90](2,3){source $Q$}
\uput[180](2.3,2.3){$1$}
\uput[90](2.8,1.5){$2$}
\uput[90](2.6,0.6){$3$}
\psline(3.8,-0.5)(5.8,-0.5)
\psline(5.8,-0.5)(5.8,1.3)
\psline(5.8,1.7)(5.8,2.1)
\psline(5.8,2.1)(3.8,2.1)
\psline(3.8,2.1)(3.8,1.7)
\psline(3.8,1.3)(3.8,0.7)
\psline(3.8,0.3)(3.8,-0.5)
\end{pspicture}

\caption{A source $Q$ that emits physical systems might be built in a way such that it only emits systems in states of a subspace.}
\label{source-figure}
\end{figure}

For example, one might think of the source $Q$ as an apparatus which is composed of a few smaller apparatuses $Q'$, $D$ and $U$ (c.f. Figure \ref{source-figure}). The apparatus $Q'$ is the \emph{actual} source of the physical systems. It emits systems of some more general kind. Say that we describe this more general kind of system by an abstract state space $(A, A_+, u_A)$. The state of a system which leaves the source $Q'$ might be in any state $\omega \in \Omega_A$. After its emission, the system enters a device $D$. This device performs a measurement on the system (the measurement has three possible outcomes, say). We might think of the device $D$ as a Stern-Gerlach device. Depending on the outcome of the measurement, the device sends the system along some path. If the measurement outcome is $1$, the system is directed towards a block where it is absorbed, so in this case, the system does not leave the source $Q$. If the outcome of the measurement is $2$ or $3$, the path that the system takes leads to a device $U$. This device ensures that the system leaves the source $Q$. We might think of this device as a beam focussing device which focusses the beam to the output hole of the source.

If a source $Q$ is built in this way, it prevents some systems from leaving the source $Q$. All systems with a state that certainly leads to the measurement result $1$ are blocked. An experimenter who performs experiments on systems that leave the source $Q$ will never see such a system. He tries to find out experimentally what the state space of the system is, the will not find out that such a system is described by the abstract state space $(A, A_+, u_A)$. Instead, he describes it by some other abstract state space $(S, S_+, u_S)$, because he only sees systems in some set of states $\Omega_S$. He describes a \emph{subspace} of the system.

Consider a quantum example. Suppose the source $Q'$ emits three-level quantum systems, i.e. the Hilbert space $\mathcal{H}$ is three-dimensional with a basis $\{ |1\rangle, |2\rangle, |3\rangle \}$. The states of the actual system are therefore described by $\Omega_A = \mathcal{S}(\mathcal{H})$. Assume that the device $D$ performs a measurement with respect to the POVM $\{ |1\rangle \langle 1|, |2\rangle \langle 2| + |3\rangle \langle 3| \}$. If the outcome of the measurement is $1$, then the system in the state $\rho = |1\rangle \langle 1|$ is blocked, otherwise leaves the source $Q$. In this case, the experimenter who uses the source $Q$ would describe the states of the system by $\Omega_S = \mathcal{S}(\mathcal{H'})$, where $\mathcal{H'} = \spa(\{|2\rangle, |3\rangle\})$. The subset $\mathcal{S}(\mathcal{H'})$ is a subspace of $\mathcal{S}(\mathcal{H})$.

Now we describe some properties that we expect from a subspace. We consider the following physical requirements:

\begin{enumerate}[(a)]

\item The states in $\Omega_S$ which correspond to maximal knowledge (i.e. the pure states of $\Omega_S$) are states of maximal knowledge in $\Omega_A$ as well. This is to be interpreted as the fact that we cannot gain knowledge by ``forgetting about the rest of the state space''.

\item The set of measurements that can be performed on the subspace arises from the convexity structure of the subspace $\Omega_S$ on its own (which allows us to consider $\Omega_S$ as a space in its own right), independently of the convex structure of the larger set of states $\Omega_A$ containing the subspace. In particular, this idea encompasses two requirements:
\begin{enumerate}[(i)]

\item The structure of $\Omega_S$ as a subspace of $\Omega_A$ should be compatible with the point of view that our ignorance about the larger space $\Omega_A$ does not change the fact that actually, we perform a measurement on $\Omega_A$ (which is \emph{the} set of states associated to the type of system in question) and not merely on $\Omega_S$. Therefore, for every measurement $\mathcal{M}_S$ on the subspace $\Omega_S$ which arises from the convex structure of $\Omega_S$, there should be a measurement $\mathcal{M}_A$ on $\Omega_A$ the restriction of which to $\Omega_S$ coincides with $\mathcal{M}_S$.

\item Each measurement $\mathcal{M}_S$ on the subspace $\Omega_S$ can be performed in a way such that the post-measurement state lies in $\Omega_S$. If this would not be the case, we could not regard $\Omega_S$ as being a subspace in its own right: A description of consecutive measurements would necessarily involve the whole space $\Omega_A$. This means that for every measurement $\mathcal{M}_S$ on $\Omega_S$, the measurement $\mathcal{M}_A$ on $\Omega_A$ that induces $\mathcal{M}_S$ has a face of possible post-measurement states that coincides with the face of possible post-measurement states for $\mathcal{M}_S$ (this face is given by repeatability).

\end{enumerate}

\end{enumerate}

Now we formulate these requirements in mathematical terms. For a simplified way of speaking, we call every subset $\Omega_S \subset \Omega_A$ that satisfies the above conditions a \emph{physical subspace}. A subset $\Omega_S \subset \Omega_A$ that induces an abstract state space but violates one of these principles is called an \emph{unphysical subspace}.

\begin{defi}[Physical subspace]
\label{subsp-def}
For an abstract state space $(A, A_+, u_A)$, a compact convex subset $\Omega_S$ of $\Omega_A$ is a \textbf{physical subspace of $\Omega_A$}\index{physical subspace}\index{subspace!physical|see{physical subspace}} if the following conditions are satisfied:
\begin{enumerate}[(a)]

\item $\ext(\Omega_S) \subset \ext(\Omega_A)$.

\item For every pure measurement $\mathcal{M}_S = \{e_1, \ldots, e_n\}$ on the abstract state space $(S, S_+, u_S)$ induced by $\Omega_S \subset \Omega_A$, there is a pure measurement $\mathcal{M}_A = \{f_1, \ldots, f_n, \ldots, f_{n+k}\}$ (for some $k$) on $(A, A_+, u_A)$ such that the following properties are satisfied:
\begin{enumerate}[(i)]
\item $f_i|_S = e_i \quad \forall i \in \{1, \ldots, n\}$,
\item $F_{f_i} = F_{e_i} \quad \forall i \in \{1, \ldots, n\}$,
\end{enumerate}
where $F_{f_i} = \{\omega \in \Omega_A \mid f_i(\omega) = 1 \}$, $F_{e_i} = \{\omega \in \Omega_S \mid e_i(\omega) = 1 \}$.

\end{enumerate}
A subset $\Omega_S \subset \Omega_A$ is an \textbf{unphysical subspace} if it induces an abstract state space but violates (a) or (b).
\end{defi}

For condition (a) of a physical subspace, we already know a simple sufficient criterion: As a corollary of Proposition \ref{face-conv-ext}, we have that faces satisfy (a).

\begin{cor}
\label{face-satisfies-a}
For an abstract state space $A$, every face $\Omega_S$ of $\Omega_A$ satisfies $\ext(\Omega_S) \subset \ext(\Omega_A)$.
\end{cor}

\begin{proof}
By Proposition \ref{face-conv-ext}, $\Omega_S$ satisfies $\Omega_S = \conv( \{ v \in \ext(\Omega_A) \mid v \in \Omega_S \})$. This implies $\ext(\Omega_S) \subset \{v \in \ext(\Omega_A) \mid v \in \Omega_S\} \subset \ext(\Omega_A)$.
\end{proof}

\begin{ex}[Unphysical and physical subspaces]
\label{subsp-ex}

We consider a few examples of induced abstract state spaces (c.f. Figure \ref{unphysical-subspaces-figure}). The first three examples are unphysical subspaces. They show that not every compact convex subset of a set of states can be regarded as a physical subset. The fourth and fifth example are physical subspaces.

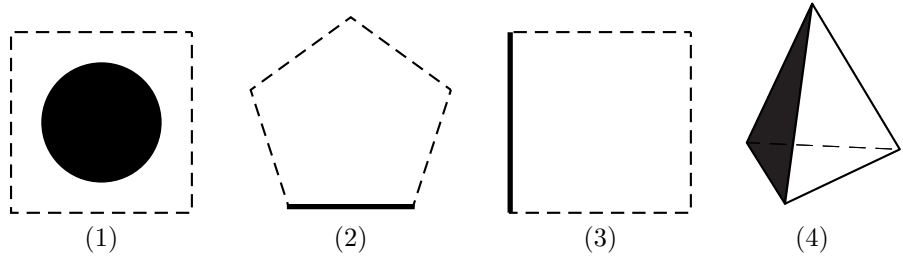
\begin{figure}[htb]
\centering

\begin{pspicture}[showgrid=false](-1.5,-1.5)(1.7,1.5)
\PstSquare[PstPicture=false, unit=1.7, linestyle=dashed]
\pscircle[PstPicture=false, fillstyle=solid, fillcolor=black](0,0){0.8}
\uput[270](0,-1.2){(1)}
\end{pspicture}
\begin{pspicture}[showgrid=false](-1.5,-1.5)(1.7,1.5)
\PstPentagon[PstPicture=false, unit=1.4, linestyle=dashed]
\psline[linewidth=0.7mm](-0.85,-1.12)(0.85,-1.12)
\uput[270](0,-1.2){(2)}
\end{pspicture}
\begin{pspicture}[showgrid=false](-1.5,-1.5)(1.7,1.5)
\PstSquare[PstPicture=false, unit=1.7, linestyle=dashed]
\psline[linewidth=0.7mm](-1.2,-1.2)(-1.2,1.2)
\uput[270](0,-1.2){(3)}
\end{pspicture}
\begin{pspicture}[showgrid=false](-1,-1.5)(1,2) 
\psset{viewpoint=8 10 10 rtp2xyz,Decran=70} 
\psSolid[Decran=4, 
object=tetrahedron, 
fcol=0 (1 setfillopacity Black),
r=3, 
action=draw*]
\uput[270](0,-1.2){(4)}
\end{pspicture}

\caption{The first three examples are unphysical subspaces: (1) a circular subset of a square, (2) a face of a pentagon, (3) a face of a square. The fourth example is a physical subspace: (4) a face of a tetrahedron.}
\label{unphysical-subspaces-figure}
\end{figure}

\begin{enumerate}[(1)]

\item A (filled) circular subset of a square is a compact convex subset, but it violates the requirement that every extreme point of the subset should correspond to a pure state of the whole set of states.

\item Let $\Omega_A$ be a regular pentagon (which is the polygon model corresponding to $n=5$). Suppose that $\Omega_S$ is a facet of $\Omega_A$. According to Corollary \ref{face-satisfies-a}, $\Omega_S$ satisfies condition (a). However, it violates condition (b) (i). To see this, note that $\Omega_S$ is a line-segment, i.e. a 1-simplex, so the induced abstract state space $(S, S_+, u_S)$ is a classical theory of two pure states. According to Proposition \ref{perf-dist-classical}, these two pure states can be perfectly distinguished by a measurement on $S$. But in the pentagon model, the two vertices of an edge cannot be perfectly distinguished by a measurement on $A$ (as one can calculate from (\ref{e_i-odd}).

\item Let $\Omega_A$ be a square (the $n=4$ polygon model). A facet $\Omega_S$ of $\Omega_A$ satisfies (a). In contrast to the previous example, it also satisfies condition (b) (i). To see this, consider the face $F_{e_3}$ associated with $e_3$ (see Figure \ref{square-figure-2}). The vertices of this 1-simplex are $\omega_2$ and $\omega_3$. They can be perfectly distinguished by the measurement $\mathcal{M}_A = \{e_2, e_4\}$ since $\omega_2 \in F_{e_2}$ and $\omega_3 \in F_{e_4}$. But $\Omega_S$ violates (b) (ii). The measurement $\mathcal{M}_A = \{e_2, e_4\}$ is the only pure measurement on $A$ that perfectly distinguishes $\omega_2$ from $\omega_3$, and for this measurement, $F_{e_2} \neq \{\omega_2\}$, $F_{e_4} \neq \{\omega_3\}$.

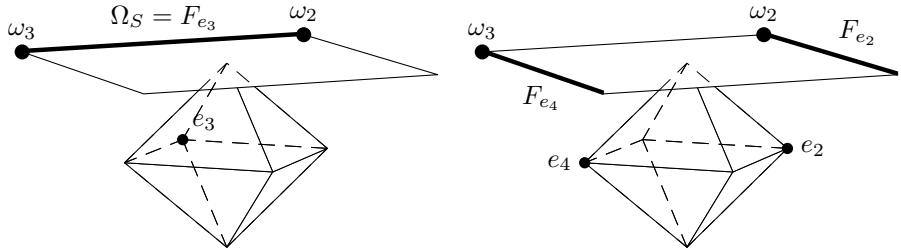
\begin{figure}[htb]
\centering

\begin{pspicture}[showgrid=false](-2,0)(4,3.5)
\psset{viewpoint=26 10 5,Decran=70}
\psset{solidmemory}
\psSolid[object=new,linewidth=0.5\pslinewidth,
action=draw*,
name=B,
sommets= 
0 0 0 
0.420448 0.420448 0.5 
-0.420448 0.420448 0.5 
-0.420448 -0.420448 0.5 
0.420448 -0.420448 0.5 
0 0 1 
0 1.18921 1 
-1.18921 0 1 
0 -1.18921 1 
1.18921 0 1, 
faces={
[1 4 0]
[0 2 1]
[0 3 2]
[0 4 3]
[1 2 5]
[2 3 5]
[3 4 5]
[4 1 5]
[6 7 8 9]}]%
\psPoint(-0.420448, -0.420448, 0.5){e3}
\psdots[dotsize=0.15](e3)
\uput[ur](e3){$e_3$}
\psSolid[object=line, linewidth=2\pslinewidth,args=-1.18921 0 1 0 -1.18921 1]
\psSolid[object=line, linewidth=2\pslinewidth,args=-1.18921 0 1.01 0 -1.18921 1.01]
\psPoint(-0.594604, -0.594604, 1){h}
\uput[u](h){$\Omega_S = F_{e_3}$}
\psPoint(-1.18921, 0, 1){w2}
\psdots[dotsize=0.2](w2)
\uput[u](w2){$\omega_2$}
\psPoint(0, -1.18921, 1){w3}
\psdots[dotsize=0.2](w3)
\uput[u](w3){$\omega_3$}
\end{pspicture}
\begin{pspicture}[showgrid=false](-2,0)(2,2.7)
\psset{viewpoint=26 10 5,Decran=70}
\psset{solidmemory}
\psSolid[object=new,linewidth=0.5\pslinewidth,
action=draw*,
name=B,
sommets= 
0 0 0 
0.420448 0.420448 0.5 
-0.420448 0.420448 0.5 
-0.420448 -0.420448 0.5 
0.420448 -0.420448 0.5 
0 0 1 
0 1.18921 1 
-1.18921 0 1 
0 -1.18921 1 
1.18921 0 1, 
faces={
[1 4 0]
[0 2 1]
[0 3 2]
[0 4 3]
[1 2 5]
[2 3 5]
[3 4 5]
[4 1 5]
[6 7 8 9]}]%
\psPoint(-0.420448, 0.420448, 0.5){e2}
\psdots[dotsize=0.15](e2)
\uput[r](e2){$e_2$}
\psPoint(0.420448, -0.420448, 0.5){e4}
\psdots[dotsize=0.15](e4)
\uput[l](e4){$e_4$}
\psSolid[object=line, linewidth=2\pslinewidth,args=0 1.18921 1 -1.18921 0 1]
\psSolid[object=line, linewidth=2\pslinewidth,args=0 1.18921 1.01 -1.18921 0 1.01]
\psPoint(-0.594604, 0.594604, 1){g}
\uput[ur](g){$F_{e_2}$}
\psPoint(0.594604, -0.594604, 1){lu}
\uput[d](lu){$F_{e_4}$}
\psSolid[object=line, linewidth=2\pslinewidth,args=0 -1.18921 1 1.18921 0 1]
\psSolid[object=line, linewidth=2\pslinewidth,args=0 -1.18921 1.01 1.18921 0 1.01]
\psPoint(-1.18921, 0, 1){w2}
\psdots[dotsize=0.2](w2)
\uput[u](w2){$\omega_2$}
\psPoint(0, -1.18921, 1){w3}
\psdots[dotsize=0.2](w3)
\uput[u](w3){$\omega_3$}
\end{pspicture}

\caption{A facet $\Omega_S$ of the $n=4$ polygon model violates condition (b) (ii) of Definition \ref{subsp-def}, so it is not a physical subspace.}
\label{square-figure-2}
\end{figure}

\item A face of a tetrahedron is an example of a physical subspace. It satisfies all the requirements of Definition \ref{subsp-def}.

\item Let $\mathcal{H}$ be a finite-dimensional Hilbert space, let $\mathcal{H}' \subset \mathcal{H}$ be a subspace of $\mathcal{H}$ (in the vector space sense). Then $\mathcal{S}(\mathcal{H}')$ is a physical subspace of $\mathcal{S}(\mathcal{H})$. \hfill $\blacksquare$

\end{enumerate}

\end{ex}

As we have mentioned above, physical subspaces naturally arise in the context of measurements. It is instructive to discuss this in more generality for the example of quantum theory. Consider a quantum system with a finite-dimensional Hilbert space $\mathcal{H}$, i.e. with $\Omega_A = \mathcal{S}(\mathcal{H})$. If we perform a projective measurement $\mathcal{M} = \{P_i\}_{i \in \mathcal{I}_\mathcal{M}}$ and obtain the outcome $k \in \mathcal{I}_\mathcal{M}$, then we know that the state of the system after this measurement is a density operator on the image of the projector, i.e. $\rho_\text{post} \in \mathcal{S}(\mathcal{H}')$ with $\mathcal{H}' = P_k \mathcal{H}$. On the other hand, $\mathcal{S}(\mathcal{H}')$ is exactly the subset of $\mathcal{S}(\mathcal{H})$ which is compatible with the repeatability of the measurement (Postulate~\ref{rep-postulate}): For every density operator $\rho \in \mathcal{S}(\mathcal{H})$, we have that $f_{P_k}(\rho) = \tr(P_k \rho) = 1$ if and only if $\rho$ is a density operator on $\mathcal{H}'$, i.e. $\rho \in \mathcal{S}(\mathcal{H}')$.

The subspace principle states that this is true for a general abstract state space. It combines the concepts of repeatability and subspaces. Repeatability requires the post-measurement state of a system to lie in a certain face of $\Omega_A$. We postulate that this subset is a physical subspace.

\begin{post}[Subspace principle]
\label{subsp-postulate}
For an abstract state space $A$, let $\Omega_S$ be a face of $\Omega_A$ associated with a pure effect $f \in E_A$, i.e. $\Omega_S = F_f$. In other words, let $\Omega_S$ be a face of $\Omega_A$ such that
\begin{equation}
\exists f \in E_A \text{ pure}: f(\omega) = 1 \Leftrightarrow \omega \in \Omega_S \ \forall \omega \in \Omega_A.
\end{equation}
Then $\Omega_S$ is a physical subspace according to Definition \ref{subsp-def}.
\end{post}

\subsection{A state discrimination principle}
\label{discrimination-section}

In this section, we introduce a state discrimination principle for abstract state spaces. It concerns the notion of perfect distinguishability of states. Recall from Definition \ref{perf-dist-st-def} that states $\omega_1, \ldots, \omega_n \in \Omega_A$ \emph{are perfectly} distinguishable if there is a measurement $\{f_1, \ldots, f_n\} \subset E_A$ such that $f_i(\omega_j) = \delta_{ij}$ for all $i, j \in \{ 1, \ldots, n \}$.

\begin{ex}[Perfectly distinguishable states]
In order to get a better feeling for the notion of perfect distinguishability, we make a few examples.
\begin{enumerate}[(1)]

\item In quantum theory, states $\rho_1, \ldots, \rho_n \in \mathcal{S}(\mathcal{H})$ are perfectly distinguishable if and only if they have support on pairwise orthogonal Hilbert subspaces:
\begin{align}
\rho_1, \ldots, \rho_n \text{ are perfectly distinguishable} \quad \Leftrightarrow \quad \supp(\rho_i) \mathrel{\bot} \supp(\rho_j) \ \forall i \neq j \,. \nonumber
\end{align}
In this case, a measurement which perfectly distinguishes between the states $\rho_1, \ldots, \rho_n$ would be given by the projectors $P_k$ onto the supports $\supp(\rho_k)$ of the density operators.\footnote{If $P_\text{total} = \sum_{i=1}^n P_k$ is a projector onto a proper Hilbert subspace of $\mathcal{H}$, we add $I_\mathcal{H} - P_\text{tot}$ to one of the projectors to get a measurement.}

\item In the polygon model with $n=3$ (c.f. Example \ref{polygon-models}), which corresponds to a triangle, the pure states $\omega_1, \omega_2$ and $\omega_3$ are perfectly distinguishable by the measurement $\{e_1, e_2, e_3\}$, as is easily verified from (\ref{omega_i}) and (\ref{e_i-odd}). 

\item In Example \ref{subsp-ex} (3), we have seen that the states $\omega_2$ and $\omega_3$ of the $n=4$ polygon model are perfectly distinguishable states. \hfill $\blacksquare$

\end{enumerate}

\end{ex}

To state the state discrimination principle properly, it is convenient to generalize the definition of perfect distinguishability to \emph{sets} of states.

\begin{defi}
For an abstract state space $A$, sets $\Lambda_1, \ldots, \Lambda_n \subset \Omega_A$ are \textbf{perfectly distinguishable sets of states}\index{perfectly distinguishable sets of states}\index{sets of states!perfectly distinguishable|see{perfectly distinguishable sets of states}} if there is a measurement $\{e_1, \ldots, e_n\}$ such that for every $i \in \{1, \ldots, n\}$, we have that
\begin{align}
e_i(\omega) = 1 \ \forall \omega \in \Lambda_i, \quad e_i(\omega) = 0 \ \forall \omega \in \Lambda_j \text{ if } j \neq i \,. \nonumber
\end{align}
If $n = 2$, we say that $\Lambda_1$ can be perfectly distinguished from $\Lambda_2$.
\end{defi}

This definition of perfectly distinguishable sets of states reduces to the Definition \ref{perf-dist-st-def} of perfectly distinguishable states in the case where set each $\Lambda_i$ only contains one state.

With this definition at hand, we can state the state discrimination principle.

\begin{post}[State discrimination principle]
\label{discr-postulate}
Let $A$ be an abstract state space, let $\Lambda_1, \Lambda_2 \subset \Omega_A$ be perfectly distinguishable sets of states. Assume that in addition, there are subsets $\Lambda_3, \Lambda_4 \subset \Lambda_2$ such that $\Lambda_3$ is perfectly distinguishable from $\Lambda_4$. Then $\Lambda_1, \Lambda_3$ and $\Lambda_4$ are perfectly distinguishable.
\end{post}

This is a natural assumption. An everyday life example: If I can distinguish black hats from colored hats and blue hats from red hats, then I can distinguish between black, blue and red hats. The state discrimination principle also holds in quantum theory, as can be seen in the following example.

\begin{ex}[State discrimination principle in quantum theory]
Let $\mathcal{H}$ be a four-dimensional Hilbert space. Let $\Lambda_1 = \{ \rho_1, \rho_2 \}$ and $\Lambda_2 = \{ \rho_3, \rho_4 \}$ be perfectly distinguishable sets of states. In quantum theory, this means that
\begin{align}
\spa(\supp(\rho_1), \ \supp(\rho_2)) \mathrel{\bot} \spa(\supp(\rho_3), \ \supp(\rho_4)) \,. \nonumber
\end{align}
Suppose that $\{ \rho_3 \}, \{ \rho_4 \} \subset \Lambda_2$ are perfectly distinguishable sets of states, i.e. $\rho_3$ and $\rho_4$ are perfectly distinguishable states. This means that
\begin{align}
\supp(\rho_3) \mathrel{\bot} \supp(\rho_4) \,. \label{supp-rho-3-rho-4}
\end{align}
Then we have that
\begin{align}
&\supp(\rho_3) \subset \spa(\supp(\rho_3), \ \supp(\rho_4)) \mathrel{\bot} \spa(\supp(\rho_1), \ \supp(\rho_2)) \nonumber \\
&\Rightarrow \supp(\rho_3) \mathrel{\bot} \spa(\supp(\rho_1), \ \supp(\rho_2)) \,, \label{supp-rho-3} \\
&\supp(\rho_4) \subset \spa(\supp(\rho_3), \ \supp(\rho_4)) \mathrel{\bot} \spa(\supp(\rho_1), \ \supp(\rho_2)) \,. \nonumber \\
&\Rightarrow \supp(\rho_4) \mathrel{\bot} \spa(\supp(\rho_1), \ \supp(\rho_2)) \,. \label{supp-rho-4}
\end{align}
Quantum theory satisfies the state discrimination principle: Equations (\ref{supp-rho-3-rho-4}), (\ref{supp-rho-3}) and (\ref{supp-rho-4}) together imply that $\Lambda_2, \{ \rho_3 \}$ and $\{\rho_4\}$ are perfectly distinguishable sets of states. \hfill $\blacksquare$
\end{ex}

While classical and quantum theory satisfy Postulate \ref{discr-postulate}, there are theories in which this is not the case, as the following example shows.

\begin{ex}[Violation of the state discrimination principle in the $n=4$ polygon model]
\label{square-discr-vio}
Once again, we consider the case where $\Omega_A$ is a square. Consider the sets $\Lambda_1 = \{ \omega_1, \omega_2 \}$ and $\Lambda_2 = \{ \omega_3, \omega_4 \}$. They can be perfectly distinguished by the measurement $\{e_2, e_4\}$ (see Figure \ref{square-figure-3}). Let $\Lambda_3$ and $\Lambda_4$ be the subsets $\Lambda_3 = \{\omega_3\}, \Lambda_4 = \{\omega_4\} \subset \Lambda_2$. They can be perfectly distinguished by the measurement $\{e_1, e_3\}$. If the square would satisfy the state discrimination principle, then the sets $\{\omega_1, \omega_2\}, \{\omega_3\}$ and $\{\omega_4\}$ would be perfectly distinguishable. In particular, the states $\omega_2, \omega_3$ and $\omega_4$ would be perfectly distinguishable. However, there is no measurement on the square which distinguishes states.

\begin{figure}[htb]
\centering

\begin{pspicture}[showgrid=false](-2,0)(4,2.7)
\psset{viewpoint=26 10 5,Decran=70}
\psset{solidmemory}
\psSolid[object=new,linewidth=0.5\pslinewidth,
action=draw*,
name=B,
sommets= 
0 0 0 
0.420448 0.420448 0.5 
-0.420448 0.420448 0.5 
-0.420448 -0.420448 0.5 
0.420448 -0.420448 0.5 
0 0 1 
0 1.18921 1 
-1.18921 0 1 
0 -1.18921 1 
1.18921 0 1, 
faces={
[1 4 0]
[0 2 1]
[0 3 2]
[0 4 3]
[1 2 5]
[2 3 5]
[3 4 5]
[4 1 5]
[6 7 8 9]}]%
\psPoint(-0.420448, 0.420448, 0.5){e2}
\psdots[dotsize=0.15](e2)
\uput[r](e2){$e_2$}
\psPoint(0.420448, -0.420448, 0.5){e4}
\psdots[dotsize=0.15](e4)
\uput[l](e4){$e_4$}
\psSolid[object=line, linewidth=2\pslinewidth,args=0 1.18921 1 -1.18921 0 1]
\psSolid[object=line, linewidth=2\pslinewidth,args=0 1.18921 1.01 -1.18921 0 1.01]
\psPoint(-0.594604, 0.594604, 1){g}
\uput[ur](g){$F_{e_2}$}
\psPoint(0.594604, -0.594604, 1){lu}
\uput[d](lu){$F_{e_4}$}
\psSolid[object=line, linewidth=2\pslinewidth,args=0 -1.18921 1 1.18921 0 1]
\psSolid[object=line, linewidth=2\pslinewidth,args=0 -1.18921 1.01 1.18921 0 1.01]
\psPoint(-1.18921, 0, 1){w2}
\psdots[dotsize=0.2](w2)
\uput[u](w2){$\omega_2$}
\psPoint(0, 1.18921, 1){w1}
\psdots[dotsize=0.2](w1)
\uput[u](w1){$\omega_1$}
\psPoint(0, -1.18921, 1){w3}
\psdots[dotsize=0.2](w3)
\uput[u](w3){$\omega_3$}
\psPoint(1.18921, 0, 1){w4}
\psdots[dotsize=0.2](w4)
\uput[u](w4){$\omega_4$}
\end{pspicture}
\begin{pspicture}[showgrid=false](-2,0)(2,3.5)
\psset{viewpoint=26 10 5,Decran=70}
\psset{solidmemory}
\psSolid[object=new,linewidth=0.5\pslinewidth,
action=draw*,
name=B,
sommets= 
0 0 0 
0.420448 0.420448 0.5 
-0.420448 0.420448 0.5 
-0.420448 -0.420448 0.5 
0.420448 -0.420448 0.5 
0 0 1 
0 1.18921 1 
-1.18921 0 1 
0 -1.18921 1 
1.18921 0 1, 
faces={
[1 4 0]
[0 2 1]
[0 3 2]
[0 4 3]
[1 2 5]
[2 3 5]
[3 4 5]
[4 1 5]
[6 7 8 9]}]%
\psPoint(-0.420448, -0.420448, 0.5){e3}
\psdots[dotsize=0.15](e3)
\uput[ur](e3){$e_3$}
\psSolid[object=line, linewidth=2\pslinewidth,args=-1.18921 0 1 0 -1.18921 1]
\psSolid[object=line, linewidth=2\pslinewidth,args=-1.18921 0 1.01 0 -1.18921 1.01]
\psPoint(-0.594604, -0.594604, 1){h}
\uput[u](h){$F_{e_3}$}
\psPoint(0, -1.18921, 1){w3}
\psdots[dotsize=0.2](w3)
\uput[u](w3){$\omega_3$}
\psPoint(0.420448, 0.420448, 0.5){e1}
\psdots[dotsize=0.15](e1)
\uput[d](e1){$e_1$}
\psPoint(1.18921, 0, 1){w4}
\psdots[dotsize=0.2](w4)
\uput[u](w4){$\omega_4$}
\psPoint(0.727673, 0.528686, 1){lele}
\uput[ur](lele){$F_{e_1}$}
\psSolid[object=line, linewidth=2\pslinewidth,args=0 1.18921 1 1.18921 0 1]
\psSolid[object=line, linewidth=2\pslinewidth,args=0 1.18921 1.01 1.18921 0 1.01]
\end{pspicture}

\caption{The square violates the state discrimination principle.}
\label{square-figure-3}
\end{figure}
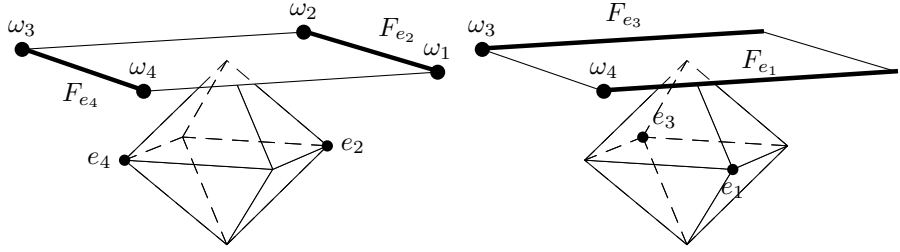
\hfill $\blacksquare$

\end{ex}

\subsection{Result: Classical theory derived from polytopic theory and three postulates}
\label{result-1-result}

In this section, we prove the first result of this thesis. It states that every \emph{polytopic} theory which satisfies Postulates \ref{rep-postulate}, \ref{subsp-postulate} and \ref{discr-postulate} is a \emph{classical} theory. Recall from Examples \ref{classical-theory-ex} and \ref{polytopic-theory-ex} that an abstract state space is a polytopic theory if $\Omega_A$ is a polytope and a classical theory if $\Omega_A$ is a simplex.

As a first step towards the result, we prove that if $A$ is a polytopic theory that satisfies Postulate \ref{subsp-postulate}, then for every face $F$ of $\Omega_A$, there is a pure effect $f \in E_A$ such that the face $F_f$ associated with $f$ coincides with $F$, i.e. $f(\omega) = 1$ if and only if $\omega \in F$. Note that without Postulate \ref{subsp-postulate}, this is not true in general. We know from Proposition \ref{rest-possible} that for a polytopic theory, the statement is always true if $F$ is a \emph{facet}, but as we have mentioned on page \pageref{counterexample-mention} in Section \ref{counterexample-mention}, in the case where $F$ is a \emph{face}, this is not true in general.

\begin{lemma}
\label{rest-always-possible}
Let $A$ be a polytopic theory which satisfies Postulate \ref{subsp-postulate}. Then, for every face $F$ of $\Omega_A$, there is a pure effect $f \in E_A$ such that for every $\omega \in \Omega_A$, we have that $f(\omega) = 1 \Leftrightarrow \omega \in F$, i.e. $F = F_f$.
\end{lemma}

\begin{proof}
We prove this Lemma by induction over the dimension of the face (c.f. Definition \ref{face-dimension-def}). Say that $\Omega_A$ is a $d$-polytope.
\begin{itemize}

\item \textbf{Base case: $F$ is a $(d-1)$-face} \\
If $F$ is a $(d-1)$-face, then $F$ is a facet of $\Omega_A$. We have proved in Proposition \ref{rest-possible} that in this case, there is a pure effect $f \in E_A$ such $F = F_f$.

\item \textbf{Inductive step: the case where $F$ is a $(d-k)$-face (for some $k \geq 2$)} \\
Assume that for every $(d-k+1)$-face $G$, there is a linear functional $g \in E_A$ such that $G = F_g$ (induction hypothesis). Let $F$ be a $(d-k)$-face of $\Omega_A$. Our goal is to show that there is a pure effect $f \in E_A$ such that $F_f = F$.

By Proposition \ref{face-sequence-prop}, there is a sequence $\Omega_A \supset F_d \supset F_{d-1} \supset \ldots \supset F_{d-k+1} \supset F_{d-k} = F$ of faces of $\Omega_A$ such that $F_{i-1}$ is a facet of $F_i$ for all $i \in \{d, \ldots, d-k+1 \}$.

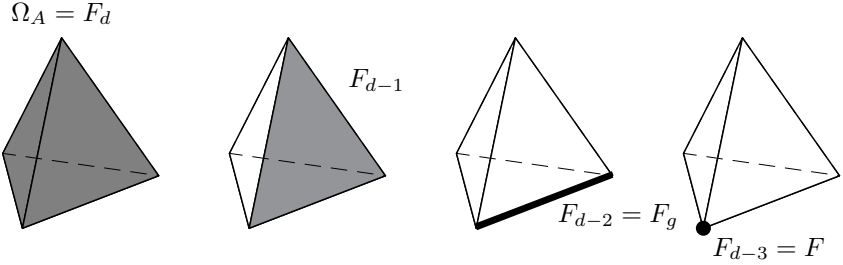
\begin{figure}[htb]
\centering

\begin{pspicture}[showgrid=false](-1,-0.2)(1.4,3.2) 
\psset{viewpoint=10 20 30 rtp2xyz,Decran=9} 
\psSolid[
object=new,  linewidth=0.7\pslinewidth,
fillcolor=gray,
sommets=
0 0 3 
-0.707107 1.22474 1 
-0.707107 -1.22474 1 
1.41421 0 1, 
faces={
[1 2 3]
[1 2 0]
[2 3 0]
[3 1 0]
},
action=draw*]%
\psPoint(0, 0, 3){lele}
\uput[u](lele){$\Omega_A = F_d$}
\end{pspicture}
\begin{pspicture}[showgrid=false](-1.5,-0.2)(1.4,3.2) 
\psset{viewpoint=10 20 30 rtp2xyz,Decran=9} 
\psSolid[
object=new,  linewidth=0.7\pslinewidth,
fcol=3 (Gray),
sommets=
0 0 3 
-0.707107 1.22474 1 
-0.707107 -1.22474 1 
1.41421 0 1, 
faces={
[1 2 3]
[1 2 0]
[2 3 0]
[3 1 0]
},
action=draw*]%
\psPoint(-0.707107, 1.02474, 2){lele}
\uput[u](lele){$F_{d-1}$}
\end{pspicture}
\begin{pspicture}[showgrid=false](-1.5,-0.2)(1.4,3.2) 
\psset{viewpoint=10 20 30 rtp2xyz,Decran=9} 
\psSolid[
object=new,  linewidth=0.7\pslinewidth,
sommets=
0 0 3 
-0.707107 1.22474 1 
-0.707107 -1.22474 1 
1.41421 0 1, 
faces={
[1 2 3]
[1 2 0]
[2 3 0]
[3 1 0]
},
action=draw*]%
\psPoint(-0.707107, 1.32474, 0.78){lele}
\psSolid[object=line, linewidth=2\pslinewidth,args=-0.707107 1.22474 1 1.41421 0 1]
\psSolid[object=line, linewidth=2\pslinewidth,args=-0.707107 1.22474 1.04 1.41421 0 1.04]
\uput[d](lele){$F_{d-2} = F_g$}
\end{pspicture}
\begin{pspicture}[showgrid=false](-1.5,-0.2)(1.4,3.2) 
\psset{viewpoint=10 20 30 rtp2xyz,Decran=9} 
\psSolid[
object=new,  linewidth=0.7\pslinewidth,
sommets=
0 0 3 
-0.707107 1.22474 1 
-0.707107 -1.22474 1 
1.41421 0 1, 
faces={
[1 2 3]
[1 2 0]
[2 3 0]
[3 1 0]
},
action=draw*]%
\psPoint(1.41421, 0, 1){lele}
\psdots[dotsize=0.2](lele)
\uput[dr](lele){$F_{d-3} = F$}
\end{pspicture}

\caption{This figure shows the sequence $\Omega_A \supset F_d \supset F_{d-1} \supset \ldots \supset F_{d-k+1} \supset F_{d-k} = F$ for the case where $\Omega_A$ is a tetrahedron ($d=3$) and where $F$ is a vertex.}
\label{face-sequence-figure-1}
\end{figure}

$F_{d-k+1}$ is a $(d-k+1)$-face of $\Omega_A$, so by the induction hypothesis, there is a pure effect $g \in E_A$ such that $F_{d-k+1} = F_g$. By virtue of Postulate \ref{subsp-postulate}, this implies that $F_{d-k+1}$ is a physical subspace of $\Omega_A$. This means that the abstract state space $(S = \spa(F_{d-k+1}), S_+ = \cone(F_{d-k+1}), u_S = u_A|_S)$ induced by $\Omega_S = F_{d-k+1}$ is an abstract state space in its own right. $F = F_{d-k}$ is a facet of $\Omega_S$. By Proposition \ref{rest-possible}, there is pure effect $e \in E_S$ on the physical subspace $S$ such that $F_e = \{ \omega \in \Omega_S \mid e(\omega) = 1 \} = F$. Proposition \ref{u-f-pure} tells us that the complementary effect $\overline e = u_S - e \in E_S$ is pure as well, so $\{ e, \overline e \}$ is a pure measurement on $S$.

As we have mentioned above, $\Omega_S = F_{k-1}$ is a physical subspace by Postulate \ref{subsp-postulate}. By the definition of a physical subspace, there is a pure measurement $\{ f, f_2, \ldots, f_l \}$ such that (in particular) $F_f = \{ \omega \in \Omega_A \mid f(\omega) = 1 \} = F_e = \{ \omega \in \Omega_S \mid e(\omega) = 1 \} = F$. This proves the claim. \qedhere

\end{itemize}
\end{proof}

Lemma \ref{rest-always-possible} allows us to prove the result of this section.

\begin{thm}
\label{result-1-thm}
Let $A$ be a polytopic theory which satisfies Postulates \ref{rep-postulate}, \ref{subsp-postulate} and \ref{discr-postulate}. Then $A$ is a classical theory\index{classical theory}.
\end{thm}

\begin{proof}

Say that $\Omega_A$ is a $d$-polytope. The idea is to prove the theorem by the following four steps.
\begin{enumerate}[(i)]
\item At first, we show that there is a sequence $\Omega_A \supset F_0 \supset F_1 \supset \ldots \supset F_d$ of faces of $\Omega_A$, where $F_{i+1}$ is a facet of $F_i$ for all $i \in \{ 0, \ldots, d-1 \}$, such that the last face consists of only one point, $F_d = \{ \widetilde \omega \}$.
\item From this sequence, we construct a sequence $\overline{F_1}, \ldots, \overline{F_d}, F_d$ of perfectly distinguishable sets of states.
\item From this sequence in turn, we construct $(d+1)$ perfectly distinguishable states $\omega_1, \ldots, \omega_d, \widetilde \omega$.
\item Finally, we show that this implies that $\Omega_A$ is a simplex.
\end{enumerate}

Now we prove each of the steps.
\begin{enumerate}[(i)]

\item Let $\omega_d \in \Omega_A$ be a pure state. In the language of convex sets, $\omega_d$ is an extreme point, so $\{ \omega_d \}$ is a face of $\Omega_A$. By Proposition \ref{face-sequence-prop}, there is a sequence $\Omega_A \supset F_0 \supset F_1 \supset \ldots \supset F_d$ of faces of $\Omega_A$, where $F_{i+1}$ is a facet of $F_i$ for all $i \in \{ 0, \ldots, d-1 \}$, such that $F_d = \{ \omega_d \}$.

\item $F_1$ is a facet of $\Omega_A$. This allows us to apply Proposition \ref{rest-possible} to see that there is a pure effect $f_1 \in E_A$ such that the face $F_{f_1}$ associated with $f_1$ coincides with $F$, i.e. $F_1 = F_{f_1} = \{ \omega \in \Omega_A \mid f_1(\omega) = 1 \}$. By Proposition \ref{u-f-pure}, we have that the complementary effect $\overline{f_1} := u_A - f_1$ is a pure effect. From Corollary \ref{nonempty-face}, we get that there is a face $F_{\overline{f_1}}$ such that $F_{\overline{f_1}} = \{ \omega \in \Omega_A \mid \overline{f_1}(\omega) = 1 \}$. By Definition \ref{opposite-face-def}, we call this the opposite face $\overline{F_1} := \overline{F_{f_1}} = F_{\overline{f_1}}$. The effects $\{ f_1, \overline{f_1} \}$ form a measurement, so
\begin{align}
&f_1(\omega) = 1 \quad \forall \omega \in F_1\,, &f_1(\omega) = 0 \quad \forall \omega \in \overline{F_1}\,, \nonumber \\
&\overline{f_1}(\omega) = 0 \quad \forall \omega \in F_1\,, &\overline{f_1}(\omega) = 1 \quad \forall \omega \in \overline{F_1}\,. \nonumber
\end{align}
Therefore,
\begin{align}
\label{level-1}
\text{the sets } F_1, \overline{F_1} \subset \Omega_A \text{ are perfectly distinguishable.}
\end{align}

$F_1 = F_{f_1}$ is a face associated with a pure effect, so by Postulate \ref{subsp-postulate}, $\Omega_S^1 = F_1$ is a physical subspace. Consider the abstract state space $(S^1 = \spa(\Omega_S^1), S_+^1 = A_+ \cap S^1, u_S^1 = u_A|_S^1)$ induced by $\Omega_S^1 = F_1$. $F_2$ is a facet of $\Omega_S^1$, so we can apply Proposition \ref{rest-possible} to see that there is a pure effect $e_2 \in E_S^1$ such that $F_2 = F_{e_2} = \{ \omega \in \Omega_S^1 \mid e_2(\omega) = 1 \}$. Like before, we can use Proposition \ref{u-f-pure} and Corollary \ref{nonempty-face} to get a pure measurement $\{ e_2, \overline{e_2} \}$ on $S^1$ and a face $\overline{F_2}$ of $\Omega_S^1$ such that $F_2$ and $\overline{F_2}$ are perfectly distinguishable:
\begin{align}
&e_2(\omega) = 1 \quad \forall \omega \in F_2\,, &e_2(\omega) = 0 \quad \forall \omega \in \overline{F_2}\,, \nonumber \\
&\overline{e_2}(\omega) = 0 \quad \forall \omega \in F_2\,, &e_2(\omega) = 1 \quad \forall \omega \in \overline{F_2}\,. \nonumber
\end{align}
$\Omega_S^1$ is a physical subspace, so there is a pure measurement $\{ g_1, g_2, \ldots, g_{2+k} \} \subset E_A$ (for some $k$) on $A$ such that $g_1|_{S^1} = e_2$, $g_2|_{S^1} = \overline{e_2}$ and $F_{g_1} = \{ \omega \in \Omega_A \mid g_1(\omega) = 1 \} = F_{e_2} = \{ \omega \in \Omega_S^1 \mid e_2(\omega) = 1 \} = F_2$ and likewise $F_{g_2} = F_{\overline{e_2}}$. Define $f_2 = g_1$, $\overline{f_2} = \sum_{i=2}^{2+k} g_i$. We have constructed a measurement $\{ f_2, \overline{f_2} \}$ on $A$ such that
\begin{align}
&f_2(\omega) = 1 \quad \forall \omega \in F_2\,, &f_2(\omega) = 0 \quad \forall \omega \in \overline{F_2}\,, \nonumber \\
&\overline{f_2}(\omega) = 0 \quad \forall \omega \in F_2\,, &f_2(\omega) = 1 \quad \forall \omega \in \overline{F_2}\,. \nonumber
\end{align}
Thus,
\begin{align}
\label{level-2}
\text{the sets } F_2, \overline{F_2} \subset F_1 \text{ are perfectly distinguishable.}
\end{align}
Now we combine (\ref{level-1}) and (\ref{level-2}) and use Postulate \ref{discr-postulate} to see that $\overline{F_1}, \overline{F_2}, F_2$ are perfectly distinguishable sets of states.

$F_3$ is a facet of $F_2$. By Proposition \ref{face-face-face}, $F_3$ is a face of $\Omega_A$. We have shown in Lemma \ref{rest-always-possible} that this implies that we have a pure effect $f_3 \in E_A$ such that $F_3 = F_{f_3}$. By completely analogous reasoning as above, we get perfectly distinguishable faces $\overline{F_1}, \overline{F_2}, \overline{F_3}, F_3$ of $\Omega_A$. We iterate this process until we get perfectly distinguishable faces $\overline{F_1}, \ldots, \overline{F_d}, F_d = \{ \widetilde \omega \}$.

\item This step is easy. We simply have to choose some state $\omega_i \in \overline{F_i}$ for every $i \in \{ 1, \ldots, d \}$ to get $(d+1)$ perfectly distinguishable states $\omega_1, \ldots, \omega_d, \widetilde \omega$.

\item The states $\omega_1, \ldots, \omega_d, \widetilde \omega$ are perfectly distinguishable, so by Proposition~\ref{perf-dist-classical}, $\Omega_A$ is a $d$-simplex. Therefore, $A$ is a classical theory. \qedhere

\end{enumerate}

\end{proof}

\newpage

\section{Main result: One simple postulate implies that every polytopic state space is classical}
\label{result-2-section}

This section is dedicated to the main result of this thesis. It states that every polytopic theory which satisfies a very simple postulate is a classical theory. The postulate claims that if the outcome of a measurement can be predicted in advance with certainty, then performing this measurement does not disturb the state. Although this postulate looks very weak, we show that it is strong enough to rule out all polytopic theories except for classical theory.

In Section \ref{non-dist-section}, we recapitulate the concept of transformations and operations. We will finish our definition of the framework of abstract state spaces by making the assumption that every pure measurement has an associated operation. As we will discuss, our postulate then becomes a preservation principle for the transformations associated with pure measurements. Then, we will examine some consequences of this principle by considering the examples of the triangle, the square and the pentagon. We will prove the main result in Section \ref{result-2-result}. At first, we will show that every polytopic theory satisfying the preservation principle has a certain property. We will say that such a polytope is \emph{uniformly pyramidal}. Finally, we show that every uniformly pyramidal polytope is a simplex, and therefore the theory in question is classical.

\subsection{Transformations and a preservation principle}
\label{non-dist-section}

In section \ref{trafo-section}, we have seen how transformations naturally arise in the context of measurements. We have defined a transformation $T$ to be a linear map $T: A \rightarrow B$ which is positive and does not increase the norm. However, we did not make an explicit assumption about how measurements are related to transformations. The reason why we did not make such an assumption in Section \ref{trafo-section} already is that we did not know at that point that we have to make the restriction to pure measurements. In Section \ref{pure-meas-section}, we have argued that whenever we make statements about post-measurement states, we have to restrict ourselves to pure measurements. Now that we know about this restriction, we resume our discussion of transformations.

We have discussed in section \ref{trafo-section} that measurements encompass two aspects: the statistic of the measurement and the disturbance of the statistic of subsequent measurements. The first aspect is covered by treating a measurement as a set of effects. In Section \ref{result-1-section}, we technically did not make explicit statements about how a post-measurement state should look  like. The only principle which explicitly said something about post-measurement states was Postulate \ref{rep-postulate}, but technically, we did not assume it. Instead, the idea of repeatability was, to some extend, implicitly contained in Postulate \ref{subsp-postulate}. The fact that we did not explicitly deal with post-measurement states allowed us to treat measurements as sets of effects. We did not have to deal with transformations. In this section, however, we \emph{do} make explicit statements about post-measurement states, so we have to set up how transformations are related to measurements in our framework. We do this by the following assumption. Since this assumption is part of the framework, we state it as an ``Assumption'' rather than as a ``Postulate''.

\begin{ass}
\label{trafo-ass}
When a pure measurement $\mathcal{M} = \{ e_1, \ldots, e_n \}$ is performed on an abstract state space $A$, then the state of the system is transformed according to an operation $\mathcal{O}_\mathcal{M} = \{ T_{e_1}, \ldots, T_{e_k} \}$ from $A$ to $A$ which induces the measurement $\mathcal{M}$ (recall from Definition \ref{trafo-def} that this means $e_k = (u_A \circ T_{e_k})$ for all $k \in \{1, \ldots, n\}$). We call this the \textbf{operation $\mathcal{O}_\mathcal{M}$ associated with the pure measurement $\mathcal{M}$}\index{operation!associated with a pure measurement}, and for every $k \in \{1, \ldots, n\}$, we say that $T_{e_k}$ is the \textbf{transformation associated with the pure effect $e_k$}.
\end{ass}

This fixes our framework. Now we are ready to explain the postulate of this section. The idea is the following. Suppose that a physical system has some property. When we say property, we mean something that we can find out by a measurement. Moreover, when we say that the system \emph{has} the property, we mean that the system is in a state such that the measurement outcome that reveals this property has probability one, i.e. we can predict the outcome of the measurement \emph{with certainty}. It is natural to assume that in this case, we can perform the measurement without altering the state. In formal terms, this reads as follows.

\begin{post}
\label{nondist-postulate}
Let $A$ be an abstract state space, let $\mathcal{M} = \{ e_1, \ldots, e_n \}$ be a pure measurement on $A$ and let $\mathcal{O}_\mathcal{M} = \{ T_{e_1}, \ldots, T_{e_n} \}$ be the associated operation. If $\omega \in \Omega_A$ is a state with a certain outcome, i.e. $e_k(\omega) = 1$ for some $k \in \{ 1, \ldots, n \}$, then the associated transformation $T_{e_k}$ satisfies $T_{e_k}(\omega) = \omega$, i.e.
\begin{align}
(u_A \circ T_{e_k})(\omega) = e_k(\omega) = 1 \quad \Rightarrow \quad T_{e_k}(\omega) = \omega \,. \nonumber
\end{align}
In other words, each transformation $T_{e_k} \in \mathcal{O}_\mathcal{M}$ leaves the face $F_{e_k}$ associated with the effect $e_k$ invariant.
\end{post}

Quantum theory obeys this postulate. Consider a spin-$\frac{1}{2}$ particle that is heading towards a Stern-Gerlach device which measures the spin of the particle with respect to the $z$-axis. We can predict the outcome of the measurement with certainty if the spin state of the particle is either $ \rho = | \uparrow_z \rangle \langle \uparrow_z |$ or $\rho = | \downarrow_z \rangle \langle \downarrow_z |$. For example, if the state is $ \rho = | \uparrow_z \rangle \langle \uparrow_z |$, then it is certain that the Stern-Gerlach measurement reveals the outcome ``up'' (which is associated to the projector $P^z_{\uparrow} = | \uparrow_z \rangle \langle \uparrow_z |$) since $\tr(P_{\uparrow}^z \rho) = 1$. In this case, the von Neumann-L\"uders projection (\ref{von-neumann-lueders}) does not alter the state:
\begin{align}
\rho_\text{post} = \frac{P_{\uparrow}^z \rho P_{\uparrow}^z}{\tr(P_{\uparrow}^z \rho)} = \rho \,. \nonumber
\end{align}

Classical theory satisfies Postulate \ref{nondist-postulate} as well. It is instructive to see this geometrically. We examine the case of a classical theory with three pure states.

\begin{ex}[Postulate \ref{nondist-postulate} in a classical theory]
Consider the polygon model corresponding to $n=3$. In this model, $\Omega_A$ is a triangle and therefore a simplex, so it is a classical theory. Consider the pure effect $\overline{e_2} = u_A - e_2$. The face $F_{\overline{e_2}}$ associated with $\overline{e_2}$ is an edge of the triangle (see Figure \ref{triangle-figure}). The effect $\overline{e_2}$ is contained in the pure measurement $\{e_2, \overline{e_2}\}$. According to Postulate \ref{nondist-postulate}, the transformation $T_{\overline{e_2}}$ associated with $\overline{e_2}$ leaves the face $F_{\overline{e_2}}$ invariant, i.e.
\begin{align}
T_{\overline{e_2}}(F_{\overline{e_2}}) = F_{\overline{e_2}} \,. \label{triangle-eq-1}
\end{align}
On the other hand, the effect $\overline{e_2}$ vanishes on the face $\overline{F_{\overline{e_2}}}$ = $F_{e_2}$ opposite to $F_{\overline{e_2}}$. The transformation $T_{\overline{e_2}}$ induces the effect $\overline{e_2}$, i.e. $\overline{e_2} = (u_A \circ T_{\overline{e_2}})$ (c.f. Definition \ref{trafo-def} and Assumption \ref{trafo-ass}). Thus, $u_A(T_{\overline{e_2}}(\omega)) = 0$ for all $\omega \in F_{e_2}$. This means that
\begin{align}
T_{\overline{e_2}}(F_{e_2}) = \{0\} \,, \label{triangle-eq-2}
\end{align}
i.e. the face $F_{e_2}$ opposite to $F_{\overline{e_2}}$ is mapped to the zero vector. The transformation $T_{\overline{e_2}}$ is linear, so Equations (\ref{triangle-eq-1}) and (\ref{triangle-eq-2}) fully determine $T_{\overline{e_2}}: A \rightarrow A$ since $\dim(\spa(F_{\overline{e_2}})) + \dim(\spa(F_{e_2})) = 3 = \dim(A)$.

\begin{figure}[htb]
\centering

\begin{pspicture}[showgrid=false](-2.2,-0.5)(4.3,3)
\psset{viewpoint=20 50 10,Decran=120}
\psset{solidmemory}
\psSolid[object=new,linewidth=0.5\pslinewidth,
action=draw*,
name=C,
fcol=6 (.15 setfillopacity Gray),
sommets=
0 0 0 
-0.707107 1.22474 1 
-0.707107 -1.22474 1 
1.41421 0 1 
-0.235702 0.408248 0.333333 
-0.235702 -0.408248 0.333333 
0.471405 0 0.333333 
0.235702 -0.408248 0.666667 
0.235702 0.408248 0.666667 
-0.471405 0 0.666667 
0 0 1, 
faces={
[0 4 8 6]
[0 6 7 5]
[0 5 9 4]
[7 6 8 10]
[8 4 9 10]
[9 5 7 10]
[1 2 3]
}]%
\psPoint(-0.235702, -0.408248, 0.333333){g}
\psdots[dotsize=0.15](g)
\uput[dl](g){$e_2$}
\psPoint(0.235702, 0.408248, 0.666667){h}
\psdots[dotsize=0.15](h)
\uput[r](h){$\overline{e_2}$}
\psSolid[object=line, linewidth=2\pslinewidth,args=-0.707107 1.22474 1 1.41421 0 1]
\psSolid[object=line, linewidth=2\pslinewidth,args=-0.707107 1.22474 1.01 1.41421 0 1.01]
\psPoint(0.653553, 0.612372, 1.02){i}
\uput[dl](i){$F_{\overline{e_2}}$}
\psPoint(0,0,0.95){a}
\uput[ur](a){$\Omega_A$}
\psPoint(-0.707107, -1.22474, 1){j}
\psdots[dotsize=0.15](j)
\uput[u](j){$F_{e_2} = \{\omega_2\}$}
\end{pspicture}
\begin{pspicture}[showgrid=false](-2,-0.5)(2.4,3)
\psset{viewpoint=20 50 10,Decran=120}
\psset{solidmemory}
\psSolid[object=new,linewidth=0.5\pslinewidth,
action=draw*,fcol=2 (0.15 setfillopacity Gray),
name=C,
sommets=
0 0 0 
-0.707107 1.22474 1 
-0.707107 -1.22474 1 
1.41421 0 1 
-0.235702 0.408248 0.333333 
-0.235702 -0.408248 0.333333 
0.471405 0 0.333333 
0.235702 -0.408248 0.666667 
0.235702 0.408248 0.666667 
-0.471405 0 0.666667 
0 0 1, 
faces={
[0 2 1]
[0 3 2]
[0 1 3]
[1 2 3]
}]%
\psSolid[object=line, linewidth=2\pslinewidth,args=-0.707107 1.22474 1 1.41421 0 1]
\psSolid[object=line, linewidth=2\pslinewidth,args=-0.707107 1.22474 1.01 1.41421 0 1.01]
\psSolid[object=line, linewidth=2\pslinewidth,args=-0.707107 1.22474 1.005 1.41421 0 1.005]
\psSolid[object=line, linewidth=2\pslinewidth,args=-0.707107 1.22474 1.015 1.41421 0 1.015]
\pstThreeDPut(0.6,0,1.8){$T_{\overline{e_2}}(\Omega_A)$}
\psPoint(0, 0, 0){t}
\psdots[dotsize=0.15](t)
\uput[d](t){$T_{\overline{e_2}}(F_{e_2}) = \{0\}$}
\psPoint(0.653553, 0.612372, 1.02){b}
\uput[ur](b){$T_{\overline{e_2}}(F_{\overline{e_2}}) = F_{\overline{e_2}}$}
\uput[90](-2,0.5){$\Omega_A^{\leq 1}$}
\end{pspicture}

\caption{The transformation $T_{\overline{e_2}}$ maps the set $\Omega_A$ (the gray triangle on the left) into the set $\conv(F_{\overline{e_2}} \cup \{0\})$ (the gray triangle on the right).}
\label{triangle-figure}
\end{figure}
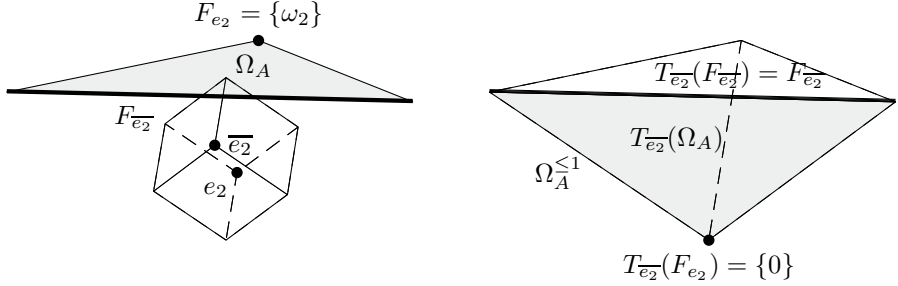

For the understanding of the proof idea in Section \ref{result-2-result}, it is worth visualizing the action of the map $T_{\overline{e_2}}$. It maps the triangle $\Omega_A$ to the triangle $\conv(F_{\overline{e_2}} \cup \{0\})$ (gray regions in Figure \ref{triangle-figure}), while the edge $F_{\overline{e_2}}$ is fixed. Therefore, it maps $\Omega_A$ to a subset of $\Omega_A^{\leq 1}$ (the upside down tetrahedron on the right of Figure \ref{triangle-figure}), so it is a positive map which does not increase the norm (c.f. Definition \ref{trafo-def}). Thus, $T_{\overline{e_2}}$ is indeed a transformation. We can say that for every face $F_e$ associated to a pure effect $e$, Postulate \ref{nondist-postulate} says that the associated transformation $T_e$ maps the set $\conv(F_e \cup \overline{F_e})$ into the set $\conv(F_e \cup \{0\})$. In the above case, $\conv(F_e \cup \overline{F_e}) = \conv(F_{\overline{e_2}} \cup F_{e_2}) = \Omega_A$. \hfill $\blacksquare$

\end{ex}

As we have seen, classical theory and quantum theory satisfy Postulate \ref{nondist-postulate}. In contrast, all polytopic theories which are not classical (i.e. where the set of normalized states $\Omega_A$ is a polytope but not a simplex) violate this postulate, as we will prove in the next Section.

Before we prove this result in full generality, we investigate some consequences of Postulate~\ref{nondist-postulate} by examining two examples. This helps to understand Postulate~\ref{nondist-postulate} geometrically.

\begin{ex}[Violation of Postulate \ref{nondist-postulate} in the square and the pentagon model]
\label{postulate-4-violation}
Once again, we consider the polygon models that we have already encountered in Examples \ref{polygon-models}, \ref{associated-faces-ex}, \ref{subsp-ex} and \ref{square-discr-vio}. We show that the square and the pentagon violate Postulate \ref{nondist-postulate}.
\begin{enumerate}[(a)]

\item Consider the polygon model corresponding to $n=4$ (the square). As we have mentioned in earlier examples, the face $F_{e_1}$ associated with the pure effect $e_1$ is an edge of the square (see Figure \ref{square-figure-4}). The pure effect $e_1$ is contained in the pure measurement $\{e_1, e_3\}$. \emph{Assume} that Postulate \ref{nondist-postulate} holds (we will see below that this leads to a contradiction). This implies that the transformation $T_{e_1}$ associated with $e_1$ leaves the face $F_{e_1}$ invariant:
\begin{align}
T_{e_1} (F_{e_1}) = F_{e_1} \,. \label{square-eq-1}
\end{align}
On the other hand, the effect $e_1$ vanishes on the face $\overline{F_{e_1}}$ opposite to $F_{e_1}$. The transformation $T_{e_1}$ induces the effect $e_1$, i.e. $e_1 = (u_A \circ T_{e_1})$. Thus, $u_A(T_{e_1}(\omega)) = 0$ for all $\omega \in \overline{F_{e_1}}$. This means that
\begin{align}
T_{e_1} (\overline{F_{e_1}}) = \{0\} \,, \label{square-eq-2}
\end{align}
i.e. the opposite face $\overline{F_{e_1}}$ is mapped to the zero vector. This leads to a contradiction: Since $T_{e_1}$ is linear, the equations (\ref{square-eq-1}) and (\ref{square-eq-2}) cannot be satisfied at the same time. Equation (\ref{square-eq-1}) specifies $T_{e_1}$ on a two-dimensional subspace (namely $\spa(F_{e_1})$). The transformation $T_{e_1}: A \rightarrow A$ is a map on a three-dimensional space $A$, so the linearity of $T_{e_1}$ only leaves the freedom to specify the image $T_{e_1}(\omega)$ for one point $\omega$ outside of $\spa(F_{e_1})$. It is not possible to map the whole edge $\overline{F_{e_1}}$ to zero (see Figure \ref{square-figure-4}).

\begin{figure}[htb]
\centering

\begin{pspicture}[showgrid=false](-2,-0.5)(4,3.5)
\psset{viewpoint=26 10 5,Decran=70}
\psset{solidmemory}
\psSolid[object=new,linewidth=0.5\pslinewidth,
action=draw*,
name=B,
sommets= 
0 0 0 
0.420448 0.420448 0.5 
-0.420448 0.420448 0.5 
-0.420448 -0.420448 0.5 
0.420448 -0.420448 0.5 
0 0 1 
0 1.18921 1 
-1.18921 0 1 
0 -1.18921 1 
1.18921 0 1, 
faces={
[1 4 0]
[0 2 1]
[0 3 2]
[0 4 3]
[1 2 5]
[2 3 5]
[3 4 5]
[4 1 5]
[6 7 8 9]}]%
\psPoint(0.420448, 0.420448, 0.5){e1}
\psdots[dotsize=0.15](e1)
\uput[dl](e1){$e_1$}
\psSolid[object=line, linewidth=2\pslinewidth,args=0 1.18921 1 1.18921 0 1]
\psSolid[object=line, linewidth=2\pslinewidth,args=0 1.18921 1.01 1.18921 0 1.01]
\psPoint(0.727673, 0.528686, 1.03){g}
\uput[ur](g){$F_{e_1}$}
\psPoint(-0.420448, -0.420448, 0.5){e3}
\psdots[dotsize=0.15](e3)
\uput[ur](e3){$\overline{e_1} = e_3$}
\psSolid[object=line, linewidth=2\pslinewidth,args=-1.18921 0 1 0 -1.18921 1]
\psSolid[object=line, linewidth=2\pslinewidth,args=-1.18921 0 1.01 0 -1.18921 1.01]
\psPoint(-0.594604, -0.594604, 1){h}
\uput[u](h){$\overline{F_{e_1}} = F_{e_3}$}
\end{pspicture}
\begin{pspicture}[showgrid=false](-2,-0.5)(2,3.5)
\psset{viewpoint=26 10 5,Decran=70}
\psset{solidmemory}
\psSolid[object=new,linewidth=0.5\pslinewidth,
action=draw*,
name=B,
sommets= 
0 0 0 
0.420448 0.420448 0.5 
-0.420448 0.420448 0.5 
-0.420448 -0.420448 0.5 
0.420448 -0.420448 0.5 
0 0 1 
0 1.18921 1 
-1.18921 0 1 
0 -1.18921 1 
1.18921 0 1, 
faces={
[6 7 8 9]}]%
\psSolid[object=line, linewidth=2\pslinewidth,args=0 1.18921 1 1.18921 0 1]
\psSolid[object=line, linewidth=2\pslinewidth,args=0 1.18921 1.01 1.18921 0 1.01]
\psSolid[object=line, linewidth=2\pslinewidth,args=0 1.18921 1.005 1.18921 0 1.005]
\psPoint(0.727673, 0.528686, 1.01){g}
\psdots[dotsize=0](g)
\uput[u](g){$T_{e_1}(F_{e_1}) = F_{e_1}$}
\psPoint(0.348311, -0.840896, -0.08){lu}
\psSolid[object=line,linestyle=dotted,args=1.18921 0 1 0.348311 -0.840896 -0.08]
\psSolid[object=line,linestyle=dotted,args=0 1.18921 1 -0.840896 0.348314 -0.08]
\psSolid[object=line,args=0.348311 -0.840896 -0.08 -0.840896 0.348314 -0.08]
\psSolid[object=line,args=0.348311 -0.840896 -0.07 -0.840896 0.348314 -0.07]
\psPoint(0.2,0,0){h}
\psPoint(0,0,0.0){h'}
\psdots[dotsize=0.18](h')
\uput[dl](h){$T_{e_1}(\overline{F_{e_1}})$}
\uput[u](h'){$T_{e_1}(\frac{1}{2} \omega_2 + \frac{1}{2} \omega_3)$}
\end{pspicture}

\caption{The equations (\ref{square-eq-1}) and (\ref{square-eq-2}) cannot be satisfied at the same time. If $F_{e_1}$ is left invariant, then the edge $\overline{F_{e_1}}$ cannot be mapped to zero. Only one point of $\overline{F_{e_1}}$ can be mapped to zero, which is chosen to be $\frac{1}{2} \omega_2 + \frac{1}{2} \omega_3$ here.}
\label{square-figure-4}
\end{figure}
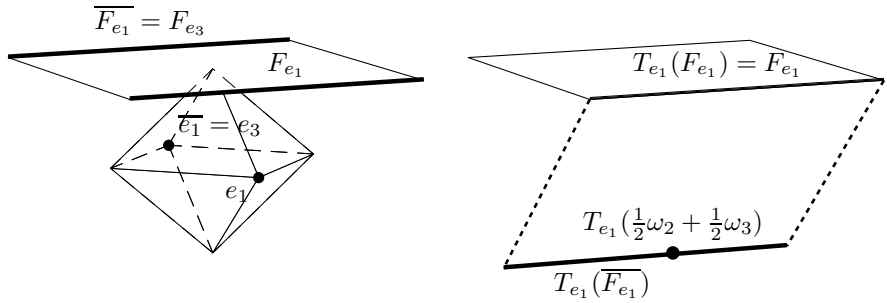

This example shows that if $\dim(\spa(F_{e})) + \dim(\spa(\overline{F_e})) = 4 > \dim(A)$ for some pure effect $e \in E_A$, then Postulate \ref{nondist-postulate} cannot be satisfied.

\item Now we consider the regular pentagon, i.e. the $n=5$ polygon model. As we have seen in Example \ref{associated-faces-ex}, the face $\overline{F_{e_3}}$ associated with the pure effect $\overline{e_3}$ is an edge of the pentagon (see Figure \ref{pentagon-figure-2}). The opposite face $\overline{\overline{F_{e_3}}} = F_{e_3}$ is a point, so in the case of the pentagon, one has that $\dim(\spa(\overline{F_{e_3}})) + \dim(\spa(F_{e_3})) = 3 = \dim(A)$. Thus, the conditions
\begin{align}
&T_{\overline{e_3}}(\overline{F_{e_3}}) = \overline{F_{e_3}} \,, \label{pentagon-eq-1} \\
&T_{\overline{e_3}}(F_{e_3}) = \{0\} \label{pentagon-eq-2}
\end{align}
do not lead to a contradiction. They can both be satisfied at the same time. However, the pentagon also violates Postulate \ref{nondist-postulate}. By the linearity of the transformation $T_{\overline{e_3}}$, the equations (\ref{pentagon-eq-1}) and (\ref{pentagon-eq-2}) fully determine $T_{\overline{e_3}}$. As one can see in Figure \ref{pentagon-figure-2}, $T_{\overline{e_3}}$ maps $\Omega_A$ to some set $T_{\overline{e_3}}(\Omega_A)$ which is not contained in $\Omega_A^{\leq 1}$ (and therefore it is not contained in $A_+$). This means that $T_{\overline{e_3}}$ is not a transformation according to Definition \ref{trafo-def}: It maps the set $\Omega_A \subset A_+$ to a set which is not contained in $A_+$, so $T_{\overline{e_3}}$ is not positive.

\begin{figure}[htb]
\centering

\begin{pspicture}[showgrid=false](-2,-0.5)(4,3.5)
\psset{viewpoint=26 10 5,Decran=70}
\psset{solidmemory}
\psSolid[object=new,linewidth=0.5\pslinewidth,
action=draw*,
name=A,
sommets= 
0 0 0 
0.152217 0.468477 0.414214 
-0.402248 0.29225 0.447214 
-0.402248 -0.29225 0.447214 
0.153645 -0.472871 0.447214 
0.497206 0 0.447214 
-0.153645 -0.472871 0.552786 
0.402248 -0.29225 0.552786 
0.402248 0.29225 0.552786 
-0.153645 0.472871 0.552786 
-0.497206 0 0.552786 
0 0 1 
0.343561 1.05737 1 
-0.899454 0.653491 1 
-0.899454 -0.653491 1 
0.343561 -1.05737 1 
1.11179 0 1, 
faces={
[0 1 8 5]
[0 5 7 4]
[0 4 6 3]
[0 3 10 2]
[0 2 9 1]
[6 4 7 11]
[7 5 8 11]
[8 1 9 11]
[9 2 10 11]
[10 3 6 11]
[12 13 14 15 16]}]%
\psPoint(-0.402248, -0.29225, 0.447214){e3}
\psdots[dotsize=0.15](e3)
\uput[u](e3){$e_3$}
\psPoint(-0.899454, -0.653491, 1){w3}
\psdots[dotsize=0.15](w3)
\uput[ur](w3){$F_{e_3} = \{ \omega_3 \}$}
\psPoint(0.402248, 0.29225, 0.552786){e3bar}
\psdots[dotsize=0.15](e3bar)
\uput[d](e3bar){$\overline{e_3}$}
\psSolid[object=line, linewidth=2\pslinewidth,args=0.343561 1.05737 1 1.11179 0 1]
\psSolid[object=line, linewidth=2\pslinewidth,args=0.343561 1.05737 1.01 1.11179 0 1.01]
\psPoint(0.727673, 0.728686, 1.03){f}
\uput[dr](f){$\overline{F_{e_3}}$}
\end{pspicture}
\begin{pspicture}[showgrid=false](-2,-0.5)(1.5,3.5)
\psset{viewpoint=26 10 5,Decran=70}
\psset{solidmemory}
\psSolid[object=new,linewidth=0.5\pslinewidth,
action=draw**,
name=A,
sommets= 
0 0 0 
0.152217 0.468477 0.414214 
-0.402248 0.29225 0.447214 
-0.402248 -0.29225 0.447214 
0.153645 -0.472871 0.447214 
0.497206 0 0.447214 
-0.153645 -0.472871 0.552786 
0.402248 -0.29225 0.552786 
0.402248 0.29225 0.552786 
-0.153645 0.472871 0.552786 
-0.497206 0 0.552786 
0 0 1 
0.343561 1.05737 1 
-0.899454 0.653491 1 
-0.899454 -0.653491 1 
0.343561 -1.05737 1 
1.11179 0 1, 
faces={
[0 16 15]
[0 15 14]
[0 14 13]
[0 13 12]
[0 12 16]
[12 13 14 15 16]}]%
\psSolid[object=line,linestyle=dotted,args=0.343561 1.05737 1 -0.323511 1.05737 0.4]
\psSolid[object=line,linestyle=dotted,args=-0.323511 1.05737 0.4 0 0 0]
\psSolid[object=line,linestyle=dotted,args=1.11179 0 1 0.905649 -0.634423 0.4]
\psSolid[object=line,linestyle=dotted,args=0.905649 -0.634423 0.4 0 0 0]
\psSolid[object=line, linewidth=2\pslinewidth,args=0.343561 1.05737 1 1.11179 0 1]
\psSolid[object=line, linewidth=2\pslinewidth,args=0.343561 1.05737 1.01 1.11179 0 1.01]
\psSolid[object=line, linewidth=2\pslinewidth,args=0.343561 1.05737 1.005 1.11179 0 1.005]
\psPoint(0.627673, 0.628686, 1.03){f}
\psdots[dotsize=0](f)
\uput[d](f){$T_{\overline{e_3}}(\overline{F_{e_3}}) = \overline{F_{e_3}}$}
\psPoint(0,0,0){e3}
\psdots[dotsize=0.15](e3)
\uput[d](e3){$T_{\overline{e_3}}(F_{e_3}) = 0$}
\psPoint(0.373561, -1.05737, 0.86){omegaleq}
\psdots[dotsize=0](omegaleq)
\uput[d](omegaleq){$\Omega_A^{\leq1}$}
\psPoint(-0.323511, 1.05737, 0.4){omegaim}
\psdots[dotsize=0](omegaim)
\uput[dl](omegaim){$T_{\overline{e_3}}(\Omega_A)$}
\psPoint(0,0,0.9){omega}
\psdots[dotsize=0](omega)
\uput[u](omega){$\Omega_A$}
\end{pspicture}

\caption{For the pure effect $\overline{e_3}$ in the pentagon model, there is a linear map $T_{\overline{e_3}}$ which satisfies $T_{\overline{e_3}}(\overline{F_{e_3}}) = \overline{F_{e_3}}$ and $T_{\overline{e_3}}(F_{e_3}) = 0$. However, this map is not a transformation, since $T_{\overline{e_3}}(\Omega_A)$ (dotted lines) is not contained in $A_+$, so $T_{\overline{e_3}}$ is not positive. \hfill $\blacksquare$}
\label{pentagon-figure-2}
\end{figure}
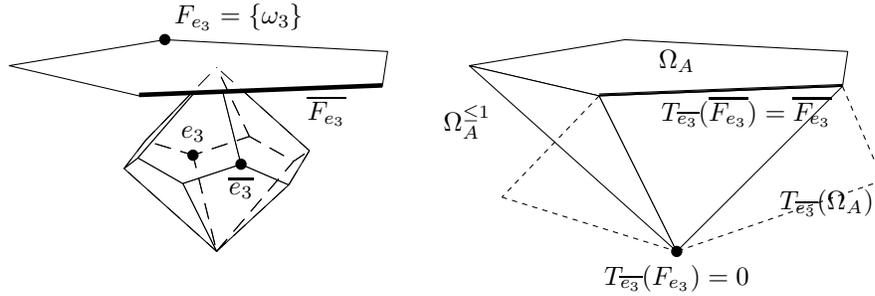

\end{enumerate}
\end{ex}
Example \ref{postulate-4-violation} shows two things that can lead to a contradiction to Postulate~\ref{nondist-postulate}. In the square example (a), we have seen that it might be that there is some pure effect $e \in E_A$ such that $T_e(F_e) = F_e$ and $T_e(\overline{F_e}) = \{0\}$ cannot be satisfied at the same time. In the pentagon example (b), the equations $T_e(F_e) = F_e$ and $T_e(\overline{F_e}) = \{0\}$ can both be satisfied, but the only map which satisfies these two equations is not positive. In Section \ref{result-2-result}, we will see that any polytopic theory either has one of these two problems or it is a simplex.

With a little thought, one can already guess from example \ref{postulate-4-violation} (b) why every polygon $\Omega_A$ that satisfies Postulate \ref{nondist-postulate} must be a simplex (i.e. a triangle). We only make a very rough argumentation here. Assume the case where $T_e(F_e) = F_e$ and $T_e(\overline{F_e}) = \{0\}$ can both be satisfied. For the map $T_e$ to be positive, it has to fit $\Omega_A$ into $\conv(F_e \cup \{0\})$. The map $T_e: A \rightarrow A$ is not bijective since $T_e(\overline{F_e}) = 0$, but the restriction of  $T_e$ to $\Omega_A$ is an affine bijection, so one can visualize that $T_e(\Omega_A) \subset \conv(F_e \cup \{0\})$ can only be satisfied if $\Omega_A = \conv(F_e \cup \overline{F_e})$. In this case, $\Omega_A$ is a triangle because $F_e$ is a facet of $\Omega_A$, i.e. a line-segment, and $\overline{F_e}$ is a point.

\subsection{The Result}
\label{result-2-result}

In this section, we prove that every polytopic theory which satisfies Postulate \ref{nondist-postulate} is a classical theory. In other words, if $A$ is an abstract state space such that $\Omega_A$ is a polytope and such that Postulate \ref{nondist-postulate} is satisfied, then $\Omega_A$ is a simplex. We prove this result in two main steps. In the first step, we define what it means for a polytope to be \emph{uniformly pyramidal}. We prove that for a polytopic theory $A$ which satisfies Postulate \ref{nondist-postulate}, the polytope $\Omega_A$ is uniformly pyramidal. This is the ``physical'' part of the proof. The second part is the proof that every uniformly pyramidal polytope is a simplex.

\begin{defi}
A polytope $P$ with a facet $B$ is called \textbf{pyramidal at $B$} if $P$ is a pyramid with base $B$ and some apex $a_B$, i.e. $P = \conv(B \cup \{a_B\})$ (c.f. Example \ref{pyramid-example}). A polytope is \textbf{uniformly pyramidal} if it is pyramidal at every facet.
\end{defi}

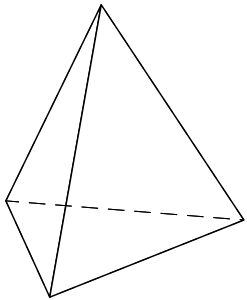
\begin{figure}[htb]
\centering

\begin{pspicture}[showgrid=false](-1.5,-1.5)(2,2.2) 
\psset{viewpoint=10 15 15 rtp2xyz,Decran=100} 
\psSolid[Decran=9, linewidth=0.7\pslinewidth,
object=tetrahedron, 
r=2.5, 
action=draw*]%
\end{pspicture}

\caption{The tetrahedron is a uniformly pyramidal polytope. Each face is a base of the tetrahedron as a pyramid, where the apex is given by the point opposite to the face.}
\label{tetrahedron-pyramid-figure}
\end{figure}

It will turn out that simplices are the only polytopes that are uniformly pyramidal, so they are the only example that we can make. The property of being uniformly pyramidal is easily visualized on a polyhedron (see Figure \ref{tetrahedron-pyramid-figure}).

The first step is to prove that every polytopic theory $A$ which satisfies Postulate \ref{nondist-postulate} has a uniformly pyramidal set of states $\Omega_A$. Before we prove this, it is convenient to introduce the following notation.

\begin{notation}
For a subset $M \subset \Omega_A$ of the set of normalized states, we define $M^{\leq 1} := \{ \omega \in A_+ \mid \exists \sigma \in M, \alpha \in [0,1]: \omega = \alpha \sigma \}$.
\end{notation}

\begin{lemma}
\label{post-4-implies-unif-pyr}
Let $A$ be a polytopic theory which satisfies Postulate \ref{nondist-postulate}. Then $\Omega_A$ is uniformly pyramidal.
\end{lemma}

\begin{proof}
We prove this lemma in three steps.
\begin{enumerate}[(i)]
\item In the first step, we consider an arbitrary facet $F$ of $\Omega_A$. We know from Proposition \ref{rest-possible} that there is a pure effect $f \in E_A$ such that $F = F_f$. We show that the opposite face $\overline F := F_{\overline f}$ (c.f. Definition \ref{opposite-face-def}) is a vertex of the polytope, i.e. $\overline F = \{a_F\}$ for some $a_F \in \Omega_A$.
\item In the second step, we consider the transformation $T_f$ associated with the effect $f$. We show that the restriction of $T_f$ to $\conv(F \cup \{a_F\})$ is a bijection from $\conv(F \cup \{a_F\})$ to $F^{\leq 1}$.
\item Finally, we show that this implies that $\Omega_A = \conv(F \cup \{a_F\})$. Since $F$ is an arbitrary facet, this is sufficient to show that $\Omega_A$ is uniformly pyramidal.
\end{enumerate}
Now we prove the three steps.
\begin{enumerate}[(i)]
\item Let $F$ be a facet of $\Omega_A$. According to Proposition \ref{rest-possible}, there is a pure effect $f \in E_A$ such that $F = F_f := \{ \omega \in \Omega_A \mid f(\omega) = 1 \}$. By Proposition \ref{u-f-pure}, the complementary effect $\overline f := u_A - f$ is pure as well, so we have that $f$ is contained in the pure measurement $\mathcal{M} = \{ f, \overline f \}$. By Assumption~\ref{trafo-ass}, $\mathcal{M}$ is associated with an operation $\mathcal{O}_\mathcal{M} = \{ T_f, T_{\overline f} \}$ which induces $\mathcal{M}$. Postulate~\ref{nondist-postulate} tells us that that the transformation $T_f$ satisfies $T_f(\omega) = \omega$ for all $\omega \in F$. By the linearity of $T_f$, we have that $T_f$ restricted on $\spa(F)$ is the identity map on $\spa(F)$:
\begin{align}
T_f|_{\spa(F)} = I_{\spa(F)} \,. \label{t_f-on-spa-f}
\end{align}
The effect $\overline f$ is pure, so by Corollary \ref{nonempty-face}, there is a non-empty face $\overline{F}$ such that $\overline{F} = F_{\overline f} := \{ \omega \in \Omega_A \mid \overline f (\omega) = 1 \}$ (in Definition \ref{opposite-face-def}, we called this the face $\overline F$ opposite to $F$). We want to show that the restriction of $T_f$ to $\spa(\overline F)$ is the zero-map. 

The effects $\mathcal{M} = \{ f, \overline f \}$ form a measurement, so the fact that $\overline f$ takes the value 1 on $\overline F$ implies that $f$ takes the value 0 on $\overline F$ since $(f + \overline f)(\omega) = u_A(\omega) = 1$ for all $\omega \in \Omega_A$. The transformation $T_f$ induces the effect $f$ (by Assumption~\ref{trafo-ass}), so $f(\omega) = u_A(T_f(\omega)) = 0$ for all $\omega \in \overline{F}$. This implies that $T_f(\omega) = 0 \in \Omega_A^{\leq 1}$ is the zero-state for all $\omega \in \overline F$. By the linearity of $T_f$, this implies that the restriction of $T_f$ to $\spa(\overline F)$ is the zero-map:
\begin{align}
T_f|_{\spa(\overline F)} = 0_{\spa(\overline F)} \,. \label{t_f-on-spa-overline-f}
\end{align}
Our assumptions require that both (\ref{t_f-on-spa-f}) and (\ref{t_f-on-spa-overline-f}) are satisfied. This is only possible if $\spa(F) \cap \spa(\overline F) = \{ 0 \}$. Let $d = \dim A$. Then $\Omega_A$ is a $(d-1)$-polytope. We have assumed that $F$ is a facet of $\Omega_A$, so $\aff(F)$ is $(d-2)$-dimensional. Thus, $\dim(\spa(F)) = d-1$. To fulfill $\spa(F) \cap \spa(\overline F) = \{ 0 \}$, $\spa(\overline F)$ must be one-dimensional. This implies that $\overline F$ consists of only one point which we call $a_F$, i.e. $\overline F = \{ a_F \}$.

\item The transformation $T_f$ is linear. By Theorem \ref{convex-linear-affine}, this implies that $T_f$ is an affine map. Thus, the restriction $T_f|_{\aff(F \cup \{a_F\})}$ of $T_f$ to $\aff(F \cup \{a_F\})$ is an affine map. We have seen in Proposition \ref{aff-inj-prop} that an affine map is injective if it maps affinely independent points to affinely independent points. This is the case for $T_f|_{\aff(F \cup \{a_F\})}$: Equation (\ref{t_f-on-spa-f}) implies that $T_f|_{\aff(F)} = I_{\aff(F)}$, and $a_F$ (which is affinely independent of $\aff(F)$ by Proposition \ref{face-cap-aff}) is mapped to the zero vector (which is affinely independent of $T_f(\aff(F)) = \aff(F)$). Thus, $T_f|_{\aff(F \cup \{a_F\})}$ is an affine injection. We have that $\conv(F \cup \{a_F\}) \subset \aff(F \cup \{a_F\})$, which implies that $T_f|_{\conv(F \cup \{a_F\})}$ is an injection. The map $T_f$ is affine, so by Proposition \ref{aff-is-arbit-aff}, one has that $T_f(\conv(F \cup \{a_F\})) = \conv(T_f(F \cup \{a_F\})) = \conv(T_f(F) \cup \{T_f(a_F)\}) = \conv(F \cup \{0\}) = F^{\leq 1}$. All in all, we have
\begin{align}
&T_f(\conv(F \cup \{a_F\})) = F^{\leq 1} \,, \label{t_f-affine} \\
&T_f|_{\aff(F \cup \{a_F\})}: \aff(F \cup \{a_F\}) \rightarrow \aff(F^{\leq 1}) \text{ is bijective} \,. \label{is-bijective}
\end{align}

\item The set of normalized states $\Omega_A$ is a polytope, so $\Omega_A^{\leq 1} = \conv(\Omega_A \cup \{ 0 \})$ is a pyramid (c.f. Example \ref{pyramid-example}). The set $F$ is a facet of $\Omega_A$, so we get from Lemma \ref{pyramid-facet-lemma} that $F^{\leq 1} = \conv(F \cup \{0\})$ is a facet of $\Omega_A^{\leq 1}$. By Proposition \ref{face-cap-aff}, this implies that
\begin{align}
F^{\leq 1} = \aff(F^{\leq 1}) \cap \Omega_A^{\leq 1}\,. \label{fleq-eq}
\end{align}
Assume that there is a $\sigma \in \Omega_A, \sigma \notin \conv(F \cup \{a_F\})$. The proof is finished if we manage to show that this is impossible (since this implies $\conv(F \cup \{a_F\}) \subset \Omega_A$, and $\conv(F \cup \{a_F\}) \subset \Omega_A$ is trivial). The subset $F$ is a facet of $\Omega_A$ (i.e. $F$ is of one dimension less than $\Omega_A$) and $a_F \in \Omega_A$ is affinely independent of $F$ (by Proposition \ref{face-cap-aff}), so $\aff(F \cup \{a_F\}) = \aff(\Omega_A)$. Thus, $\sigma \in \Omega_A$ implies
\begin{align}
\sigma \in \aff(F \cup \{a_F\})\,. \label{sigma-in-aff-cup}
\end{align}
Note that for any subset $M$ of a vector space, one has that
\begin{align}
\aff(M) = \aff(\conv(M)) \,. \label{affconvaff}
\end{align}
From (\ref{sigma-in-aff-cup}), we get that
\begin{align}
T_f(\sigma) \in T_f(\aff(F \cup \{a_F\})) \overset{(\ref{affconvaff})}{=} T_f(\aff(\conv(F \cup \{a_F\}))) \,. \label{laulum}
\end{align}
The map $T_f$ is affine, so by Proposition \ref{aff-is-arbit-aff}, we have that
\begin{align}
\begin{split}T_f(\aff(\conv(F \cup \{a_F\}))) &= \aff(T_f(\conv(F \cup \{a_F\}))) \\
&\overset{(\ref{t_f-affine})}{=} \aff(F^{\leq1}) \,. \end{split} \label{lelaum}
\end{align}
Equations (\ref{laulum}) and (\ref{lelaum}) imply
\begin{align}
T_f(\sigma) \in \aff(F^{\leq 1}) \,. \label{t_f-sigma-in-aff}
\end{align}
The map $T_f: A \rightarrow A$ is a transformation (c.f. Definition \ref{trafo-def}), so $\sigma \in \Omega_A$ implies
\begin{align}
T_f(\sigma) \in \Omega_A^{\leq 1} \,. \label{t_f-sigma-in-om}
\end{align}
(\ref{t_f-sigma-in-om}), (\ref{t_f-sigma-in-aff}) and (\ref{fleq-eq}) imply that
\begin{align}
T_f(\sigma) \in F^{\leq 1} \,. \label{t_f-sigma-in-fleq}
\end{align}
Now we combine (\ref{t_f-affine}), (\ref{is-bijective}), (\ref{sigma-in-aff-cup}) and (\ref{t_f-sigma-in-fleq}) to see that $\sigma \in \conv(F \cup \{ a_F \})$. This is a contradiction to the assumption that $\sigma \notin \conv(F \cup \{a_F\})$. This completes the proof.
\hfill \qedhere

\end{enumerate}

\end{proof}

We have just done the first of the two main steps to prove our main result. The second step is the following lemma.

\begin{lemma}
\label{unif-pyr-implies-simplex}
Every uniformly pyramidal polytope is a simplex.
\end{lemma}

\begin{proof}
We prove this lemma by induction over the dimension of the polytope.

\begin{itemize}

\item \textbf{Base case: $P$ is a uniformly pyramidal 1-polytope} \\
This case is trivial. Every 1-polytope is a 1-simplex.

\item \textbf{Inductive step: $P$ is a uniformly pyramidal $d$-polytope} \\
Let $d \in \mathbb{N}, d \geq 2$ and assume that every uniformly pyramidal $(d-1)$-polytope is a simplex (induction hypothesis). Let $P$ be a uniformly pyramidal $d$-polytope, let $B \subset P$ be a facet of $P$. We show that $B$ is a simplex. By the definition of a simplex (c.f. Example \ref{simplex-example}), this is sufficient to show that $P$ is a simplex since the apex is $a_B$ is affinely independent of $B$.

The polytope $P$ is uniformly pyramidal. Therefore, there is an apex $a_B \in P$ such that $P = \conv(B \cup \{a_B\})$. This implies that the number of vertices of $P$ is $n+1$, where $n$ is the number of vertices of $B$. Let $F \subset B$ be a facet of $B$. By Lemma \ref{pyramid-facet-lemma}, $G := \conv(F \cup \{a_B\})$ is a facet of $P$. The polytope $P$ is uniformly pyramidal, so $P = \conv(G \cup \{a_G\})$ for some $a_G \in P$. $P$ has $n+1$ vertices, so $G$ has $n$ vertices. Thus, $F$ has $n-1$ vertices, which means that $B$ has only one more vertex than $F$. This implies that $B = \conv(F \cup \{a_F\})$ for some $a_F \in B$. The set $F$ is an arbitrary face of $B$, so we have just shown that $B$ (which is a $(d-1)$-polytope) is uniformly pyramidal. By the induction hypothesis, every uniformly pyramidal $(d-1)$-polytope (in particular $B$) is a simplex. \hfil \qedhere

\end{itemize}
\end{proof}

\begin{thm}[Main Result]
Let $A$ be a polytopic theory that satisfies Postulate \ref{nondist-postulate}. Then $A$ is a classical theory.
\end{thm}

\begin{proof}
This theorem is a corollary of Lemma \ref{post-4-implies-unif-pyr} and Lemma \ref{unif-pyr-implies-simplex}.
\end{proof}

\newpage
\section{Conclusion and Outlook}
\label{outlook-section}

In the first part of this thesis, we introduced the mathematics of convex sets, polytopes and generalized probabilistic theories, and we provided a derivation of the framework of abstract state spaces. To our knowledge, such a derivation is new. Along this introduction and framework derivation, we developed most of the techniques that we used to infer the results of Part \ref{result-part} of this thesis.

In Section \ref{result-1-section}, we have seen that within polytopic theories, we can characterize classical theory by Postulates \ref{rep-postulate}, \ref{subsp-postulate} and \ref{discr-postulate}, i.e. by repeatability, a subspace principle and a state discrimination principle. To our knowledge, such an interpretation has not been made before. From a technical viewpoint, the inside into the mathematics of abstract state spaces gained by the proof in Section \ref{result-1-result} is arguably small since, in particular, Postulate \ref{subsp-postulate} is quite strong. From a physical viewpoint, however, the ideas presented in Section \ref{result-1-section} could provide interesting ideas in future attempts to infer quantum theory from physical postulates (c.f. Section \ref{the-role-section} and the beginning of Part \ref{result-part}).

In Section \ref{result-2-section}, we have seen that the very reasonable and seemingly weak Postulate \ref{nondist-postulate} is in fact strong enough to rule out all polytopic theories except for classical theories. This might be an interesting starting point in future attempts to derive quantum theory from physical assumptions. In particular, it is a new approach insofar as to our knowledge, there has not been any consideration of post-measurement states in generalized probabilistic theories so far. Within polytopic theories, we have seen that it is sufficient to assume Postulate~\ref{nondist-postulate} alone to rule out ``unreasonable theories'' (assuming that the only reasonable theories are classical theory and quantum theory). In a more general class of theories, it is presumably necessary to assume a few more postulates to rule out ``unreasonable theories''. This is where Postulates \ref{rep-postulate}, \ref{subsp-postulate} and \ref{discr-postulate} might come into play.

However, from the technical side, if one wants to generalize the results of this thesis to broader classes of abstract state spaces (i.e. to non-polytopic theories), one needs to develop other techniques than those presented in this thesis. Our derivations rely quite strongly on the fact that we are dealing with polytopes. For example, the concept of a facet of a polytope is very central in our derivations. In general, convex sets do not have something like a facet. The largest proper face of the Bloch sphere, for example, is a single point and therefore a zero-dimensional face of a two-dimensional convex set. All our techniques involving facets do not apply in this case. Nonetheless, it might be that the same Postulates, by the use of other techniques, have strong implications in broader classes of generalized probabilistic theories.

\newpage

\begin{appendix}

\section*{Appendix}
\addcontentsline{toc}{section}{Appendix}

\section{Compact convex sets in infinite-dimensional topological vector spaces}
\label{appendix-a}

In Section \ref{general-convex-section}, we have seen that by a theorem of Minkowski (Theorem \ref{minkowski}), every compact convex subset of a finite-dimensional vector space is the convex hull of its extreme points. This theorem can be generalized to infinite-dimensional topological vector spaces, where it is known as the Krein-Milman Theorem. However, in that case, the statement is weaker: one needs to take the closure. 

\begin{thm}[Krein-Milman, see {\cite[Theorem VIII.4.4]{Werner}}]
\label{krein-milman}\index{Krein Milman Theorem}
Let $V$ be a locally convex Hausdorff topological vector space, let $C$ be a compact convex subset of $V$. Then $C$ is the closure of the convex hull if its extreme points,
\begin{align}
C = \overline{\conv(\ext(C))} \,. \nonumber
\end{align}
\end{thm}

The Krein-Milman Theorem allows us to consider polytopes in a more general setting. In Section \ref{polytope-section}, we have defined polytopes as compact convex subsets of \emph{finite}-dimensional vector spaces with finitely many extreme points. It is very convenient to assume that the vector space is finite-dimensional since every finite-dimensional real vector space is essentially the same as $\mathbb{R}^n$ for some $n$. As we mentioned in Section \ref{general-convex-section}, there is only one (Hausdorff) topology in finite dimensions. Now we want to turn to the question how restrictive it is to assume that the vector space is finite-dimensional. It turns out that as long as we stick to ``well-behaved'' topologies, this assumption does not cause a loss of generality. When we say ``well-behaved'', we mean a locally convex Hausdorff topology. The following proposition states this formally.

\begin{prop}
\label{polytope-finite-dim-v}
Let $P$ be a compact convex subset of a Hausdorff locally convex topological vector space $V$ with finitely many extreme points. Then $P$ is contained in a finite-dimensional subspace of $V$.
\end{prop}

\begin{proof}
By the Krein-Milman Theorem \ref{krein-milman}, $P = \overline{\conv(\ext(P))}$. Obviously, $\overline{\conv(\ext(P))} \subset \overline{\spa(\ext(P))}$. We have assumed that $\ext(P)$ is finite, so $\spa(\ext(P))$ is a finite-dimensional subspace of $V$. Finite-dimensional subspaces of locally convex Hausdorff spaces are always closed \cite[Aufgabe VIII.6.5 (b)]{Werner}, so $P \subset \spa(\ext(P))$ which is finite-dimensional.
\end{proof}

Thus, defining polytopes as subsets of finite-dimensional vector spaces causes no loss of generality (as long as we ignore topologies which are not Hausdorff and locally convex). This has an interesting consequence for our framework. In Section \ref{abs-st-sp-section}, we have made Assumption \ref{finite-dim-ass} which states that the vector space containing the convex subset of normalized states is finite-dimensional. In Sections \ref{result-1-section} and \ref{result-2-section}, we have considered polytopic theories, i.e. theories with finitely many pure states. Proposition \ref{polytope-finite-dim-v} allows us to weaken our assumptions when we deal with theories with only finitely many pure states: If we make Assumptions \ref{canc-conv-ass} and \ref{compact-ass}, we have that the set of normalized states of a theory is a compact convex subset of a real vector space. When we restrict to theories with finitely many pure states, we do not have to make Assumption \ref{finite-dim-ass} since Proposition \ref{polytope-finite-dim-v} tells us that such a theory always takes place in a finite-dimensional subspace of a real vector space. This is interesting insofar as pure states have quite a direct physical interpretation (c.f. Section \ref{maximal-knowledge-section}), whereas the physical interpretation of the dimension of the vector space is not so clear.

\section{The equivalence of compact convex sets and abstract state spaces}
\label{appendix-b}

In Section \ref{abs-st-sp-section}, we have listed three assumptions concerning the set of normalized states. These three assumptions state that the set of normalized states is a compact convex subset of a finite-dimensional vector space. Then we have sketched how such a compact convex set gives rise to an abstract state space. We mentioned that there is a one-to-one correspondence between compact convex subsets of finite-dimensional vector spaces and abstract state spaces. In this appendix, we formulate and prove this correspondence mathematically. This needs some preparation.

At first we need to know when two cones are equivalent from the viewpoint of ordered vector spaces. Such an equivalence is established by an ``isomorphism in the cone sense''. Such an isomorphism is called an \emph{order isomorphism} and is defined as follows.

\begin{defi}
Let $V$ and $W$ be two ordered vector spaces. A linear operator $\phi: V \rightarrow W$ is called an \textbf{order-isomorphism}\index{order-isomorphism} if
\begin{enumerate}[(a)]
\item $\phi$ is bijective and
\item  $v \geq 0$ in $V$ if and only if $\phi(v) \geq 0$ in $W$, i.e. $v \geq_{V_+} 0 \Leftrightarrow \phi(v) \geq_{W_+}$.
\end{enumerate}
In other words, a linear operator $V$ is an order isomorphism if $\phi$ maps the cone $V_+$ bijectively onto the cone $W_+$. If there exists an order isomorphism between $V$ and $W$, we say that $V$ and $W$ are \textbf{order-isomorphic}\index{order-isomorphic}.
\end{defi}

In Section \ref{cones-and-ovs}, we have introduced the notion of a base $\mathcal{B}$ of a cone $V_+$ as a convex subset of $V_+$ which fully characterizes the cone structure of $V_+$. Thus, one would expect that in the case where two cones are equivalent in the cone sense (i.e. where the two cones are order-isomorphic), one can find bases of the cones which are equivalent in the convex set sense (i.e. two bases which are convex-isomorphic) and vice versa. This is indeed the case.

\begin{prop}
\label{order-isom-convex-isom}
Let $(V, \leq_{V_+})$, $(W, \leq_{W_+})$ be finite-dimensional ordered vector spaces with positive cones $V_+$, $W_+$, respectively. Assume that $V_+$ admits a base. Then the following are equivalent:
\begin{enumerate}[(a)]
\item $V$ and $W$ are order-isomorphic.
\item There is a base $\mathcal{B}_V$ of $V_+$ and a base $\mathcal{B}_W$ of $W_+$ such that $\mathcal{B}_V$ and $\mathcal{B}_W$ are convex-isomorphic, and $\dim V = \dim W$.
\end{enumerate}
\end{prop}

\begin{proof}
We prove the two implications (a) $\Rightarrow$ (b) and (b) $\Rightarrow$ (a) separately.

\begin{itemize}

\item (a) $\Rightarrow$ (b): Let $V$ be an ordered vector space such that $V_+$ has a base $\mathcal{B}_V$. Let $\phi: A \rightarrow B$ be an order isomorphism from $V$ to some other ordered vector space $W$. $V_+$ is a base of $V$, so  by Theorem \ref{base-strictly-positive}, there is a strictly positive linear functional $f \in V^*$ and a $\alpha > 0$ such that
\begin{align}
\mathcal{B}_V = \{ v \in V_+ \mid f(v) = \alpha \} \,. \label{blulu}
\end{align}
Let $g$ be the linear functional on $W$ given by $g(w) = (f \circ \phi^{-1})(w)$ for all $w \in W$. The functional $g$ is obviously strictly positive on $W$, so $\mathcal{B}_W = \{ w \in W_+ \mid g(w) = \alpha \}$ is a base for $W_+$ by Theorem \ref{base-strictly-positive}. Now we show that $\mathcal{B}_V$ and $\mathcal{B}_W$ are convex-isomorphic. We have that
\begin{align}
&\phi(\mathcal{B}_V) \overset{(\ref{blulu})}{=} \{ \phi(v) \mid v \in V_+, f(v) = \alpha \} \nonumber \\
&= \{ w \in W_+ \mid \underbrace{\exists v \in V_+: v = \phi^{-1}(w)}_{(\star)}, f(v) = f(\phi^{-1}(w)) = \alpha \}\,. \nonumber
\end{align}
By the definition of an order-isomorphism, $\phi|_{V_+}: V_+ \rightarrow W_+$ is bijective, so $(\star)$ is trivially satisfied for all $w \in W_+$. This means that
\begin{align}
\phi(\mathcal{B}_V) = \{ w \in W_+ \mid (f \circ \phi^{-1})(w) = g(w) = \alpha \} = \mathcal{B}_W \,. \nonumber
\end{align}
The order-isomorphism $\phi$ is linear and therefore convex-linear, so $\mathcal{B}_V$ and $\mathcal{B}_W$ are convex-isomorphic. The spaces $V$ and $W$ have the same dimension by the definition of an order-isomorphism.

\item (b) $\Rightarrow$ (a): Let $V_+$ be the positive cone of an ordered vector space $V$ with base $\mathcal{B}_V$. Let $W_+$ be the positive cone of some other ordered vector space $W$ with base $\mathcal{B}_W$. Assume that there is a convex-linear map $\chi: V \rightarrow W$ such that $\chi(\mathcal{B}_V) = \mathcal{B}_W$ and $\chi_{\spa(\mathcal{B}_V)}$ is bijective (i.e. $\mathcal{B}_V$ and $\mathcal{B}_W$ are convex-isomorphic). We have assumed that $\dim V = \dim W$, so we can assume that $\chi$ is bijective. According to Theorem \ref{convex-linear-affine}, there is a linear map $L: V \rightarrow W$ and a vector $c \in W$ such that $\chi$ is given by $\chi(v) = L(v) + c$ for all $v \in V$. The map $\chi$ is bijective, so $L$ must be invertible. In the following, we construct a map $\phi: V \rightarrow W$ and show that it is an order-isomorphism.

By Theorem \ref{base-strictly-positive}, there is a linear functional $f \in V^*$ such that
\begin{align}
\mathcal{B}_V = \{ v \in V_+ \mid f(v) = 1 \} \,. \label{bvbasis}
\end{align}
Let $b_V \in \mathcal{B}_V$ be any vector in the base of $V_+$. We have that $f(b_V) = 1$, so $b_V \notin \ker(f)$. Thus, any vector $v \in V$ can be expressed as a unique linear combination $v = v_\text{ker} + \alpha b_V$, where $v_\text{ker} \in \ker(f)$. We define the map $\phi: V \rightarrow W$ by
\begin{align}
\phi(v) = \phi(v_\text{ker} + \alpha b_V) = L(v_\text{ker}) + \alpha (L(b_V) + c)\,. \nonumber
\end{align}
The map $\phi$ is linear since $\phi$ is the linear extension of the map
\begin{align}
\left\{ \begin{array}{cll} v_\text{ker} & \mapsto & L(v_\text{ker}) \text{ for all } v_\text{ker} \in \ker(f) \\ b_V & \mapsto & L(b_V) + c \text{ for the particular vector } b_V \end{array} \right. \nonumber
\end{align}
To show that $\phi$ is an order isomorphism, we have to show that (i) $\phi$ is bijective and that (ii) $v \geq_{V_+} 0$ if and only if $\phi(v) \geq_{W_+} 0$. To show these two properties, it is useful to notice that the restriction $\phi|_{\mathcal{B}_V}$ of $\phi$ to $\mathcal{B}_V$ coincides with $\chi|_{\mathcal{B}_V}$. To see this, let $v \in \mathcal{B}_V$. We said that any vector $v \in V$ can be decomposed as $v = v_\text{ker} + \alpha b_V$, where $v_\text{ker} \in \ker(f)$. We assume that $v \in \mathcal{B}_V$ so by (\ref{bvbasis}), we have that $1 = f(v) = f(\alpha b_V + v_\text{ker}) = \alpha f(b_V) = \alpha$ and thus $v = b_V + v_\text{ker}$. This implies
\begin{align}
\phi(v) &= \phi(v_B + v_\text{ker}) = \phi(v_B) + \phi(v_\text{ker}) = L(v_B) + c + L(v_\text{ker}) \nonumber \\
&= L(v_B + v_\text{ker}) + c = L(v) + c \nonumber \\
&= \chi(v) \,. \nonumber
\end{align}
Now we show that
\begin{enumerate}[(i)]
\item $\phi$ is bijective: Note that by Theorem \ref{base-strictly-positive}, there is a linear functional $g \in W^*$ such that $\mathcal{B}_W = \{ w \in W_+ \mid g(w) = 1 \}$. From $\phi|_{\mathcal{B}_V} = \chi|_{\mathcal{B}_V}$ it follows that the functional $f$ satisfying (\ref{bvbasis}) can be chosen to be $f = g \circ \phi$.\footnote{Note that if $\spa(\mathcal{B}_V) \neq V$, the functional $f$ is not unique.} For any $v_\text{ker} \in \ker(f)$, we have that $g(\phi(v_\text{ker})) = f(v_\text{ker}) = 0$, so $\phi$ maps $\ker(f)$ bijectively to $\ker(g)$ (since $L$ is bijective). It remains to be shown that $\phi(b_V) \notin \ker(g)$. This follows from $g(\phi(b_V)) = f(b_V) = 1$, so $\phi$ is bijective.
\item $v \geq_{V_+} 0 \Leftrightarrow \phi(v) \geq_{W_+} 0$: Trivially, we have that $v = 0 \Leftrightarrow \phi(v) = 0$. For $v >_{V_+} 0 \Rightarrow \phi(v) >_{W_+} 0$, note that we have shown above that $\phi|_{\mathcal{B}_V} = \chi|_{\mathcal{B}_V}$. The set $\mathcal{B}_V$ is a base of $V_+$, so if $v >_{V_+} 0$ (i.e.$ v \in V_+ \backslash \{0\})$, then $v = \alpha_v b_v$ for some $\alpha_v > 0$, $b_v \in \mathcal{B}_V$. Therefore, $v >_{V_+} 0$ implies
\begin{align}
\phi(v) = \underbrace{\alpha_v}_{>0} \underbrace{\chi(b_v)}_{\in \mathcal{B}_W} >_{W_+} 0 \quad \text{by property (\ref{ovs1}).} \nonumber
\end{align}
By the bijectivity of $\phi$ we can make the same argumentation in the reverse direction to get that $\phi(v) >_{W_+} 0 \Rightarrow v >_{V_+} 0$. This completes the proof. \hfill \qedhere
\end{enumerate}

\end{itemize}

\end{proof}

The idea we want to develop in the following is that a set of normalized states, i.e. a compact convex subset of a finite-dimensional real vector space, gives rise to a ``unique'' abstract state space. In this context, ``unique'' means unique up to order-isomorphism. Vice versa, we will see that every abstract state space gives rise to a unique compact convex subset of a finite-dimensional vector space. Thus, the two structures are in a one-to-one-correspondence. Before we can prove this correspondence, we need to prove three lemmas.

\begin{lemma}
\label{hyperplane-zero}
Let $H$ be an affine hyperplane in a vector space $V$. Then the following two statements are equivalent:
\begin{enumerate}[(a)]
\item $\spa(H) = V$.
\item The zero vector is not contained in $H$.
\end{enumerate}
\end{lemma}

\begin{proof}
We prove the two implications (a) $\Rightarrow$ (b) and $(b) \Rightarrow (a)$ separately.
\begin{itemize}

\item (a) $\Rightarrow$ (b): Suppose that the zero vector is contained in $H$. By the definition of an affine hyperplane, there is a linear functional $f \in V^*$ and a $k \in \mathbb{R}$ such that $H = \{ v \in V \mid f(v) = k \}$. The zero vector is in $H$, so $k = 0$ since $f(0) = 0$ by the linearity of $f$. Let $x \in \spa(H)$, i.e.
\begin{align}
x = \sum\limits_i \alpha_i v_i \quad \text{for some numbers } \alpha_i \in \mathbb{R} \text{ and some vectors } v_i \in H. \nonumber
\end{align}
Then
\begin{align}
f(x) = \sum\limits_i \alpha_i \underbrace{f(v_i)}_{0} = 0 \,, \nonumber
\end{align}
so $x \in H$ and therefore $\spa(H) = H \neq V$.

\item (b) $\Rightarrow$ (a): Suppose that the zero vector is not in $H$, i.e. the zero vector is affinely independent of $H$. It is easy to verify that
\begin{align}
\aff(H \cup \{0\} ) = \aff(\aff(H) \cup \{0\}) = \{ \alpha v \mid \alpha \in \mathbb{R}, v \in \aff(H) \} = \spa(H) \,. \nonumber
\end{align}
The zero vector is affinely independent of $H$, so $\aff(H \cup \{ 0 \} ) = V$ since $H$ is an affine hyperplane in $V$ (and therefore has codimension one). This implies that $\spa(H) = V$. \hfill \qedhere

\end{itemize}
\end{proof}

\begin{lemma}
\label{formalism-lemma-1}
Let $C \subset V$ be a convex subset of a finite-dimensional real vector space $V$. Then there is a convex subset $C'$ of some finite-dimensional real vector space $W$ such that the following properties are satisfied:
\begin{align}
&\bullet \quad C \text{ and } C' \text{ are convex-isomorphic,} \nonumber \\
&\bullet \quad \aff(C') \text{ is an affine hyperplane of } W, \nonumber \\
&\bullet \quad \spa(C') = W. \nonumber
\end{align}
\end{lemma}

\begin{proof}
We perform a proof by cases.
\begin{enumerate}

\item \emph{$\aff(C)$ is an affine hyperplane of $V$}:
\begin{enumerate}
\item \emph{$\aff(C)$ does not contain the zero vector}: In this case, Lemma \ref{hyperplane-zero} applies and we see that $\spa(C) = V$. This means we can simply choose $C' = C$ in $W = V$ and all required properties are satisfied.
\item \emph{$\aff(C)$ contains the zero vector}: In this case, we translate $C$ by a nonzero vector normal to $\aff(C)$ to obtain a set $C'$ which does not contain the zero vector and for which $\aff(C')$ is an affine hyperplane. The set $C'$ is convex-isomorphic to $C$ because a translation is convex-linear by Theorem \ref{conv-lin-aff}. Lemma \ref{hyperplane-zero} applies to $C'$ and we see that $\spa(C') = W$, so $C'$ satisfies all the desired properties.
\end{enumerate}

\item \emph{$\aff(C) = V$}: In this case, consider $C$ as a subset of $\widetilde V = V \oplus \mathbb{R}$, i.e. consider $\widetilde C = \{ (v, 0) \in V \oplus \mathbb{R} \mid v \in C \} \subset \widetilde V$. The set $\widetilde C$ is convex-isomorphic to $C$ and the affine hull $\aff(\widetilde C)$ of $\widetilde C$ is an affine hyperplane of $\widetilde V$. For $\widetilde C$, either case 1. (a) or 1. (b) is on hand, for which we have proved the claim.

\item \emph{$\dim(\aff(C)) < \dim V - 2$.} In this case, consider $C$ as a subset of the vector space $\spa(C)$. Then $C$ applies to case 1. or 2., for which we have already proved the claim. \hfill \qedhere

\end{enumerate}
\end{proof}

\begin{lemma}
\label{formalism-lemma-2}
Let $\mathcal{B}$ be a convex subset of a finite-dimensional real vector space $V$. Assume that there exists a linear functional $f \in V^*$ and a $k > 0$ such that $f(v) = k$ for all $v \in \mathcal{B}$.
Then the following hold:
\begin{enumerate}[(a)]
\item  $K := \{ \alpha v \mid \alpha \geq 0, v \in \mathcal{B} \}$ is a cone and $\mathcal{B}$ is a base for $K$.
\item If in addition $\spa(\mathcal{B}) = V$, then $K$ is generating.
\end{enumerate}
\end{lemma}

\begin{proof}
We prove the two claims separately.
\begin{enumerate}[(a)]

\item We verify properties (\ref{cone1}), (\ref{cone2}) and (\ref{cone3}) to show that $K$ is a cone.
\begin{itemize}
\item (\ref{cone1}): Let $v, w \in K$. We have that $v = a \hat v$, $w = b \hat w$ for some $a, b > 0$ and some $\hat v, \hat w \in \mathcal{B}$. Then
\begin{align}
v + w = a \hat v + b \hat w = (a + b) \underbrace{\left( \frac{a}{a+b} \hat v + \frac{b}{a+b} \hat w \right)}_{\in \mathcal{B} \text{ since } \mathcal{B} \text{ is convex}} \in K\,. \nonumber
\end{align}
\item (\ref{cone2}): For any $\alpha > 0$ and $v \in K$, we have that $\alpha K \in K$ by the definition of $K$, so (\ref{cone2}) is satisfied.
\item (\ref{cone3}): Let $v \in K$ be nonzero and let $\hat x \in K$ such that $x = c \hat x$ for some $c > 0$. If $-x \in K$, then $- \hat x \in \mathcal{B}$. This would contradict the assumption that there is a linear functional $f \in V^*$ and a $k > 0$ such that $f(v) = k$ for all $v \in \mathcal{B}$. This implies $K \cap (-K) = \{0\}$.
\end{itemize}
Therefore, $K$ is a cone. To see that $\mathcal{B}$ is a base of $K$, note that for every $y \in K \ \{0\}$, there is a $\beta > 0$ and a $\hat y \in \mathcal{B}$ such that $y = \beta \hat y$. If moreover $y = \gamma \hat z$, we have
\begin{align}
0 = f(y) - f(y) = \beta f(\hat y) - \gamma f(\hat z) = (\beta - \gamma) k, \quad k > 0 &\quad \Rightarrow \quad \beta = \gamma \nonumber \\
&\quad \Rightarrow \quad \hat y = \hat z\,. \nonumber
\end{align}
Therefore, the representation $y = \beta \hat y$ is unique and we have that $\mathcal{B}$ is a basis for the cone $K$.

\item Assume that $\spa(\mathcal{B}) = V$. It is easily verified that $K - K = \spa(K) = \spa(\mathcal{B})$. Thus, $K$ is generating. \qedhere

\end{enumerate}
\end{proof}

With these three lemmas, we are ready to prove one direction of the equivalence between compact convex subsets of finite-dimensional vector spaces and abstract state spaces. It reads as follows.

\begin{prop}
\label{formalism-prop}
Let $\widetilde{\Omega_A}$ be a set of normalized states which satisfies Assumptions \ref{canc-conv-ass}, \ref{finite-dim-ass} and \ref{compact-ass}. Then there is an abstract state space $(A, A_+, u_A)$ such that $\Omega_A := \{ \omega \in A_+ \mid u_A(\omega) = 1 \}$ is convex-isomorphic to $\widetilde{\Omega_A}$. This abstract state space is unique up to order-isomorphism.
\end{prop}

\begin{proof}
Assumptions \ref{canc-conv-ass}, \ref{finite-dim-ass} and \ref{compact-ass} imply that $\widetilde \Omega_A$ is a compact convex subset of a finite-dimensional vector space. By virtue of Lemma \ref{formalism-lemma-1}, there is a convex subset $\Omega_A$ of a finite-dimensional real vector space $A$ which is convex-isomorphic to $\widetilde{\Omega_A}$ such that $\spa(\Omega_A) = A$ and such that $\aff(\Omega_A)$ is an affine hyperplane of $A$. A convex-linear bijection is obviously a homeomorphism, so $\Omega_A$ is compact. It holds that $\spa(\aff(\Omega_A)) = \spa(\Omega_A) = A$, so by Lemma \ref{hyperplane-zero}, the zero vector is not contained in $\aff(\Omega_A)$. This implies that there exists a linear functional $f \in V^*$ and a $k > 0$ such that $f(v) = k$ for all $\omega \in \Omega_A$. This allows us to apply Lemma \ref{formalism-lemma-2} to get a generating cone $A_+$ with base $\Omega_A$. By Theorem \ref{compact-to-closed-thm} and the compactness of $\Omega_A$, we have that $A_+$ is closed. Theorem \ref{base-strictly-positive} implies the existence of a strictly positive linear functional $u_A$ with $\Omega_A = \{ \omega \in A_+ \mid u_A(\omega) = 1 \}$. By Theorem \ref{strictly-positive-order-unit}, $u_A$ is an order unit. This shows the existence of an abstract state space $(A, A_+, u_A)$ with the claimed properties. The uniqueness up to order isomorphism follows from Proposition \ref{order-isom-convex-isom} and the fact that being convex-isomorphic is a transitive relation.
\end{proof}

This establishes that the structure of a set of normalized states that satisfies Assumptions \ref{canc-conv-ass}, \ref{finite-dim-ass} and \ref{compact-ass} gives rise to the structure of an abstract state space (unique up to order isomorphism). To have an equivalence of the two structures, we also want the converse. This is provided by the following theorem.

\begin{thm}[{\cite[Chapter 3.1]{Aliprantis-Tourky}}]
If $V_+$ is a closed and generating cone of a finite-dimensional real vector space $V$ and $u_V$ is a strictly positive linear functional on $V$, then the set $\{ v \in V \mid u_V(v) = 1 \}$ is a compact base for $V_+$.
\end{thm}

This shows that there is a one-to-one-correspondence between
\begin{enumerate}[(a)]
\item the structure of a compact convex subset of a finite-dimensional real vector space and
\item the structure of an abstract state space. 
\end{enumerate}
In section \ref{abs-st-sp-section}, the physical and mathematical assumptions for the set of normalized states defined structure (a), but the one-to-one-correspondence allows us to treat them as structure (b). This is why we can deal with abstract state spaces when we talk about generalized probabilistic theories.

\end{appendix}

\newpage
\quad
\newpage

\section*{Symbols and abbreviations}
\addcontentsline{toc}{section}{Symbols and abbreviations}

\vspace{1cm}

\begin{tabular}{lp{9.5cm}}

\multicolumn{2}{l}{{\large \textbf{Quantum theory}}} \\

\hline

$\mathcal{H}$ & The Hilbert space associated with a quantum system \\
$\Herm(\mathcal{H})$ & The vector space of Hermitian operators on Hilbert space $\mathcal{H}$ \\
$\rho$ & A density operator on $\mathcal{H}$ \\
$\mathcal{S}(\mathcal{H})$ & Set of density operators on Hilbert space $\mathcal{H}$ \\
$f_P$ & The linear functional $\rho \mapsto \tr(P \rho)$ for some Hermitian operator $P$ on $\mathcal{H}$ \\
$\supp{\rho}$ & The support of a density operator \\

\hline

\vspace{0.5cm} \\

\multicolumn{2}{l}{{\large \textbf{Convex sets, polytopes and cones}}} \\

\hline

$L_{x, y}$ & The line segment connecting the points $x$ and $y$ \\
$\ext(C)$ & The set of extreme points of a convex subset $C$ of a real vector space \\
$\aff(M)$ & The affine hull of a subset $M$ of a real vector space \\
$\conv(M)$ & The convex hull of a subset $M$ of a real vector space \\
$\spa(M)$ & The linear span of a subset $M$ of a vector space \\
$\cone(M)$ & The conical hull of a subset $M$ of a vector space \\
$\overline M$ & The closure of a subset of a topological space \\
$(V, \leq)$ & An ordered vector space \\
$V_+$ & The positive cone of an ordered vector space $V$ \\
$\leq_{K}$ & The cone order over a vector space $V$ induced by a cone $K \subset V$ \\
$V_+^*$ & The dual cone of a cone $V_+ \subset V$ \\

\hline

\end{tabular}

\newpage

\begin{tabular}{lp{9.5cm}}

\multicolumn{2}{l}{{\large \textbf{Abstract state spaces}}} \\

\hline

$(A, A_+, u_A)$ & An abstract state space \\
$A, B$ & The vector space associated with an abstract state space (often used as a symbol for the whole triple $(A, A_+, u_A)$ or $(B, B_+, u_B)$, respectively) \\
$A_+$ & The positive cone in an abstract state space $A$ \\
$u_A$ & The distinguished order unit on an abstract state space $A$ \\
$\Omega_A$ & The set of normalized states in an abstract state space $A$ \\
$\Omega_A^{\leq 1}$ & The set of subnormalized states in an abstract state space $A$ \\
$E_A$ & The set of effects on an abstract state space $A$ \\

\hline

\end{tabular}

\newpage
\addcontentsline{toc}{section}{Index}
\printindex

\clearpage
\addcontentsline{toc}{section}{References}

\bibliography{master_thesis-bib}
\bibliographystyle{amsalpha}

\end{document}